\documentclass[a4paper,11pt]{article}

\usepackage{jheppub} 

\usepackage[T1]{fontenc} 

\usepackage{relsize}
\usepackage{pifont}
\usepackage{xcolor}
\usepackage[utf8]{inputenc}
\usepackage{cleveref}
\usepackage{amsmath,bm}
\usepackage{amssymb}
\usepackage{pifont}
\usepackage{graphicx}
\usepackage{multirow}
\usepackage[normalem]{ulem}
\usepackage{comment}
\usepackage{booktabs}
\usepackage{slashed}
\usepackage{makecell}

\usepackage{tikz-feynman}
\tikzfeynmanset{compat=1.1.0}

\newcommand{\braket}[1]{\ensuremath{\left\langle #1 \right\rangle}}

\def\dcs{{\Phi^{\pm \pm}}}
\def\tcs{{\Phi^{\pm \pm \pm}}}
\def\qcs{{\Phi^{\pm \pm \pm \pm}}}

\def\r {\rightarrow}

\newdimen\arrowsize
\newdimen\mylw
\pgfkeys{/my arrows/chemeq/.style={draw,thick,double distance=2pt,onearc-onearc}}
\pgfkeys{/my arrows/size/.code={\pgfsetarrowoptions{onearc}{#1}}}
\def\myalw{.4pt}
\pgfarrowsdeclare{onearc}{onearc}{%
  \mylw=\myalw
  \pgfarrowsleftextend{-\pgfgetarrowoptions{onearc}-.5\mylw}
  \pgfarrowsrightextend{0pt}
}{%
  \pgfsetdash{}{0pt}
  \mylw=\pgflinewidth
  \pgfsetlinewidth{\myalw}
  \advance\arrowsize by.5\pgflinewidth
  \pgfpathmoveto{\pgfpoint{-\pgfgetarrowoptions{onearc}}{-\pgfgetarrowoptions{onearc}-.5\mylw}}%
  \pgfpatharc{180}{90}{\pgfgetarrowoptions{onearc}}
  \pgfusepathqstroke
}

\title{\boldmath  
	 Neutrino masses from new Weinberg-like operators: Phenomenology of TeV scalar multiplets}
	 
\author[a,b]{Alessio Giarnetti,}
\author[c,d]{Juan Herrero-García,}
\author[a,b]{Simone Marciano,}
\author[a,b]{Davide Meloni,}
\author[c,d]{and Drona Vatsyayan}

\affiliation[a]{Dipartimento di Matematica e Fisica, Universit\'a di Roma Tre,
Via della Vasca Navale 84, 00146, Roma, Italy}

\affiliation[b]{INFN Sezione di Roma Tre, Via della Vasca Navale 84, 00146, Roma, Italy}

\affiliation[c]{Departament de Física Teòrica, Universitat de València, 46100 Burjassot, Spain}

\affiliation[d]{Instituto de Física Corpuscular (CSIC-Universitat de València),
Parc Científic UV, C/Catedrático José Beltrán, 2, E-46980 Paterna, Spain}

\emailAdd{alessio.giarnetti@uniroma3.it}
\emailAdd{juan.herrero@ific.uv.es}
\emailAdd{simone.marciano@uniroma3.it}
\emailAdd{davide.meloni@uniroma3.it}
\emailAdd{drona.vatsyayan@ific.uv.es}

\abstract{The unique dimension-$5$ effective operator, $LLHH$, known as the Weinberg operator, generates tiny Majorana masses for neutrinos after electroweak spontaneous symmetry breaking. If there are new scalar multiplets that take vacuum expectation values (VEVs), they should not be far from the electroweak scale. Consequently, they may generate new dimension-$5$ Weinberg-like operators which in turn also contribute to Majorana neutrino masses. In this study, we consider scenarios with one or two new scalars up to quintuplet $SU(2)$ representations. We analyse the scalar potentials, studying whether the new VEVs can be induced and therefore are naturally suppressed, as well as the potential existence of pseudo-Nambu-Goldstone bosons. Additionally, we also obtain general limits on the new scalar multiplets from direct searches at colliders, loop corrections to electroweak precision tests and the $W$-boson mass. 
}


\preprint{IFIC/23-53}

\makeatletter
\gdef\@fpheader{}
\makeatother

\begin{document} 
\maketitle
\flushbottom

\section{Introduction} \label{sec:intro}

The standard model (SM), although an extremely successful theory, leaves several problems without a solution, such as the origin of the (tiny) neutrino masses, the nature of dark matter and the generation of  the matter-antimatter asymmetry. In particular, in the SM neutrino masses are absent due to the particle content and the requirement of renormalisability. If the SM is understood as an Effective Field Theory (EFT) valid at the electroweak scale, new higher-dimensional effective operators $\mathcal{O}_i$ beyond the renormalisable interactions included in $\mathcal{L}_{\rm SM}$, should be considered, suppressed by high-energy scales,
\begin{equation} \label{eq:EFT}
 \mathcal{L}=\mathcal{L}_{\rm SM}+ \left(\sum^d_{n=5}\sum_i C^{(i)}_n\, \mathcal{O}_i + {\rm H.c.}\right)\,,
 \end{equation}
  where $n=5,\ldots, d$ runs over the dimension of the effective operator labeled by $i=0,1,\ldots$ and $C^{(i)}_n$ are dimensionful ($[C^{(i)}_n]=4-n$) Wilson Coefficients (WCs). Notice that the latter are matrices in flavor space, whose indices we do not explicitly show in the following.

At the lowest order, dimension-$5$, there is a unique operator with the SM Higgs doublet ($H$) with hypercharge $Y=1/2$ and the lepton doublet ($L$) with hypercharge $Y=-1/2$ \cite{Weinberg:1979sa} 
\begin{align} 
-\mathcal{L}^{(0)}_5 &= \frac{C^{(0)}_{5,a}}{2} \mathcal{O}_{5,a}^{(0)}+ {\rm H.c.}  \nonumber\\
&= \frac{C^{(0)}_{5,a}}{2} (\overline{\tilde L} H)_\mathbf{1} (\tilde H^\dagger L)_\mathbf{1} + {\rm H.c.}\,, \label{eq:Weinberg}
\end{align}
where $\tilde H=i \sigma_2 H^*$ and $\tilde L= i \sigma_2 L^c$, with $\sigma_2$ the second Pauli matrix, and the charge-conjugated field being $L^c=C \overline L^T$. We have chosen a \textit{fermion-like} contraction and $\mathbf{1}$ would correspond to an $SU(2)$ singlet representation of the UV completion.  Alternatively, by using a Fierz identity, one can also write a \textit{fermion-like} triplet contraction, $(LH)_\mathbf{3} (LH)_\mathbf{3}$, or a \textit{scalar-like} triplet contraction, $(LL)_\mathbf{3} (HH)_\mathbf{3}$. At tree level, the UV completions of the Weinberg operator are the usual seesaws \cite{Minkowski:1977sc,Yanagida:1980xy,Gell-Mann:1979vob,Mohapatra:1979ia,Schechter:1980gr,Schechter:1981cv,Lazarides:1980nt,Mohapatra:1980yp,Foot:1988aq}, discussed in detail below. For an EFT approach to lepton number violation (LNV), see Refs.~\cite{Babu:2001ex, deGouvea:2007qla,deGouvea:2014lva,Angel:2012ug,delAguila:2012nu,Gargalionis:2020xvt,Herrero-Garcia:2019czj}.

\begin{figure}[!htb]
\centering
\begin{tikzpicture}[scale=0.7, transform shape, every text node part/.style={align=center}]
    \begin{feynman}
        \node [blob,draw=black, pattern color=black] (a);
        \vertex [above left=3cmof a] (b) {\Large{$\braket{H}$}};
         \vertex [below left=3cmof a] (c) {\Large{$\nu_L$}};
        \vertex [above right=3cmof a] (e) {\Large{$\braket{H}$}};
        \vertex [below right=3cmof a] (f) {\Large{$\nu_L^c$}};
        \diagram* {
            (c) -- [fermion,thick] (a) -- [scalar,thick,insertion=1] (b),
            (f) -- [anti fermion,thick] (a) -- [scalar,thick,insertion=1] (e)
        };
    \end{feynman}
\end{tikzpicture}
\caption{Contribution to Majorana neutrino masses from the dimension-5 Weinberg operator $\mathcal{O}^{(0)}_5$.}
\label{fig:numsm}
\end{figure}
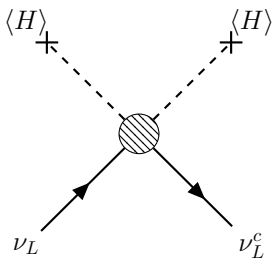

After Electroweak Spontaneous Symmetry Breaking (EWSSB), $\mathcal{O}_{5,a}^{(0)}$ generates Majorana neutrino masses via the Feynman diagram of Fig.~\ref{fig:numsm}, suppressed by the large scale of LNV,
\begin{equation}  \label{eq:mnuW}
    \mathcal{L}\supset -\frac{m_\nu^{(0)}}{2}\overline{\nu^c_L}\nu_L + \text{ H.c.}\,\quad{\rm with}\quad{m^{(0)}_\nu}=C^{(0)}_{5,a}\,v^2\,.
\end{equation}

Considering that $C^{(i)}_n \equiv c^{(i)}_n/\Lambda$, with $c^{(i)}_n$ a dimensionless parameter and $\Lambda$ the new physics scale,
Eq.~\eqref{eq:mnuW} yields an explanation for the smallness of neutrino masses, which scale as $v^2/\Lambda$, where $v$ is the Vacuum Expectation Value (VEV) of the SM Higgs field, $\braket{H}\equiv v=174$ GeV. For order one couplings, $c^{(0)}_{5,a}\sim\mathcal{O}(1)$, reproducing the atmospheric mass scale, $m^{(0)}_\nu \gtrsim \mathcal{O}(0.05)$ eV, demands $\Lambda \lesssim 10^{14}$ GeV, close to the scale of Grand Unified Theories. This is the seesaw mechanism \cite{Yanagida:1980xy}. If this is the origin of neutrino masses, it will therefore be very difficult to probe the new physics responsible for lepton number violation. 

In this work, we postulate the presence of new low-energy $SU(2)$ scalar multiplets $\Phi_i$ and study their implications for neutrino masses. The new scalar masses are bounded from above by perturbativity and unitarity from the fact that their masses are $\propto \lambda\, v^2$ ({where $\lambda$ is some quartic scalar coupling}), implying that the SM is augmented by low-energy ($\mathcal{O}$ (TeV)) degrees of freedom with which one can construct new SM EFT. 

The possible new Weinberg-like operators with up to two new scalars $\Phi_1, \Phi_2$ are 
\begin{align} 
-\mathcal{L}_5 &= \frac{1}{2} \sum_i C^{(i)}_{5}\mathcal{O}^{(i)}_{5} + {\rm H.c.}  \nonumber \,
\end{align}
where
\begin{equation}
\begin{aligned}
\mathcal{O}^{(0)}_{5,a}=(LH)_\mathbf{1}(LH)_\mathbf{1}\,,\qquad {\mathcal{O}}^{(1)}_{5}&=(LH)_\mathbf{N}(L\Phi_1)_\mathbf{N}\,,\\
\mathcal{O}^{(2)}_{5}=(L\Phi_1)_\mathbf{N} (L\Phi_1)_\mathbf{N}\,,\qquad \mathcal{O}^{(3)}_{5}&=(L\Phi_1)_\mathbf{N} (L\Phi_2)_\mathbf{N}\,.
\end{aligned}\label{eq:newW}
\end{equation}
These have to be understood as classes of operators, encoding all the allowed possible \emph{conjugations} of the scalar fields. For instance, ${\mathcal{O}}^{(1)}_{5}$ may involve $\widetilde H$ instead of $H$.\footnote{Note that, if considered individually, operators with for example $\Phi^c_1$ instead of $\Phi_1$ would correspond to just a redefinition of the new field. However, in class-$\bf B$ scenarios (see later) more than one operator is generated and therefore whether the field or its conjugate appears is relevant.}

Note that $C_{5,a}^{(0)}$ and $C_{5}^{(2)}$ are symmetric in flavor space, i.e. $C_{5,a}^{(0)T}=C_{5,a}^{(0)}$ and $C_5^{(2)T}=C_5^{(2)}$. These operators are obtained at low energy by integrating out a heavy mediator at tree level. Once again, in writing Eq.~\eqref{eq:newW}, we have chosen a \textit{fermion-like} contraction for our basis of operators, with $\mathbf{N}$ denoting the highest $SU(2)$ representation of the UV completion. Below, we provide the relevant factors for all the other possible Lorentz contractions.
As long as the new scalars take a VEV, $\langle \Phi_i\rangle \equiv v_i \neq 0$, Majorana neutrino masses are generated, which in this case will scale as $v^2/\Lambda$, $v v_i/\Lambda$, $v_i^2/\Lambda$ and $v_i v_j/\Lambda$ corresponding to $\mathcal{O}^{(0)}_{5,a}$, $\mathcal{O}^{(1)}_5$, $\mathcal{O}^{(2)}_5$ and $\mathcal{O}^{(3)}_5$, respectively. 

 Moreover, as the VEVs of the new multiplets ($v_i$) in which we will be interested violate the SM custodial symmetry ($\rho \neq 1$), they are bounded to be quite small, typically $v_i\lesssim \mathcal{O}(1)$ GeV. Such a suppression in the VEVs is natural if they are induced by the Higgs one, and is precisely what is needed to have tiny neutrino masses with not so large lepton number-violating scales, i.e., $\Lambda$ is parametrically suppressed. We will therefore be interested in scenarios that do \emph{not} generate $\mathcal{O}^{(0)}_{5,a}$, but only $\mathcal{O}^{(1)}_5$, or $\mathcal{O}^{(2)}_5$ and/or $\mathcal{O}^{(3)}_5$. For instance, if $\mathcal{O}^{(2)}_5$ is generated, compared to the SM Weinberg operator, we have
\begin{equation}
\frac{C_{5,a}^{(0)}}{C_5^{(2)}} \sim \frac{v^2_i}{v^2}\, \lesssim 10^{-4}\, \,.
\end{equation}
Possible extra suppression in the WCs $C_{5,a}^{(0)}/C_5^{(2)}$ may lower the scale even further. The presence of new scalars at the EW scale, along with the Electroweak Precision Tests (EWPTs) and collider searches of heavy scalars makes these scenarios more testable than the usual high-scale seesaws.

In this work, we study and provide a catalogue of the new scenarios generated by these new Weinberg-like operators constructed with extra scalar multiplets that take a VEV. We consider \emph{natural} scenarios, in which the new scalar multiplets obtain suppressed induced VEVs, i.e., $v_i \sim v^3/m_\Phi^2$. The most appealing scenarios are the ones in which the dominant contribution to neutrino masses is provided by the new operators. The UV models that generate those at tree level are considered \emph{genuine} and will be studied in a companion paper \cite{Giarnetti:2023osf}.

Similar cataloguing works have been done in the literature. Refs.~\cite{Bonnet:2009ej,Cepedello:2017lyo,Anamiati:2018cuq} discuss the generation of neutrino masses from operators higher than dimension 5. Minimal tree-level seesaw models and radiative neutrino mass models with additional scalar multiplets are discussed in Refs.~\cite{McDonald:2013kca} and \cite{Law:2013saa}, respectively. Dirac neutrino masses from $SU(2)$ multiplet fields are studied in Ref.~\cite{Wang:2016lve}. Generic constraints on scalar eletroweak multiplets have been discussed in Refs.~\cite{Hally:2012pu,Dawson:2017vgm}, while constraints on the Higgs couplings due to triplets and quadruplets were studied in Refs.~\cite{Logan:2010en} and \cite{Durieux:2022hbu,Kannike:2023bfh}, respectively. 

The rest of the paper is structured as follows. In Section \ref{sec:Weinbergs} we consider neutrino masses generated by Weinberg-like operators constructed by one or two new extra scalar multiplets. In Section \ref{sec:SMEFT} neutrino masses are considered in the pure SMEFT with just the Higgs doublet. In Sections \ref{sec:extensions_one} and \ref{sec:extensions_two} we extend the low-energy content of the SM with one or two new extra scalars, respectively, and list the new Weinberg-like operators they generate. In Section \ref{sec:scalar_pot} we analyse the scalar sector and the constraints from EWPTs on the induced VEVs of the scalars. In Section \ref{sec:pheno} we focus on the phenomenology. We further study in detail the production of the scalars at colliders, as well as the loop corrections to EWPTs. These latter constraints are general and may be applied to models with new scalar multiplets, irrespective of neutrino masses. Finally, we conclude in Section \ref{sec:conc}. 

\section{Neutrino masses from Weinberg-like operators} \label{sec:Weinbergs}

\subsection{Neutrino masses in the Standard Model Effective Field Theory} \label{sec:SMEFT}
At dimension $5$, there is only one operator, the Weinberg $\mathcal{O}^{(0)}_{5,a}$ in Eq.~\eqref{eq:Weinberg}. The other possible Lorentz contractions of the fields are
\begin{equation}
\begin{aligned}
\mathcal{O}^{(0)}_{5,a}=(HL)_\mathbf{1}(HL)_\mathbf{1} \,,\quad &\mathcal{O}^{(0)}_{5,b}=(HL)_\mathbf{3}(HL)_\mathbf{3}\\
\mathcal{O}^{(0)}_{5,c}=(HH)_\mathbf{3}(LL)_\mathbf{3} \,,\quad &
\mathcal{O}^{(0)}_{5,d}=(HH)_\mathbf{1}(LL)_\mathbf{1}.
\end{aligned}\label{smops}
\end{equation}
Since the singlet $(HH)_1=0$, the last operator identically vanishes. The other cases yield non-zero neutrino masses and their UV completions at tree level correspond to the seesaws: the operators $\mathcal{O}_{5,a}^{(0)}$, $\mathcal{O}_{5,c}^{(0)}$ and $\mathcal{O}_{5,b}^{(0)}$ are associated with the Type-I (fermion singlet), Type-II (scalar triplet) and Type-III (fermion triplet) seesaws, respectively. Also, they may be generated at the loop level \cite{Ma:2006km} (see Ref.~\cite{Cai:2017jrq} for a review of radiative neutrino mass models). Note that there exists only one independent contraction among the fields in the Weinberg operator, though every possible contraction can be obtained from the others via a Fierz transformation \cite{Fierz:1937wjm}. Indeed, using 
$ \left(\sigma_i\right)_{ab}\left(\sigma_i\right)_{cd}=2\delta_{ad}\delta_{bc}-\delta_{ab}\delta_{cd}$, with $\sigma_i$ ($i=1,2,3$) the Pauli matrices (generators of the $SU(2)$ group) and $(a,b,c,d)$ the $SU(2)$ indices, the different operators in Eq.~\eqref{smops} are related as
\begin{equation}
        \mathcal{O}_{5,b}^{(0)}=-\mathcal{O}_{5,a}^{(0)} = -\frac{1}{2} \mathcal{O}_{5,c}^{(0)}\, .
\end{equation}

In the following, we will focus on the case in which $C_{5,a}^{(0)}$ is very suppressed, so that its contribution to neutrino mass is irrelevant. Also, to alleviate notation, we will denote the fermion singlet Lorentz contraction ($\mathcal{O}^{(0)}_{5,a}$) by $\mathcal{O}^{(0)}_5$.

\subsection{Extensions with one new scalar multiplet} \label{sec:extensions_one}

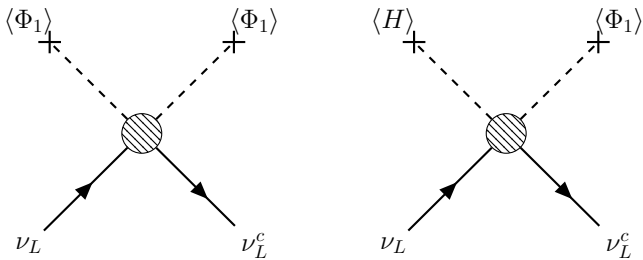
\begin{figure}[!htb]\centering
\begin{tikzpicture}[scale=0.7, transform shape, every text node part/.style={align=center}]
    \begin{feynman}
        \node [blob,draw=black, pattern color=black] (a);
        \vertex [above left=3cmof a] (b) {\Large{$\braket{\Phi_1}$}};
         \vertex [below left=3cmof a] (c) {\Large{$\nu_L$}};
        \vertex [above right=3cmof a] (e) {\Large{$\braket{\Phi_1}$}};
        \vertex [below right=3cmof a] (f) {\Large{$\nu_L^c$}};
        \diagram* {
            (c) -- [fermion,thick] (a) -- [scalar,thick,insertion=1] (b),
            (f) -- [anti fermion,thick] (a) -- [scalar,thick,insertion=1] (e)
        };
    \end{feynman}
\end{tikzpicture}
\hspace{2em}
\begin{tikzpicture}[scale=0.7, transform shape,every text node part/.style={align=center}]
    \begin{feynman}
        \node [blob,draw=black, pattern color=black] (a);
        \vertex [above left=3cmof a] (b) {\Large{$\braket{H}$}};
         \vertex [below left=3cmof a] (c) {\Large{$\nu_L$}};
        \vertex [above right=3cmof a] (e) {\Large{$\braket{\Phi_1}$}};
        \vertex [below right=3cmof a] (f) {\Large{$\nu_L^c$}};
        \diagram* {
            (c) -- [fermion,thick] (a) -- [scalar,thick,insertion=1] (b),
            (f) -- [anti fermion,thick] (a) -- [scalar,thick,insertion=1] (e)
        };
    \end{feynman}
\end{tikzpicture}
\caption{Contribution to Majorana neutrino masses from the possible new Weinberg-like operators $\mathcal{O}_5^{(1,2)}$ with a new scalar multiplet, $\Phi_1$.} \label{feyn1insertion}
\end{figure}

In this section we construct Weinberg-like operators with a single new scalar multiplet, $\Phi_1$. 
We will denote such scenarios as ``class-\textbf{A}'' scenarios. In presence of both the SM Higgs $H$ and the new multiplet $\Phi_1$, the possible new Weinberg-like operators involving $\Phi_1$ are $\mathcal{O}_5^{(1)}$ and $\mathcal{O}_5^{(2)}$, see Eq.~\eqref{eq:newW}. As argued before, we will be interested in the scenarios where the $\mathcal{O}_5^{(0)}$ contribution is suppressed, so that the dominant contribution to neutrino masses comes from the Feynman diagrams in Fig. \ref{feyn1insertion}.
Let us discuss first the $\mathcal{O}_5^{(1)}$ term. In order to determine the $SU(2)$ representations for $\Phi_1$, we first observe the transformation $H \, L_\alpha \,L_\beta\sim \bf 2\otimes2\otimes2=4\oplus2\oplus2^\prime$ under SU(2). However, if $\Phi_1$ is an $SU(2)$ doublet,  a two-Higgs-Doublet Model \cite{Lee:1973iz} (see Ref.~\cite{Branco:2011iw} for a review) is obtained, which has already been widely studied in literature \cite{Oliver:2001eg,Hernandez-Garcia:2019uof}. Therefore, $\Phi_1 \sim {\bf 4}$ is the only interesting possibility for our purposes. The hypercharge of the new scalar quadruplet is fixed by the Weinberg operator, and we take the negative hypercharge assignment without loss of generality, $-1/2$ or $-3/2$.\footnote{{Note that both positive and negative hypercharge assignments, $\pm1/2$ or $\pm3/2$, are allowed, depending on the definition of the effective operator: $\Phi_1\sim(4,-1/2)$ for $ \overline{L} \Phi_1 \widetilde H^T L^c $ or $\Phi_1\sim(4,+1/2)$ for $ \overline{L} \Phi_1^c \widetilde H^T L^c $; similarly, $\Phi_1\sim(4,-3/2)$ for $ \overline{L} \Phi_1 H^T L^c $, or $\Phi_1\sim(4,+3/2)$ for $\overline{L} {\Phi}^c_1 H^T L^c $.}}
Let us now consider $\mathcal{O}_5^{(2)}$. Whatever the representation of $\Phi_1$ is, say $\mathbf{N_1}$, it is always possible to construct a singlet with two identical scalar fields and two lepton doublets. However, in this case, the hypercharge of the new multiplet is forced to be $-1/2$. Thus, in order to have a neutral scalar component, $\Phi_1$ must be in an even $SU(2)$ representation. If we restrict our study to $\mathbf{N_1}\leq\mathbf{5}$ in order to avoid problems with unitarity and non-perturbativity close to the EW scale, due to their large RGE running \cite{Hally:2012pu,Earl:2013jsa}, the only viable representations are the doublet (which corresponds to the already mentioned 2HDM) and, again, the quadruplet. Neglecting the 2HDM-like models, the only viable possibilities in $\mathcal{O}_5^{(1)}$ and $\mathcal{O}_5^{(2)}$ are the following:
\begin{enumerate}
    \item $\mathbf{A_I}$: $\Phi_1=(4,-1/2) \equiv \mathbf{4}_{-1/2}$  
    \item $\mathbf{A_{II}}$: $\Phi_1=(4,-3/2) \equiv \mathbf{4}_{-3/2}$  
\end{enumerate}

\subsection{Extensions with two new scalar multiplets} \label{sec:extensions_two}

Let us now consider the new Weinberg-like operators  $\mathcal{O}_5^{(3)}$, see Eq.~\eqref{eq:newW}, which may arise due to the presence of two additional scalar multiplets, $\Phi_1$ and $\Phi_2$.
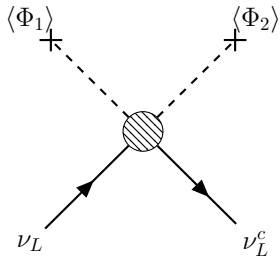
\begin{figure}[!htb]\centering
\begin{tikzpicture}[scale=0.7, transform shape, every text node part/.style={align=center}]
    \begin{feynman}
        \node [blob,draw=black, pattern color=black] (a);
        \vertex [above left=3cmof a] (b) {\Large{$\braket{\Phi_1}$}};
         \vertex [below left=3cmof a] (c) {\Large{$\nu_L$}};
        \vertex [above right=3cmof a] (e) {\Large{$\braket{\Phi_2}$}};
        \vertex [below right=3cmof a] (f) {\Large{$\nu_L^c$}};
        \diagram* {
            (c) -- [fermion,thick] (a) -- [scalar,thick,insertion=1] (b),
            (f) -- [anti fermion,thick] (a) -- [scalar,thick,insertion=1] (e)
        };
    \end{feynman}
\end{tikzpicture}
\caption{Feynman diagram associated with the new Weinberg-like operator $\mathcal{O}^{(3)}_5$, which may arise due to the presence of two new scalar multiplets, $\Phi_1$ and $\Phi_2$.} \label{feyn2}
\end{figure}
We will denote such scenarios as ``class-\textbf{B}'' scenarios. As before, we are interested in the case in which neutrino masses are \emph{not} generated via the $\mathcal{O}_5^{(0)}$ operator which involves only the SM Higgs. The new operator $\mathcal{O}_5^{(3)}$ produces neutrino masses via the Feynman diagram shown in Fig. \ref{feyn2}.

In this framework, given the two scalar multiplets representations $\mathbf{N_1}$ and $\mathbf{N_2}$, we need to construct a singlet among $\Phi_1$, $\Phi_2$ and two lepton doublets. Using $SU(2)$ properties, we require either {$\mathbf{N_1}\otimes\mathbf{N_2}\subset\mathbf{1}$} or {$\mathbf{N_1}\otimes\mathbf{N_2}\subset\mathbf{3}$}. The former is possible only for $\mathbf{N_1}=\mathbf{N_2}$, whereas, in the latter case, we have $\mathbf{N_1}\otimes\,\mathbf{3}=(\mathbf{N_1-2})\oplus\mathbf{N_1}\oplus(\mathbf{N_1+2})$. Therefore, in addition to the case of being equal, $\mathbf{N_1}$ and $\mathbf{N_2}$ may be either two consecutive even or two consecutive odd representations. In order to construct the Weinberg-like operator $\mathcal{O}_5^{(3)}$ and account for the possible presence of conjugated scalar fields, we also require that the modulus of the hypercharges of $\Phi_1$ and $\Phi_2$ differ by one unit, \emph{i.e.}, $|Y(\Phi_1)|\pm|Y(\Phi_2)|=1$. As before, we consider representations $\mathbf{N_{(1,2)}}\leq \mathbf{5}$, neglect scenarios containing doublets or two identical representations with identical hypercharges and require that each scalar multiplet contains a neutral component. This leaves us with the following six possibilities:
\begin{enumerate}       
    \item $\mathbf{B_I}$: $\Phi_1=(4,1/2)\equiv \mathbf{4}_{1/2}$ and $\Phi_2=(4,-3/2)\equiv \mathbf{4}_{-3/2}$  
    \item $\mathbf{B_{II}}$: $\Phi_1=(3,0)\equiv \mathbf{3}_0$ and $\Phi_2=(5,-1)\equiv \mathbf{5}_{-1}$      
    \item $\mathbf{B_{III}}$: $\Phi_1=(5,1)\equiv \mathbf{5}_{1}$ and $\Phi_2=(5,-2)\equiv \mathbf{5}_{-2}$  
    \item $\mathbf{B_{IV}}$: $\Phi_1=(5,0)\equiv \mathbf{5}_{0}$ and $\Phi_2=(5,-1)\equiv \mathbf{5}_{-1}$
    \item $\mathbf{B_V}$: $\Phi_1=(3,0)\equiv \mathbf{3}_0$ and $\Phi_2=(3,-1)\equiv \mathbf{3}_{-1}$
    \item $\mathbf{B_{VI}}$: $\Phi_1=(3,-1)\equiv \mathbf{3}_{-1}$ and $\Phi_2=(5,0)\equiv \mathbf{5}_{0}$ 
\end{enumerate}
Once again, a change in sign for the hypercharge of each scalar multiplet will not affect the results of our analyses; thus, one can also write these scenarios with opposite hypercharges. Notice that, given the cut on the $\Phi_i$ representations, there are no scenarios where the two scalar multiplets have the same hypercharge. Further, it can be seen that the $\mathbf{B_I}$ scenario contains multiplets from the class-\textbf{A} scenarios, hence, in this case neutrino masses may also obtain a contribution from $\mathcal{O}_5^{(1,2)}$ apart from $\mathcal{O}_5^{(3)}$. Finally, it is worth mentioning that in the $\mathbf{B_{V,VI}}$ scenarios containing a scalar triplet with hypercharge $-1$, a Type-II seesaw contribution may be induced. In these cases, in order to have $C^{(3)}_5 \gtrsim C^{(0)}_5$, the trilinear coupling of the triplet with the Higgs doublets would have to be suppressed. Despite being in principle {\it non-genuine}, we will consider also these two scenarios in the rest of the paper under the hypothesis that $C^{(0)}_5$ is negligible compared to $C^{(3)}_5$.

\subsection{Wilson Coefficients for neutrino masses}

Here, we quantify neutrino masses at tree level in the different scenarios containing new scalar multiplets. In order to compute their values, the four fields appearing in the Weinberg-like operators need to be contracted in a singlet, so the possible contractions are
\begin{equation*}
    (\overline{L}\Phi_i)_{\mathbf N} ( \Phi_j^T L^c)_{\mathbf N} \quad{\rm or}\quad ( \overline{L}L^c)_\mathbf{N} ( \Phi_i^T \Phi_j)_\mathbf{N}\,,
\end{equation*}
where the UV complete model contains a fermion (scalar) mediator in the $SU(2)$ representation $\mathbf{N}$ in the case of former (latter). As already mentioned before, the different contractions are related by Fierz transformations.

We first discuss the case of a \textit{fermion-like} contraction $(\overline{L}\Phi_i)_{\mathbf N} ( \Phi_j^T L^c)_{\mathbf N}$, for which the neutrino masses can be obtained from Eq.~\ref{eq:newW} as
\begin{equation}\label{eq:mnueft}
m_\nu\simeq {C^{(0)}_5}\,\omega_{\mathbf{N}}^{00}\,v^2+\sum_{i=1,2}[{ {\Tilde{C}^{(1)}_5}}]_i\,\omega_{\mathbf{N}}^{0i}\,v v_i+\sum_{i=1,2}[{ {C^{(2)}_5}}]_i\,\omega_{\mathbf{N}}^{ii}\, v_i^2+{ {\Tilde{C}^{(3)}_5}}\,\omega_{\mathbf{N}}^{12}\,v_1v_2\,,
\end{equation}
where we define $ {\tilde{C}^{(1,3)}_5 = C^{(1,3)}_5+[C^{(1,3)}_5]^T}$ and denote $\omega^{ij}_{\textbf{N}}$ as the numerical factor corresponding to this contraction, where the index $i,j=0$ depicts the SM Higgs. 

The representation $\mathbf{N}$ of the fermion mediator can be determined from the representation of the two scalar multiplets ($\mathbf{N_1}$ and $\mathbf{N_2}$) entering a particular Weinberg operator (including the SM Higgs doublet). The contraction $\left( \overline{L}\Phi_i\right)$ gives $\mathbf{N_i}\otimes\mathbf{2}=(\mathbf{N_i-1})\oplus(\mathbf{N_i+1})$. Thus, if the representations of the two scalar multiplets are identical, the possible contractions of the full Weinberg operator are either from two $\mathbf{(N_i-1)}$-tuplets or from two $\mathbf{(N_i+1)}$-tuplets, hence $\mathbf{N}=\mathbf{N_i \pm 1}$. On the other hand, if $\mathbf{N_1}\neq\mathbf{N_2}$, for instance, when $\mathbf{N_1}$ and $\mathbf{N_2}$ are two consecutive odd or even representations, the only possible contraction is between two $(\mathbf{N_1+1})$-tuplets, where $\mathbf{N_1}$ is the smaller of the two representations, thus $\mathbf{N}=\mathbf{N_1}+1$ for $\mathbf{N_1}<\mathbf{N_2}$. 

In the case of a \textit{scalar-like} contraction $\left( \overline{L}L^c\right)_{\textbf{N}} \left( \Phi_i^T \Phi_j\right)_{\textbf{N}}$, the two lepton doublets can only give a singlet or a triplet. As the singlet between the two doublets does not contribute to neutrino masses at tree level, the only possible contraction for this case is when the two scalar multiplets also form a triplet. Hence, the only possible UV scalar mediator is a triplet, i.e., $\mathbf{N}=\mathbf{3}$, as in the Type-II seesaw model. Neutrino masses can be written as
\begin{equation}\label{eq:mnueft1}
m_\nu\simeq {C^{(0)}_5}\,z_{\mathbf{3}}^{00}\,v^2+\sum_{i=1,2}[{ {\Tilde{C}^{(1)}_5}}]_i\,z_{\mathbf{3}}^{0i}\,v v_i+\sum_{i=1,2}[{ {C^{(2)}_5}}]_i\,z_{\mathbf{3}}^{ii}\, v_i^2+{ {\Tilde{C}^{(3)}_5}}\,z_{\mathbf{3}}^{12}\,v_1v_2\,,
\end{equation}
where $z^{ij}_{\mathbf{3}}$ is the numerical factor coming from the contraction $( \overline{L}L^c)_\mathbf{3} ( \Phi_i^T \Phi_j)_\mathbf{3}$.
\begin{table}[!htb]
\centering
\begin{tabular}{l|cc|ccc|}
\cline{2-6}
\cline{2-6} 
& \multicolumn{1}{c|}{$\omega^{01}_\mathbf{3}$} & $z^{01}_\mathbf{3}$ & \multicolumn{1}{c|}{$\omega^{11}_\mathbf{3}$} & \multicolumn{1}{c|}{$\omega^{11}_\mathbf{5}$} & $z^{11}_\mathbf{3}$ \\ \hline\hline
\multicolumn{1}{|c|}{$\mathbf{A_I}$} & \multicolumn{1}{c|}{$2/\sqrt{3}$}        & $4/\sqrt{3}$        & \multicolumn{1}{c|}{$-8/3$}                   & \multicolumn{1}{c|}{$1/2$}                   & $-2/3$                    \\ \hline
\multicolumn{1}{|c|}{$\mathbf{A_{II}}$} & \multicolumn{1}{c|}{$-1$}                & $1$                 & \multicolumn{1}{c|}{\ding{56}}                & \multicolumn{1}{c|}{\ding{56}}                & \ding{56}                \\ \hline
\end{tabular}
\caption{\label{WilsonA} Numerical factors for all the possible contractions appearing in Weinberg-like operators of class-\textbf{A} scenarios.}
\end{table}

\begin{table}[!htb]
\centering
\begin{tabular}{c|cccccc|}
\cline{2-7} 
\multicolumn{1}{l|}{} & \multicolumn{1}{c|}{$\omega^{12}_\mathbf{2}$} & \multicolumn{1}{c|}{$\omega^{12}_\mathbf{3}$} & \multicolumn{1}{c|}{$\omega^{12}_\mathbf{4}$} & \multicolumn{1}{c|}{$\omega^{12}_\mathbf{5}$} & \multicolumn{1}{c|}{$\omega^{12}_\mathbf{6}$} & $z^{12}_\mathbf{3}$ \\ \hline\hline
\multicolumn{1}{|c|}{$\mathbf{B_{I}}$} & \multicolumn{1}{c|}{\ding{56}}                & \multicolumn{1}{c|}{$1/\sqrt{3}$}             & \multicolumn{1}{c|}{\ding{56}}                & \multicolumn{1}{c|}{$-\sqrt{3}/4$}            & \multicolumn{1}{c|}{\ding{56}}                & $-1/\sqrt{3}$            \\ \hline
\multicolumn{1}{|c|}{$\mathbf{B_{II}}$}      & \multicolumn{1}{c|}{\ding{56}}                & \multicolumn{1}{c|}{\ding{56}}                & \multicolumn{1}{c|}{$-1/\sqrt{2}$}            & \multicolumn{1}{c|}{\ding{56}}                & \multicolumn{1}{c|}{\ding{56}}                & $1/\sqrt{2}$             \\ \hline
\multicolumn{1}{|c|}{$\mathbf{B_{III}}$}     & \multicolumn{1}{c|}{\ding{56}}                & \multicolumn{1}{c|}{\ding{56}}                & \multicolumn{1}{c|}{$1/2$}                      & \multicolumn{1}{c|}{\ding{56}}                & \multicolumn{1}{c|}{$-2/5$}                   & $1/2$                    \\ \hline
\multicolumn{1}{|c|}{$\mathbf{B_{IV}}$}      & \multicolumn{1}{c|}{\ding{56}}                & \multicolumn{1}{c|}{\ding{56}}                & \multicolumn{1}{c|}{$-1/(2\sqrt{6})$}              & \multicolumn{1}{c|}{\ding{56}}                & \multicolumn{1}{c|}{$\sqrt{6}/5$}             & $-\sqrt{3/8}$            \\ \hline
\multicolumn{1}{|c|}{$\mathbf{B_{V}}$}      & \multicolumn{1}{c|}{$-1/\sqrt{2}$}            & \multicolumn{1}{c|}{\ding{56}}                & \multicolumn{1}{c|}{$-\sqrt{2}/3$}            & \multicolumn{1}{c|}{\ding{56}}                & \multicolumn{1}{c|}{\ding{56}}                & $1/\sqrt{2}$             \\ \hline
\multicolumn{1}{|c|}{$\mathbf{B_{VI}}$}      & \multicolumn{1}{c|}{\ding{56}}                & \multicolumn{1}{c|}{\ding{56}}                & \multicolumn{1}{c|}{$1/\sqrt{6}$}             & \multicolumn{1}{c|}{\ding{56}}                & \multicolumn{1}{c|}{\ding{56}}                & $-1/\sqrt{6}$            \\ \hline
\end{tabular}
\caption{\label{WilsonB} Numerical factors for all the possible contractions appearing in Weinberg-like operators of class-\textbf{B} scenarios.}
\end{table}

In Tables~\ref{WilsonA} and \ref{WilsonB}, we list the numerical values of $\omega$ and $z$ corresponding to different representations and operators for class-\textbf{A} and  \textbf{B} scenarios, respectively. Notice that in the scenario containing two quadruplets ($\mathbf{B_{I}}$), there also exist neutrino mass contributions from $\mathcal{O}_5^{(1)}$ and $\mathcal{O}_5^{(2)}$ Weinberg-like operators, with the former potentially dominating over the others due to the large SM Higgs VEV. The corresponding numerical factors are exactly the same as the ones shown in Table~\ref{WilsonA} for the class-\textbf{A} scenarios. Note that the only contraction which exists for every scenario is the one where both the lepton doublets and the scalars are in a triplet, namely $(\bar{L}L^c)_\mathbf{3}(\Phi_1^T \Phi_2)_\mathbf{3}$.  

\section{The scalar sector} \label{sec:scalar_pot}

In this section we want to underline the physics of the scalar sector, analyzing the scalar potentials for all the possible cases as well as the implications of the couplings of new multiplets with the SM Higgs.

\subsection{The scalar potential} \label{sec:scalarV}

Given the new scalar multiplets in our scenarios, we discuss the possible potential terms that may be built involving the new scalars and the SM Higgs doublet. First, we consider the simple case of the class-\textbf{A} scenarios involving only one additional scalar multiplet. If the new scalar $\Phi_1$ carries lepton number, some of the potential terms may violate the additional $U(1)_L$ symmetry. Thus, we can divide the full scalar potential in two parts: lepton-number-conserving and LNV terms
\begin{equation}     \label{eq:potentialA1}
    V^{\mathbf{A}}(H,\Phi_1) = V_L^{\mathbf{A}}(H,\Phi_1) + V^{\mathbf{A}}_{\slashed L}(H,\Phi_1)\,.
\end{equation}
In the case of a quadruplet with hypercharge $-1/2$, we have
\begin{align}     
   V_L^{\mathbf{A_I}}(H,\Phi_1)=&-\mu_H^2\,(H^\dagger H)+\mu_{\Phi_1}^2\,(\Phi_1^\dagger \Phi_1)+\lambda_1 (H^\dagger H)^2+\lambda_2 (\Phi_1^\dagger \Phi_1)^2+ \nonumber \\
        &+\lambda_3\, H^\dagger H\Phi_1^\dagger \Phi_1+\lambda_4\,H^\ast H \Phi_1^\ast \Phi_1+\lambda_5\,\Phi_1^\ast\Phi_1\,\Phi_1^\ast\Phi_1\,, \label{eq:potentialA1L}
\end{align}
and
  \begin{equation}     \label{eq:potentialA1X}
  V^{\mathbf{A_I}}_{\slashed L}(H,\Phi_1)=
\lambda_6 \,\Phi_1^\ast H \Phi_1 \Phi_1+\lambda_7\,H\Phi_1 H\Phi_1+\lambda_8\,H^\ast\Phi_1 HH+{\rm H.c.}\,,
    \end{equation}
where the $\lambda_2$ and $\lambda_5$ as well as $\lambda_3$ and $\lambda_4$ terms, despite having the same field content, act as different potential terms due to difference in tensor contractions. For example, the tensor contractions of the quartic coupling $\lambda_6 \Phi_1^\ast\,H\,\Phi_1\,\Phi_1$ can be written in full glory as $\left( \Phi_1^\ast\right)^{ijk}H_d \left(\Phi_1\right)_{d^\prime m^\prime i}\left(\Phi_1\right)_{m j k }\epsilon^{d d^\prime}\epsilon^{m m^\prime}$. 

For the quadruplet with hypercharge $-3/2$, the scalar potential has a different form
\begin{equation}
    \begin{aligned}
    V^{\mathbf{A_{II}}}_L(H,\Phi_1)=&\;\,\mu_H^2(H^\dagger H)+\mu_{\Phi_1}(\Phi_1^\dagger\Phi_1)+\lambda_1(\Phi_1^\dagger\Phi_1)^2+\lambda_2(H^\dagger H)^2+\lambda_3(H^\dagger H)(\Phi_1^\dagger \Phi_1)+\\
    +&\lambda_4H^\ast\Phi_1\Phi_1^\ast H+\lambda_6 \Phi_1^\ast\Phi_1\Phi_1^\ast\Phi_1\,,
    \end{aligned}
    \label{eq:potentialA2l}
\end{equation}
and
\begin{equation}
    V^{\mathbf{A_{II}}}_{\slashed{L}}(H,\Phi_1)=\;\,\lambda_6\Phi_1 HHH+{\rm H.c.}\,,
    \label{eq:potentialA2X}
\end{equation}
where the only LNV term is linear in the new scalar multiplet.

Let us now consider the class-\textbf{B} scenarios, in which we have two different scalar multiplets. In this case, there is a plethora of scalar potential terms. For simplicity, we show here only the LNV terms, which are relevant for the following discussion. Looking only at the scalar potential, we might recognize also another accidental $U(1)_X$ symmetry which arises due to the presence of a second scalar multiplet. Hence, considering the lepton number violating terms only, 
we can divide them in $U(1)_X$-conserving and violating parts as
\begin{equation}     \label{eq:potentialB1}
    V^{\mathbf{B}}_{\slashed{L}}(H,\Phi_1,\Phi_2)\supset V_{X}^{\mathbf{B}} (H,\Phi_1,\Phi_2) + V^{\mathbf{B}}_{\slashed X}(H,\Phi_1,\Phi_2)\,.
\end{equation}
where the $U(1)_X$-conserving term is
\begin{equation}\label{eq:u1xcon}
   V_{X}^{\mathbf{B}}(H,\Phi_1,\Phi_2) =\lambda_1 H\, H\, \Phi_1 \,\Phi_2\,,
\end{equation} 
provided that the condition $X_1=-X_2$ holds, where $X_{1,2}$ are the $U(1)_X$ charges of $\Phi_{1,2}$ and the SM Higgs doublet is uncharged. Such a term is present in all the class-$\mathbf{B}$ scenarios. On the other hand, the $U(1)_X$-violating terms are listed in Table~\ref{tab:potentials}.
\begin{table}[!htb]
\centering
\begin{tabular}{c|c|}
\cline{2-2}
&$V_{\slashed X}(H,\Phi_1,\Phi_2)$\\ \hline\hline
\multicolumn{1}{|c|}{\multirow{3}{*}{$\mathbf{B_{I}}$}} &$\lambda_2 \Phi_2^\ast\Phi_1^\ast H\,\Phi_2 + \lambda_3(\Phi_2^\ast\Phi_1^\ast H\,\Phi_2)^\prime+\lambda_4 \Phi_1^\ast\Phi_1^\ast H\,H+$\\
\multicolumn{1}{|c|}{}                  &$\lambda_5\Phi_1^\ast H^\ast H\,H\,+\lambda_6H\,H\,H\Phi_2+\lambda_7\Phi_1^\ast\Phi_1^\ast   H\Phi_1+$ \\
\multicolumn{1}{|c|}{} & $ \lambda_{8} H\Phi_1\Phi_1\Phi_2+\lambda_{9} \Phi_1\Phi_1\Phi_1\Phi_2+{\rm H.c.}$\\ \hline
\multicolumn{1}{|c|}{\multirow{2}{*}{$\mathbf{B_{II}}$}}  & $\mu_1\Phi_2^\ast\Phi_2\Phi_1+\mu_2\Phi_1H^\ast H+{\mu_3^2\, \Phi_1^2}+\lambda_2\Phi_2^\ast\Phi_2\Phi_1\Phi_1$       \\
\multicolumn{1}{|c|}{}                                                                                                                       & $+\lambda_3H^\ast H\Phi_1\Phi_1+\lambda_4\Phi_1\Phi_1\Phi_1\Phi_1+{\rm H.c.}$                                           \\ \hline
\multicolumn{1}{|c|}{$\mathbf{B_{III}}$}                  & $\mu_1 \Phi_1\Phi_1\Phi_2+{\rm H.c.}$                                  \\ \hline
\multicolumn{1}{|c|}{\multirow{2}{*}{$\mathbf{B_{IV}}$}} & $\mu_1\Phi_1\Phi_1\Phi_1 + \mu_2\Phi_2^\ast\Phi_2\Phi_1+{\mu_3^2\,\Phi_1^2}$              \\
\multicolumn{1}{|c|}{}                                                                                                                     & $+ \lambda_2H^\ast H \Phi_1\Phi_1+\lambda_3\Phi_1\Phi_1\Phi_1\Phi_1 + {\rm H.c.}$                  
                                \\ \hline
\multicolumn{1}{|c|}{{\multirow{2}{*}{$\mathbf{B_{V}}$}} }&$\mu_1 \Phi_1 \Phi_2^\ast \Phi_2+\mu_2\Phi_1 H^\ast H+\mu_3 \Phi_2 H H +{\mu_4^2\,\Phi_1^2}$\\
\multicolumn{1}{|c|}{}& $+\lambda_2 \left(\Phi_1 \Phi_1 \Phi_1 \Phi_1\right) + \lambda_3 \left(H^\ast H \Phi_1 \Phi_1\right)+{\lambda_4\,\Phi_1^2 \Phi_2^\ast \Phi_2}+{\rm H.c.}$\\ \hline 
\multicolumn{1}{|c|}{{\multirow{2}{*}{$\mathbf{B_{VI}}$}} }       &    $\mu_1 \Phi_1 \Phi_1^\ast \Phi_2+\mu_2\Phi_1 H
H+{\mu_3^2\,\Phi_2^2}+{\mu_4\,\Phi_2^3}+$\\
\multicolumn{1}{|c|}{}                                     &
$+ \lambda_2 \left(\Phi_2 \Phi_2 \Phi_1^\ast \Phi_1\right) +\lambda_3 \left(H^\ast H \Phi_2 \Phi_2\right) +\lambda_4 \Phi_2\Phi_2\Phi_2\Phi_2+{\rm H.c.}     $\\ \hline
\end{tabular}
\caption{\label{tab:potentials}Scalar potential terms violating $U(1)_L$ and $U(1)_X$ accidental symmetries in the class-$\textbf{B}$ scenarios. We denote the term $\lambda_3$ in $\mathbf{B_{I}}$ by a $'$ as there are two independent contractions associated with the fields involved in the operator.}
\end{table}
We use $\lambda_i$ to denote the dimensionless coupling in the quartic terms and $\mu_i$ for the dimensionful coupling in the trilinear terms.\footnote{These couplings can be constrained from the stability of the potential. As the full scalar potential is very cumbersome, a complete study is beyond the scope of the present work.} The definition of the potential terms in tensor notation is shown in Appendix \ref{sec:contractions} for the $\mathbf{B_{I}}$ scenario.

As we will point out in Section~\ref{sec:majo}, the scalar potential symmetries are needed to recognize the presence of two different massive pseudo-Nambu-Goldstones in our class-\textbf{B} scenarios. Another interesting discussion about the scalar potential involves the terms linear in the new scalar multiplets. These terms will be important in Section~\ref{sec:inducedvevs}. In order to build a linear invariant term, we need at least two or three Higgs doublets. The $SU(2)$ transformation properties of the relevant products of Higgs doublets are:
\begin{itemize}
\item $H^2=\mathbf{2}\otimes \mathbf{2}=\mathbf{3}\oplus \mathbf{1}$\,,
\item $H^3=\mathbf{2}\otimes \mathbf{2} \otimes \mathbf{2}=\mathbf{4}\oplus \mathbf{2}\oplus \mathbf{2}^\prime$\,.
\end{itemize}
Let us start with the class-\textbf{A} scenarios, in which we add to the SM only one scalar quadruplet with hypercharge $-1/2$ or $-3/2$. Given the products shown above, the only allowed potential term linear in a quadruplet $\Phi_1$ is with three SM Higgs. For both hypercharge assignments $-1/2$ and $-3/2$, an invariant mixed potential term of the form $\Phi_1 H^3$ is allowed (see terms with $\lambda_8$ and $\lambda_6$ in Eqs. \eqref{eq:potentialA1L} and \eqref{eq:potentialA2X}, respectively). In the class \textbf{B} scenarios, with quadruplets ($\mathbf{B_{I}}$) we can again have $\Phi_i H^3$ terms ($\lambda_5$ and $\lambda_6$ terms in Table~\ref{tab:potentials}); with triplets, on the other hand, it is possible to construct invariant terms of the form $\mu \Phi_i H^2$, where $\mu$ is some dimensionful coupling. This is the case of $\mathbf{B_{II,V,VI}}$ scenarios. $SU(2)$ quintuplets do not allow for linear terms. 

However, there is an interesting scalar potential operator which can always be constructed in class-\textbf{B} scenarios, the mixed linear term $\lambda_1 \Phi_1\Phi_2H^2$. Indeed, as shown in Sec. 2.4, the contraction $(LL)_3(\Phi_1\Phi_2)_3$ is always allowed in our scenarios. Since the combination $(HH)$ has the same $SU(2)$ structure of the combination $(LL)$, the scalar potential must contain a term coming from the contraction $(HH)_3 (\Phi_1\Phi_2)_3$.

Given the scalar potential, the scalar masses can be computed after EWSB. If a linear term in $\Phi_i$ of the form $\Phi_i H^3$ is present in the potential (see Eq.~\eqref{eq:potentialA2X}), the scalar masses have an upper bound given by
\begin{equation}
    M_{\Phi_i} \simeq \sqrt{\lambda'}\cdot v\left( 1+\sqrt{\frac{\lambda^{\prime\prime}}{\lambda^\prime}\frac{v}{v_i}}\,\right)\,,
\end{equation}
where $v_{i}$ is the VEV of the new multiplet, and $\lambda'\text{ and }\lambda^{\prime\prime}$ are two of the scalar potential couplings, associated with $\Phi_i^2\,H^2$ and $\Phi_i\, H^3$, respectively. Constraints from perturbativity $\lambda^{\prime(\prime\prime)}<\sqrt{4\pi}$ imply $M_\Phi<10^3$ TeV for small values of the scalar VEV. The new VEVs are expected to be small and are discussed in Section~\ref{sec:rhoSec}. The above bound holds only for scenarios containing quadruplets, namely $\mathbf{A_I}$, $\mathbf{A_{II}}$ and $\mathbf{B_{I}}$. In the absence of the enhancement given by the $\mathcal{O}(v/v_i)$ term, we obtain a much more stringent bound, $M_\Phi<500$ GeV. Therefore, these scalars may be produced at colliders.

\subsection{The Pseudo-Nambu-Goldstone bosons}
\label{sec:majo}

In our scenarios, as already mentioned, new scalars can have non-zero lepton number. Subsequently, when they take a VEV, the spontaneous breaking of lepton number may give rise to a Nambu-Goldstone boson, i.e., the \emph{Majoron} \cite{Chikashige:1980ui,Gelmini:1980re,Diaz:1998zg,Escribano:2021uhf}. To illustrate this, in this section we analyse the scalar potential of one of the class-\textbf{A} scenarios as well as of one of the class-\textbf{B} scenarios as an example. The other scenarios belonging to a given class may be treated similarly, and they will provide similar results.

\subsubsection{The $\mathbf{A_I}$ scenario}
\label{majoA1}
In scenario $\mathbf{A_I}$, the SM Higgs doublet $H$ is accompanied by a scalar quadruplet $\Phi_1 = \mathbf{4}_{-1/2}$. The full scalar potential for this scenario is given in Eqs.~\eqref{eq:potentialA1L} and \eqref{eq:potentialA1X}. In this scenario, neglecting the scalar potential, we can identify two $U(1)$ symmetries: $U(1)_Y$ of hypercharge and $U(1)_L$ of lepton number in the full Lagrangian. Therefore, we expect two Nambu-Goldstone bosons after the scalars take VEVs: the would-be  one, $z$ associated with the spontaneous symmetry breaking of hypercharge gets $eaten$ by the $Z$-boson, while the other one is the so called \emph{Majoron}. When we consider the scalar potential terms, while the former remains massless since the potential is invariant under $U(1)_Y$, the latter can acquire a mass. The quartic terms $(\lambda_6,\lambda_7,\lambda_8)$ in Eq.~\eqref{eq:potentialA1X} play an important role: in the limit in which they vanish, lepton number is conserved in the scalar potential. Therefore, the breaking of the $U(1)_L$ is only spontaneous and the $Majoron$ is a massless Nambu-Goldstone boson. On the other hand, if those lepton number violating (LNV) terms are different from zero, the scalar potential presents an explicit breaking, so that the \emph{Majoron} acquires a non-zero mass, i.e., it becomes a \emph{pseudo-Majoron} $J$.

If there is no CP-violation in the scalar sector, we can assume all the parameters and VEVs
to be real and split the neutral scalar fields into their real and imaginary parts:
\begin{equation}
        H^0=v+S_H+i\,\chi\,,\quad \Phi_1^0=v_1+S_{\Phi_1}+i\,\eta\,.
    \label{eq:A1neutral}
\end{equation}
We focus only on the imaginary part of the neutral components of the scalar fields, since we are interested in discussing the Nambu-Goldstone bosons' mass spectrum. The CP-odd mass eigenstates $\{z,J \}$ are related to the corresponding weak eigenstates $\{ \chi,\eta\}$ as
\begin{equation}
    \begin{pmatrix}
        z\\
        J
    \end{pmatrix}=O \begin{pmatrix}
        \chi\\
        \eta
    \end{pmatrix}\,,
\end{equation}
where $O$ is the $2\times 2$ orthogonal matrix that diagonalizes the CP-odd neutral states squared mass matrix $\mathcal{M}_I^2$
\begin{equation}
    O \mathcal{M}_I^2 O^T=\text{diag}(m_z^2,m_J^2)\,,\quad \text{with}\, \quad O=\begin{pmatrix}
        \cos\alpha&\sin\alpha\\
        -\sin\alpha&\cos\alpha
    \end{pmatrix}\,.
\end{equation}
The masses of the $z$ and $J$ CP-odd mass eigenstates are
\begin{equation}
    m_z^0=0\,,\quad m_J^2=\dfrac{v\,(v^2+v_1^2)}{9v_1}\left(-3\sqrt{3}\,\lambda_8 +24\,\dfrac{v_1}{v}\lambda_7 +2\sqrt{3}\,\dfrac{v_1^2}{v^2}\lambda_6\right)\,.
    \label{eq:majoA1}
\end{equation}
As expected, we obtain a pure massless would-be Nambu-Goldstone boson $z$, associated with the spontaneous breaking of the hypercharge, and a massive pseudo-Nambu-Goldstone boson $J$, related to the explicit breaking of the lepton number in one unit, that we identify as the \emph{pseudo-Majoron}. Notice that its mass depends on the LNV quartic couplings of the potential and, as already mentioned, in the limit $(\lambda_6,\lambda_7,\lambda_8)\rightarrow0$, it becomes massless. The mixing angle $\alpha$ is given by
\begin{equation}
    \tan\alpha=\dfrac{v_1}{v}\, .
\end{equation}
Due to the strong hierarchy ($v_i\ll v$), which we will discuss in the next sections, the mixing between the two CP-odd scalar eigenstates is very suppressed. The same applies to the mixing between the CP-even ones.

\subsubsection{The $\mathbf{B_{I}}$ scenario}
\label{majoB1}

We now study the case of the $\mathbf{B_{I}}$ scenario as an example. We choose this specific class-\textbf{B} scenario since it is a natural extension of the $\mathbf{A_I}$ scenario discuss above. Indeed, in $\mathbf{B_{I}}$ we add to $\mathbf{A_I}$ a second multiplet, $\Phi_2 = \mathbf{4}_{-3/2}$. Moreover, the study of such a scenario is particularly interesting since we can infer the results on the Majoron of the scenario $\mathbf{A_{II}}$ setting to zero all the scalar potential couplings in which $\Phi_1$ appears. In this scenario, as in all class-\textbf{B} scenarios, there are three $U(1)$ symmetries: $U(1)_Y$ of hypercharge, $U(1)_L$ of lepton number, and an additional $U(1)_X$ accidental symmetry. 
In the limit in which all the violating terms in the potential go to zero, we expect to have three massless Nambu-Goldstone bosons; if instead we turn on $\lambda_1$ (Eq.~\eqref{eq:u1xcon})  we get a massive Nambu-Goldstone boson \emph{i.e.}, the \emph{pseudo-Majoron} $J$, associated with the explicit breaking of the lepton number. If $\lambda_i$ for $i=2,...,9$ are also different from zero (see Table~\ref{tab:potentials}), a second massive \emph{pseudo-Nambu-Goldstone} boson $\Omega$ is obtained, related to the $U(1)_X$ explicit accidental symmetry breaking.
As in Eq.~\eqref{eq:A1neutral}, we can split the neutral scalar fields into their real and imaginary parts:
\begin{equation}
    \begin{aligned}
        H^0=v+S_H+i\,\chi\,,\\
        \Phi_1^0=v_1+S_{\Phi_1}+i\,\eta\,,\\
        \Phi_2^0=v_2+S_{\Phi_2}+i\,\rho\,.
    \end{aligned}
\end{equation}
Due to the large number of operators in the scalar potential, the squared masses of the three CP-odd mass eigenstates $\{ z,J,\Omega\}$ are quite cumbersome. Therefore, we expand at leading order in $v^3$. Under such an assumption, the squared masses of $\{ z,J,\Omega\}$ become
\begin{equation}
    \begin{aligned}
        m_z^2=0\,,\quad m_J^2\simeq\dfrac{\lambda_5 v^3}{2\sqrt{3}v_1}\,,\quad m_\Omega^2 \simeq\dfrac{\lambda_6 v^3}{2v_2}\,.
    \end{aligned}
\end{equation}
As expected, we obtain a pure massless would-be Nambu-Goldstone boson $z$, associated with the spontaneous breaking of hypercharge, and two massive pseudo-Nambu-Goldstone bosons associated with the explicit breaking of the $U(1)_L$ and $U(1)_X$ symmetries, respectively. Even if at the leading order there is no dependence on the LNV potential coupling $\lambda_1$, in the limit of $U(1)_X$ conservation, namely when $\lambda_i$ with $i=2,...,9$ are zero, the pseudo-Majoron $J$ remains massive with $m_J^2\simeq \lambda_1 v^2 (v_1^2+v_2^2) /(2\,v_1\,v_2\sqrt{3})$. Defining $(\tan\alpha)_{uv}$ as the tangent of mixing between the CP-odd scalars $u$ and $v$, we obtain:
\begin{equation}
(\tan\alpha)_{\Phi_1\Phi_2}=\dfrac{\lambda_1\,v_1\,v_2}{\sqrt{3}\lambda_6\,v\,v_1+\lambda_5\,v\,v_2}\,,\quad (\tan\alpha)_{H\Phi_1}=2\dfrac{v_1}{v}\,,\quad (\tan\alpha)_{H\Phi_2}=6\dfrac{v_2}{v}\,.
\end{equation}
Given the strong hierarchy between the VEVs of the new scalars and SM Higgs one, it follows that the mixing among the CP-odd scalars is again very suppressed.

It is interesting to notice that whenever a massive pseudoscalar is included in our scenario, we might set lower bounds on the LNV couplings. Indeed, such particles cannot have a mass which is below 45 GeV, otherwise we would have observed them in $Z$-boson decays \cite{CMS:2022ett}.
If we have terms proportional to $\lambda{''}\Phi_i H^3$ in the potential ($\mathbf{A_I}$, $\mathbf{A_{II}}$, $\mathbf{B_{I}}$ scenarios), neglecting small corrections proportional to $v_{\Phi}$, we have
\begin{equation}
    m_J\simeq v\cdot \sqrt{\lambda''\frac{v}{v_{i}}+\mathcal{O}\left(\frac{v_i}{v}\right)}\,.
\end{equation}
This means that the lower limit on $\lambda''$, in the case of $v_\Phi\sim\mathcal{O}(1 \, \, \rm{GeV})$ is $\lambda''>10^{-4}$. However, this limit can be further lowered taking a smaller $v_i$. In other scenarios, where we can have terms like $\lambda_1 \Phi_1 \Phi_2 H^2$ and $\lambda' \Phi_i^2 H^2 \,(i=1,2)$, there is no enhancement from $v_i$ in the denominator, and the requirement $m_J>45$ GeV gives $\lambda_1,\lambda'>10^{-1}$.

\subsection{The VEVs of new scalars and the $\rho$ parameter}\label{sec:rhoSec}

The $\rho$ parameter is defined as $\rho=m_W^2/(c_W^2m_Z^2)$, where $m_W$ and $m_Z$ are the mass of the $W$ and the $Z$ bosons, respectively, and $c_W$ is the cosine of the Weinberg's angle. In the SM, $\rho=1$ due to {\it custodial symmetry} \cite{Veltman:1977kh,Sikivie:1980hm}; this is in agreement with the current experimental data,  $\rho=1.00017\pm0.00025$ \cite{ParticleDataGroup:2022pth}.\footnote{We do not take into account here the recent measurement of the $W$-boson mass from CDF \cite{CDF:2022hxs}, which will be discussed later.} When a new scalar multiplet that takes a VEV, as in our scenarios, is added to the SM, custodial symmetry may be broken and therefore there may be a modification of the $\rho$ parameter. Thus, the allowed parameter space of such scalar extensions is highly restricted, see for example Ref.~\cite{Dawson:2017vgm}. Each scalar multiplet with  weak isospin $I_j$, hypercharge $Y_j$ and VEV $v_j$ contributes at tree level as \cite{Langacker:1980js}
\begin{equation}
    \rho=\frac{\sum_j [(I_j(I_j+1)-Y^2_j]v_j^2}{2\sum_j Y^2_j v^2_j}\, .
    \label{rho}
\end{equation}
In order to not spoil the electroweak precision measurements, we need $\Delta\rho=\rho-1\ll1$. In addition, from the extraction of Fermi's constant, $G_F$, we have that $2\sum_j[I_j(I_j+1)-Y_j^2]v_j^2=(2\sqrt{2}G_F)^{-1}= (174 \text{ GeV})^2$. Therefore, it is clear that the VEV of any new scalar multiplet added to the SM must be rather small because the top mass and perturbativity require that the SM Higgs VEV is not much smaller than $174$ GeV.\footnote{Note that there are representations which maintain $\rho=1$ and have a neutral component, i.e. those that satisfy $3Y_j^2 = I_j(I_j+1)$, for example, a $\mathbf{2}_{1/2}$ (a 2HDM) and a $\mathbf{7}_2$.}  In the following, we quantify the constraints on the VEVs from the experimental value of the $\rho$ parameter, working in the regime $v \gg v_i$.
\begin{table}[!htb]
\centering
\begin{tabular}{|c|c|c|}
\cline{1-3}
\textbf{scenario}&$v_1^{\rm max} (\text{GeV})$&\textbf{Induced VEVs}\\ \hline\hline
$\mathbf{A_I}$&3.3&\ding{52}\\ \hline
$\mathbf{A_{II}}$&2.6&\ding{52}\\ \hline
\end{tabular}
\caption{95\% CL upper limits on the VEVs of the new scalar multiplets from the $\rho$ parameter and the possibility of inducing small VEVs in class-\textbf{A} scenarios.}\label{tab:rhoclass1}
\end{table}
First, let us consider the scenarios belonging to class-\textbf{A}, namely the ones in which we add a single scalar multiplet: $\Phi_1=\mathbf{4}_{-1/2}$ for $\mathbf{A_I}$ and $\Phi_1=\mathbf{4}_{-3/2}$ for $\mathbf{A_{II}}$. Recalling the general formula, Eq.~\eqref{rho}, we get
\begin{eqnarray}
    \Delta \rho_{\mathbf{A_I}}&=& \frac{3 v_1^2}{(4\sqrt{2}G_F)^{-1}-3v_1^2}\quad \mathrm{for} \, \, \mathbf{A_I}\,,\\
    \Delta \rho_{\mathbf{A_{II}}}&=& \frac{-3 v_1^2}{(4\sqrt{2}G_F)^{-1}+3v_1^2}\quad \mathrm{for} \, \, \mathbf{A_{II}} \,.
\end{eqnarray}
Using the experimental bound, we obtain the limits on the new VEVs, which are listed in Table~\ref{tab:rhoclass1}. As expected, the BSM VEVs are always much smaller than the SM Higgs one. Using this result for which $v_1\sim\mathcal{O}(1 \,\mathrm{GeV})$, it is also possible to estimate the energy scale of the UV complete theory, since from the Weinberg-like operators the neutrino mass can be written as $m_\nu\sim v v_1/\Lambda$ and $m_\nu\sim v_1^2/\Lambda$ (if the dimension-5 operators $\mathcal{O}_5^{(1)}$ and $\mathcal{O}_5^{(2)}$ are generated). Imposing $m_\nu\gtrsim\mathcal{O}(0.05 \, \mathrm{eV})$, we can estimate the order of magnitude of the energy scale $\Lambda$. Since we do not know in principle which of the operators gives the leading contribution to neutrino masses, we show here the mass scales $\Lambda^{(i)}$ one would obtain if only one of the operators exists. Taking the Wilson coefficients to be $\mathcal{O}(1)$, we obtain for perturbative Wilson coefficients $\Lambda^{(1)}\lesssim 10^{12}$ GeV for $\mathcal{O}_5^{(1)}$, which is present for both scenarios $\mathbf{A_{I,II}}$ and $\Lambda^{(2)}\lesssim10^{10}$ GeV for $\mathcal{O}_5^{(2)}$, which is present for the scenario $\mathbf{A_I}$ but not for the scenario $\mathbf{A_{II}}$.

Let us now consider the more complicated situation of class-$\mathbf{B}$ scenarios, where we add two additional scalars, namely $\Phi_1=\mathbf{N_1}_{Y_1}$ and $\Phi_2=\mathbf{N_2}_{Y_2}$. In this case, the isospin of the scalars can be deduced by the dimension of the $SU(2)$ representation using $\mathbf{N_j}=2 I_j+1$. Thus, Eq.~\eqref{rho} becomes:
\begin{equation}
\Delta\rho(\mathbf{N_1},\mathbf{N_2},v_1,v_2)=\dfrac{\left[\left(\frac{\mathbf{N_1}^2-1}{4}\right)-3Y_1^2\right]v_1^2+\left[\left(\frac{\mathbf{N_2}^2-1}{4}\right)-3Y_2^2\right]v_2^2}{\left(4\sqrt{2}G_F\right)^{-1}-v_1^2\left[\left(\frac{\mathbf{N_1}^2-1}{4}\right)-3Y_1^2\right]-v_2^2\left[\left(\frac{\mathbf{N_2}^2-1}{4}\right)-3Y_2^2\right]}\,.
    \label{conic}
\end{equation}
\begin{figure}[!htb]
\centering
{\includegraphics[width=.49\textwidth]{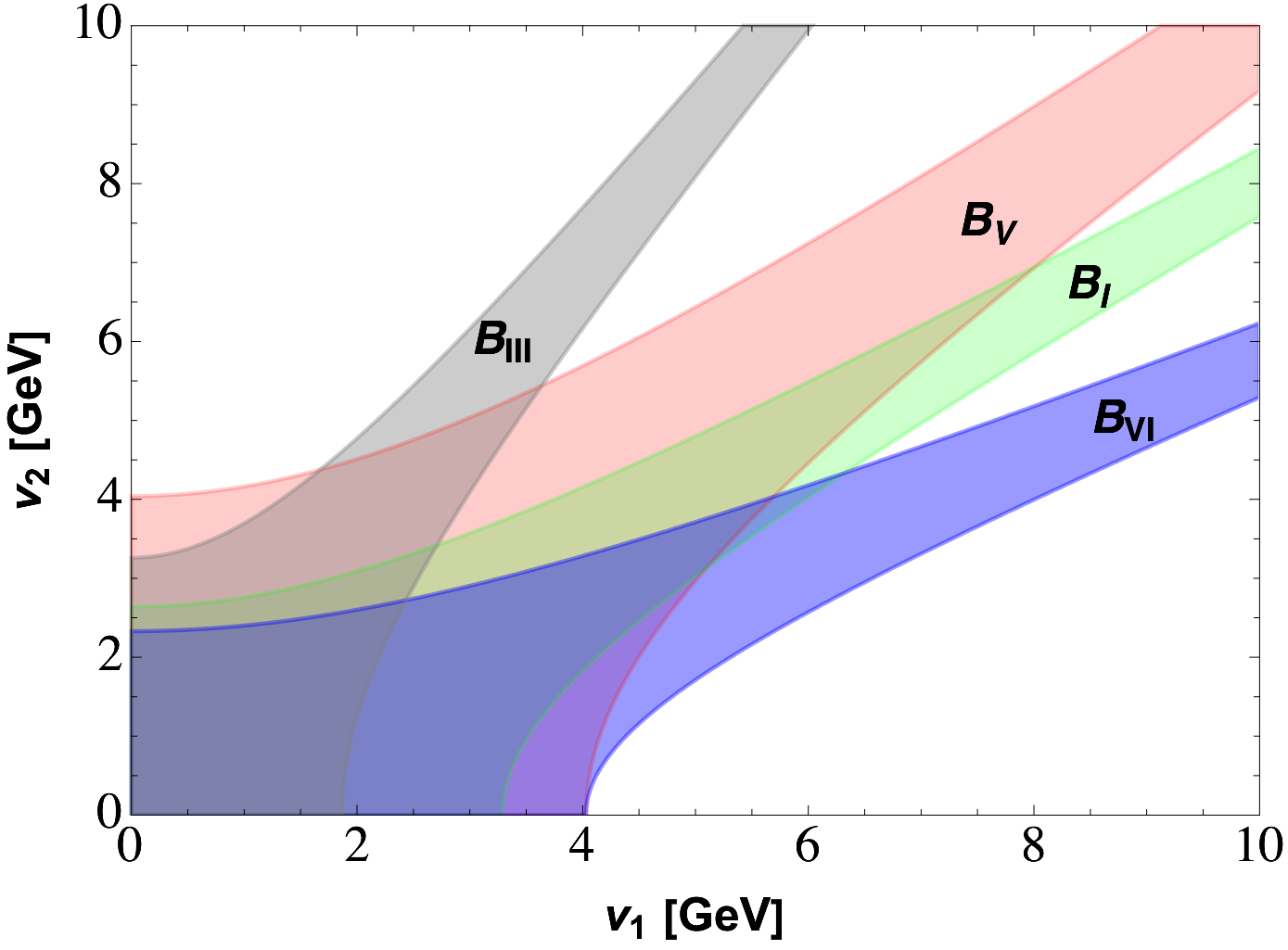}}
{\includegraphics[width=.49\textwidth]{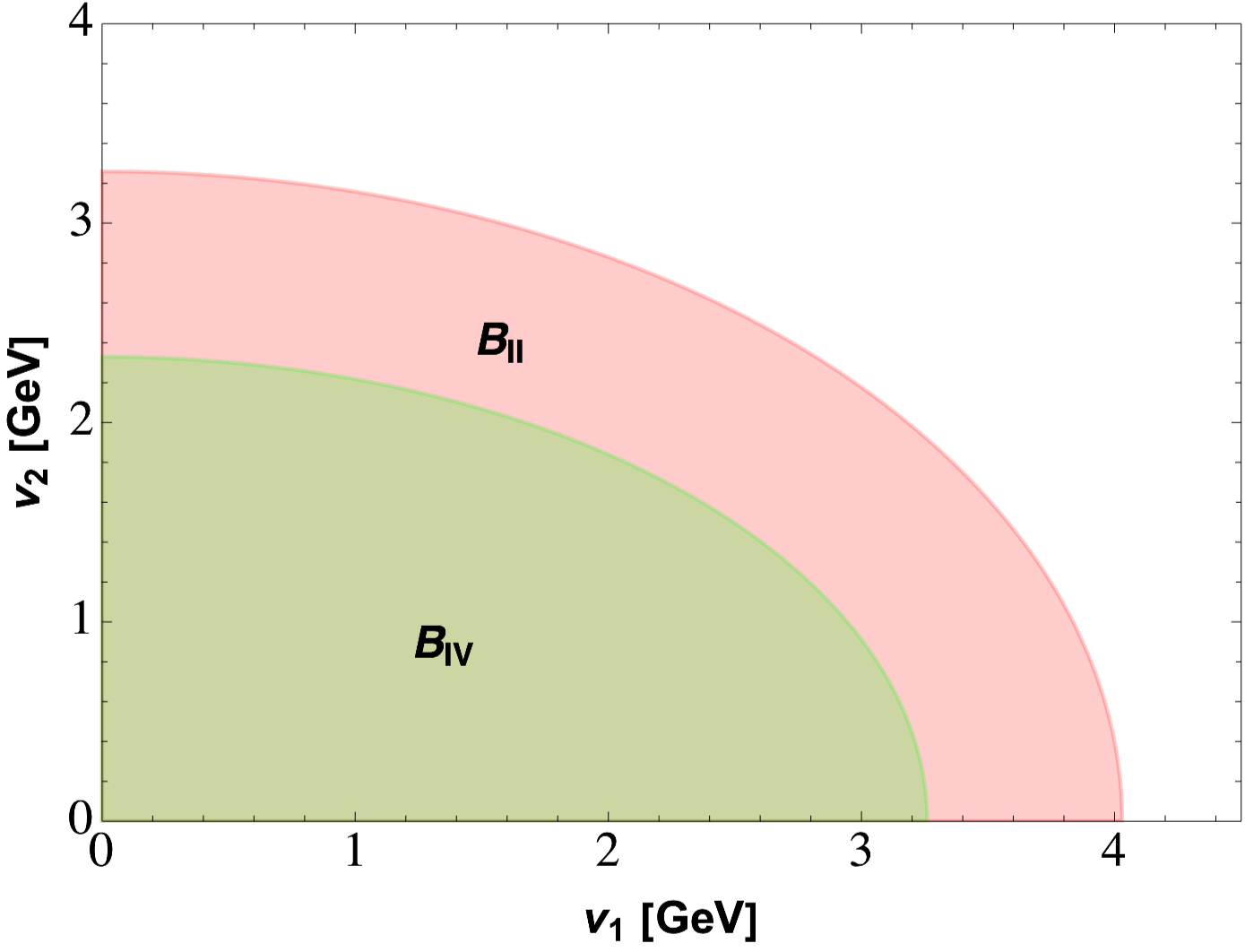}}
\caption{Allowed regions for the new scalar multiplet VEVs in class-$\mathbf{B}$ scenarios computed using the 95\% CL limits on the $\rho$ parameter, $\rho=1.00017\pm0.00025$ \cite{ParticleDataGroup:2022pth}. Two different structures emerge: hyperbolae for $\mathbf{B_{I,III,V,VI}}$ and ellipses for $\mathbf{B_{II,IV}}$, see text for details.} 
\label{fig:subfigrho}
\end{figure}
The allowed regions for each scenario $\mathbf{B}_i$ are displayed in Fig.~\ref{fig:subfigrho}. As can be observed, two shapes for the allowed regions in the plane $(v_1,v_2)$ emerge. We can understand them as follows. If the condition 
\begin{equation}
\sqrt{\frac{\mathbf{N_i}^2-1}{12}}>|Y_i|
\label{cond}
\end{equation}
holds for both $i=1$ and $i=2$, the allowed region is the area included in an ellipse of equation $v_1^2/a_1^2+v_2^2/a_2^2=1$. On the other hand, if the condition of Eq.~\eqref{cond} holds only for $i=1$ or $i=2$, the allowed region is the area delimited by two hyperbolae of equations $v_1^2/a_1^2-v_2^2/a_2^2=\pm 1$. In both cases, the coefficient $a_i$ is a function of the $\rho$ parameter, the Fermi constant $G_F$ and the quantum numbers of the new scalar field $\Phi_i$.
\begin{table}[!htb]
\centering
\begin{tabular}{|c|c|c|c|c|}
\cline{1-5}
\textbf{Scenario}&\textbf{Region}&  $a_1 (\text{GeV})$ &$a_2 (\text{GeV})$ &\textbf{Induced VEVs}\\ \hline\hline
$\bf B_I$&Hyperbola&3.3&2.6&\ding{52}\\ \hline
$\bf B_{II}$&Ellipse&4.0&3.3&\Large{$\bm{\sim}$}\\ \hline
$\bf B_{III}$&Hyperbola&1.9&3.3&\ding{56}\\ \hline
$\bf B_{IV}$&Ellipse&3.3&2.3&\ding{56}\\ \hline
$\bf B_V$&Hyperbola&4.0&4.0&\ding{52}\\ \hline
$\bf B_{VI}$&Hyperbola&4.0&2.3&\Large{$\bm{\sim}$}\\ \hline
\end{tabular}
\caption{Parameters of the 95\% CL allowed regions for the VEVs of the new scalar multiplets and the possibility for inducing a small VEV for the ${\bf B_i}$ scenarios. Scenarios marked with a \ding{52}, $\bm{\sim}$ and \ding{56} correspond to both, only one and no VEVs induced by linear terms in the scalar potential, respectively.  For discussion, see Section~\ref{sec:inducedvevs}.}
\label{tab:rhoclassB}
\end{table}
Using this information, we can infer bounds on the VEV values. 

When the allowed region is an ellipse (scenarios $\mathbf{B_{II}}$ and $\mathbf{B_{IV}}$), the values of the two semi-axes $a_{1,2}$ correspond to the maximum values that the VEVs can assume. On the other hand, when the allowed region is the space between two hyperbolae (scenarios $\mathbf{B_{I}}$, $\mathbf{B_{III}}$, $\mathbf{B_{V}}$ and $\mathbf{B_{VI}}$), the BSM VEVs can in principle also have large values if there is some partial cancellation among the contributions. However, they cannot exceed the limit imposed by the Fermi constant, namely 174 GeV. Moreover, given that the mass of the top quark is proportional to the SM Higgs VEV $v$, by perturbativity the latter is not allowed to be too small. Thus, the new VEVs $v_1$ and $v_2$ in the $\mathbf{B_{I,III,V,VI}}$ scenarios also cannot exceed roughly $\mathcal{O}(10\, \text{GeV})$. 

Analogous to the case of class-\textbf{A} scenarios, we can deduce the energy scale of the new physics by imposing that neutrino masses are $m_\nu\gtrsim\mathcal{O}(0.05 \, \mathrm{eV})$ and that the Wilson coefficients are of $\mathcal{O}(1)$. Since for class-\textbf{B} scenarios we can have the dimension-5 operator $\mathcal{O}_5^{(3)}$ from which $m_\nu\sim v_1 v_2/\Lambda$, we obtain for all the scenarios, an energy scale $\Lambda \lesssim \mathcal{O}(10^{10})$ GeV. The semi-axes of the conics which represent the allowed region for the new scalar VEVs obtained from EWPT for the class-\textbf{B} scenarios are listed in Table~\ref{tab:rhoclassB}. Notice that in scenario $\mathbf{B_{I}}$, the Wilson coefficients for {$\mathcal{O}_5^{(1)}$} and  $\mathcal{O}_5^{(2)}$ related to neutrino masses are not zero; given that the particle content of scenario $\mathbf{B_{I}}$ is the same as the class-\textbf{A} scenarios, the new physics energy scale related to {$\mathcal{O}_5^{(1)}$} and  $\mathcal{O}_5^{(2)}$ for $\mathbf{B_{I}}$ can be read in the discussion above. In this scenario, if all the operators are generated with $\mathcal{O}(1)$ Wilson coefficients, $\Lambda$ should be the largest, i.e., $10^{12}$ GeV.

\subsection{Naturally-induced small VEVs}
\label{sec:inducedvevs}

As studied above, EWPT constrain the VEVs of the new scalar multiplets to be much smaller than the Higgs doublet one. A natural way to have suppressed VEVs for the new multiplets is that they are induced by the Higgs doublet one through linear terms of the new fields in the potential. For the scenarios in which a linear term $\mu\, \Phi_i H^2$ (with $\mu$ a trilinear dimensionful coupling)  or $\lambda\, \Phi_i H^{3}$ (with $\lambda$ a dimensionless quartic coupling) is present, a small induced VEV will be generated,
\begin{align}
     v_i & \simeq \mu \,\dfrac{v^2}{2m_{\Phi_i}^2} \label{eq:H2}\,,\\
     v_i &\simeq \lambda\, \dfrac{v^3}{2m_{\Phi_i}^2}\,, \label{eq:H3}
\end{align}
where $m_{\Phi_i}$ is the mass of the scalar multiplet $\Phi_i$. In Section~\ref{sec:scalarV} we have defined the conditions under which these terms exist in the potential: $\mu\, \Phi_i H^2$ requires that $\Phi_i$ is a triplet, while $\lambda\, \Phi_i H^{3}$ requires a quadruplet. We also showed that, regardless of the $SU(2)$ representation, in our class-\textbf{B} scenarios there always exists the mixed term $\lambda_{1}\Phi_1\Phi_2 H^2$ (see Eq.~\eqref{eq:u1xcon}). This would allow to induce a VEV for $\Phi_2$ (or $\Phi_1)$ which will be suppressed with respect to $v_1$ ($v_2$),
\begin{equation}
    v_{2(1)}\simeq \lambda_{1} \,v_{1(2)}\,\dfrac{v^2}{2m^2_{\Phi_{2(1)}}}\,.
    \label{eq:mixed}
\end{equation}

Summarizing, we have (see also Table~\ref{tab:potentials}): 
\begin{itemize}
\item In the class-$\mathbf{A}$ scenarios, there exists a linear term in $\Phi_1$ with three Higgs doublets. These depend on the two couplings $\lambda_8$ and $\lambda_6$ for $\mathbf{A_I}$ and $\mathbf{A_{II}}$ (see Eq.~\eqref{eq:potentialA1X}), respectively. These terms can induce a naturally small VEV for $\Phi_1$, see Eq.~\eqref{eq:H3}.
\item In the $\mathbf{B_{I}}$ scenario, where we have two different quadruplets, there are both quartic linear terms ($\lambda_5$ and $\lambda_6$ terms in Table~\ref{tab:potentials} for $\Phi_1$ and $\Phi_2$) and a mixed term proportional to $\lambda_1$.
Again, in this case both VEVs can be induced by the two quartic terms, see Eq.~\eqref{eq:H3}, and are naturally suppressed. Another possibility is to induce one of the two VEVs with the first two terms and the second one with the mixed term. As already mentioned, it is expected that the VEV induced by the last term, Eq.~\eqref{eq:mixed}, is subdominant.
\item In our $\mathbf{B_{II}}$ and $\mathbf{B_{VI}}$ scenarios, the triplet $\Phi_1$ may have a linear term proportional to $\mu_2$ in both cases, but not the quintuplet $\Phi_2$. However, a mixed term between the triplet and the quintuplet is always contained in the potential, proportional to $\lambda_1$.
Thus, $v_1$ may be induced by the first term, Eq.~\eqref{eq:H2}, while $v_2$ by the mixed one, Eq.~\eqref{eq:mixed}, therefore both are naturally suppressed.
\item When none of the scalars are in the doublet, triplet or quadruplet representations, namely for the $\mathbf{B_{III}}$ and $\mathbf{B_{IV}}$ scenarios, we only have the mixed term with $\lambda_1$ in Eq.~\eqref{eq:u1xcon}. Thus, we can write a relation between the two BSM VEVs suppressing one of the two, but both of them cannot be naturally suppressed. In these scenarios, therefore, it is less natural to have small VEVs.
\item In scenario $\mathbf{B_{V}}$, where $\Phi_1$ and $\Phi_2$ are triplets, the scalar potential contains linear terms with two SM Higgses (with $\mu_2$ and $\mu_3$ as couplings for $\Phi_1$ and $\Phi_2$) and a mixed term with both new multiplets proportional to $\lambda_1$, see Eq.~\eqref{eq:u1xcon}.
In this scenario the two small VEVs can be directly induced by the SM Higgs via the first two potential terms proportional to $\mu_2$ and $\mu_3$. Another possibility is to induce the VEV for one of the scalars via the linear term first and the second one using the mixed term. Typically, in the latter case the second VEV is smaller than the first.
\end{itemize}
We point out the presence of small induced VEVs in the last column of the Tables~\ref{tab:rhoclass1} and \ref{tab:rhoclassB}. We use the symbols \ding{52}, $\bm{\sim}$ and \ding{56} to signal that both, only one and no VEVs are induced by the linear terms in the scalar potential, respectively. In Appendix \ref{sec:d>5} we report for each scenario the exact expressions for neutrino masses from the small induced VEVs.

\section{Phenomenology} \label{sec:pheno} 

In this section we study the phenomenological signatures associated to the different scalars present in the new operators. Given that perturbative unitarity constraints and some signals like $h\rightarrow \gamma \gamma$ depend on couplings of the scalar potential are mostly independent of neutrino masses and have been studied elsewhere for similar scenarios \cite{Arhrib:2011vc,Chiang:2012qz,Picek:2012ei,Ghosh:2018drw}, we will mainly focus on the signals of the new scalar multiplets at colliders and in EWPTs at loop level, which are directly related to neutrino masses.

\subsection{Collider searches of multiply-charged scalars}

The presence of multiple charged scalars belonging to different multiplets can lead to interesting phenomenological signatures at colliders. All the scenarios contain doubly-charged scalars, and except for $\mathbf{A_I}, \mathbf{B_{V}}$ and $\mathbf{B_{VI}}$, all the scenarios also contain a triply-charged scalar. Furthermore, $\mathbf{B_{III}}$ contains a quadruply-charged scalar. We are mainly interested in the phenomenology of scalars with charges 2 and above ($Q>2$), the decays of which can lead to multi-lepton events. Therefore, we do not consider the triplet $\mathbf{3}_0$ in the analysis below. The other triplet in our scenarios, $\mathbf{3}_{-1}$, has been widely studied in the literature in the context of Type-II seesaw (see Refs.~\cite{Garayoa:2007fw,FileviezPerez:2008jbu,Aoki:2011pz}), and is also not discussed in the analysis below. Also, we will assume that the scalars do not mix among themselves or with any other scalars in the scenario, which is justified given that the new VEVs are much smaller than the SM Higgs one. 

\subsubsection{Production}

The multiply-charged scalars can be pair-produced at the LHC via the Drell-Yann (DY) mechanism with an $s$-channel $\gamma/Z$ exchange, whereas the associated production of the scalars involves an $s$-channel $W^\pm$ boson exchange.\footnote{The scalars can also be photo-produced by photon fusion, e.g. via $\gamma \gamma \r \Phi^{\pm\pm\pm}\Phi^{\mp\mp\mp}\,, \Phi^{\pm\pm}\Phi^{\mp\mp}$. The squared matrix element for these processes is proportional to the fourth power of the charge and thus the cross-section is enhanced for scalars with large charges. These processes can be dominant for large masses and enhance the cross section by an order of magnitude for multiply charged scalars. However, given the small parton density of photon and the large uncertainty associated with these processes, we do not consider them in the following.} It can take place from a quark-antiquark initial state as
\begin{align}
    &q\bar{q} \r {\gamma, Z} \r \Phi^{\pm\pm\pm\pm}\Phi^{\mp\mp\mp\mp}\,,\Phi^{\pm\pm\pm}\Phi^{\mp\mp\mp}\,, \Phi^{\pm\pm}\Phi^{\mp\mp}\,,\Phi^{\pm}\Phi^{\mp}\,,\nonumber\\
    &q\bar{q'} \r {W^{\pm}} \r \Phi^{\pm\pm\pm\pm}\Phi^{\mp\mp\mp}\,,\Phi^{\pm\pm\pm}\Phi^{\mp\mp}\,, \Phi^{\pm\pm}\Phi^{\mp}\,.
\end{align}
These cross sections depend mainly on the mass and quantum numbers of the scalars (hypercharge and isospin of the multiplet), and due to being $s$-channel they are suppressed for heavier masses. 
\begin{figure}[!htb]
\centering
\includegraphics[width=0.48\linewidth]{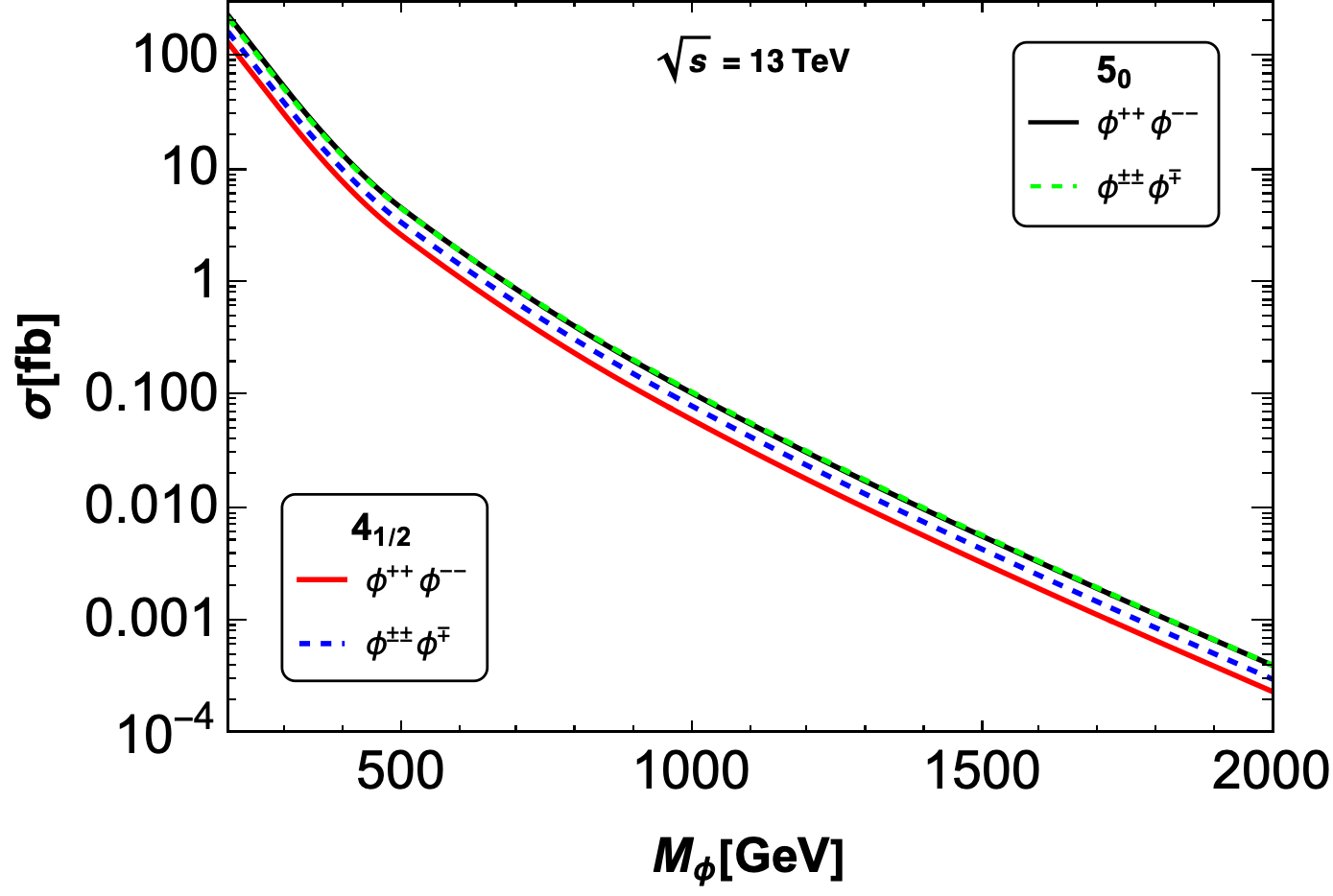}
\includegraphics[width=0.48\linewidth]{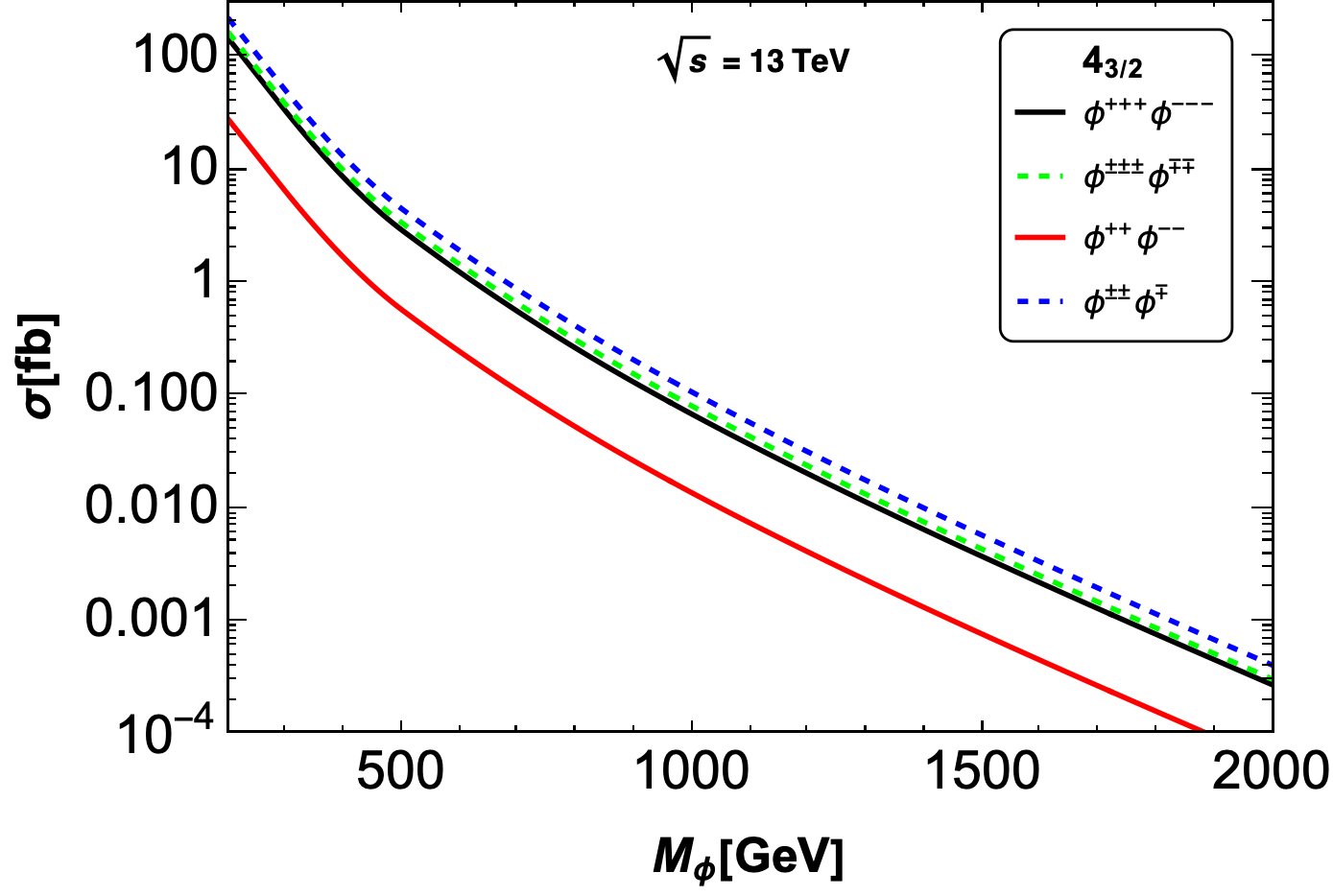}
\includegraphics[width=0.48\linewidth]{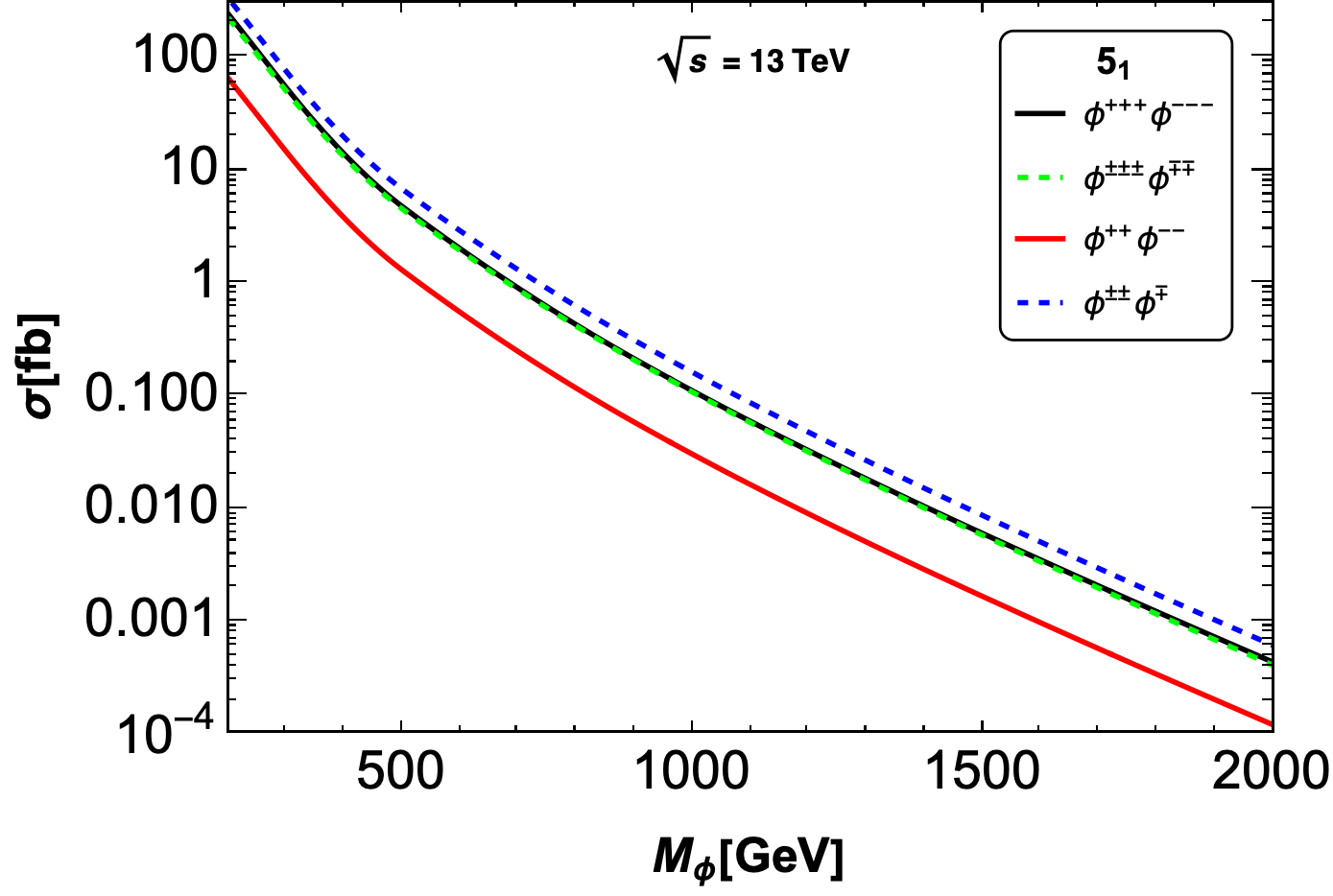}
\includegraphics[width=0.48\linewidth]{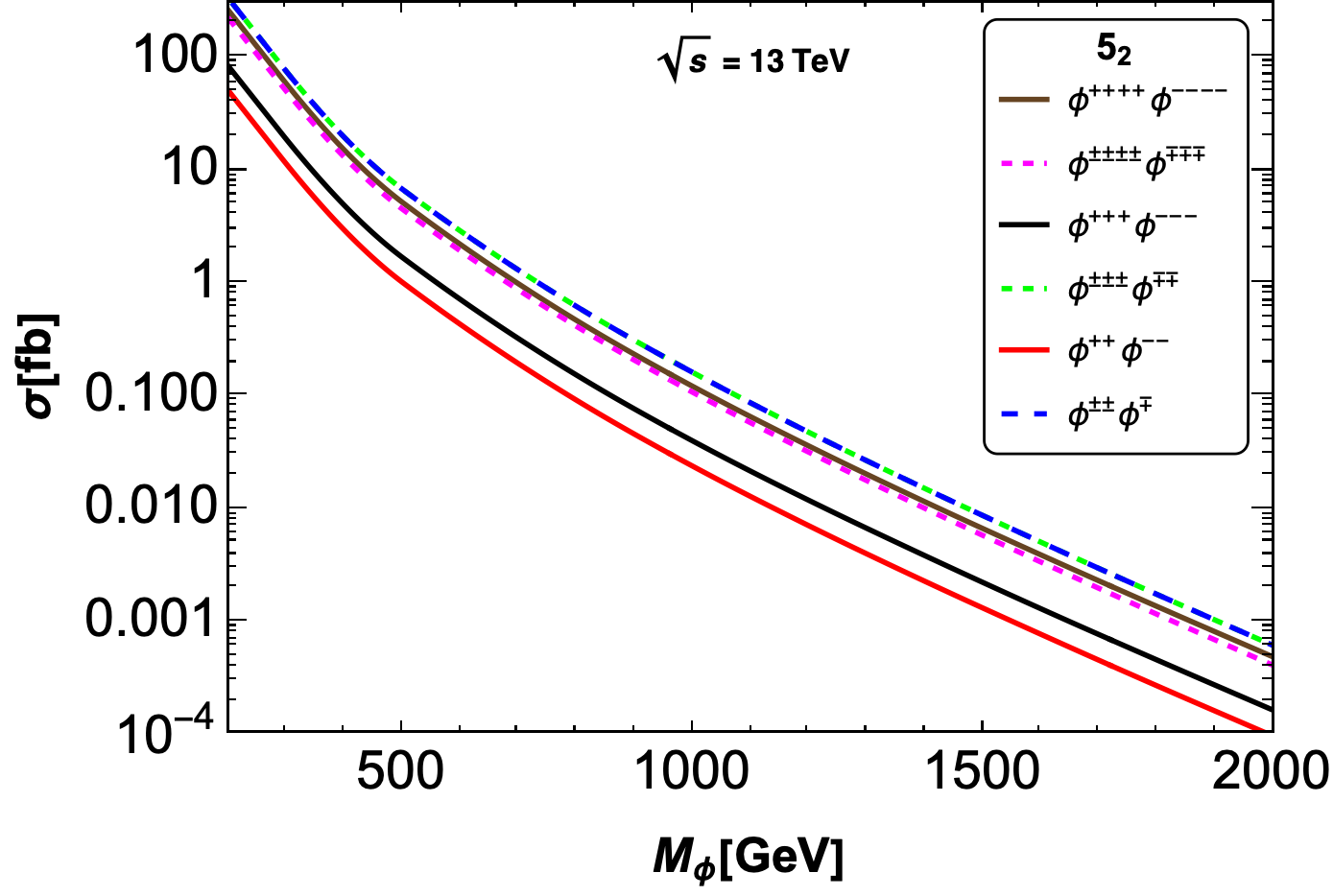}
\caption{Associated (dashed curves) and pair production (solid curves) of multi-charged scalars present in our scenarios at the $\sqrt{13}$ TeV LHC.}
\label{fig:collprod}
\end{figure}
We implement the models in \texttt{SARAH} \cite{Staub:2013tta} and numerically compute the cross-sections in \texttt{MadGraph5\_amC@NLO} \cite{Alwall:2014hca} using the \texttt{NNPDF23\_nlo\_as\_0119} PDF set \cite{Ball:2013hta}. The production cross sections for the charged scalars belonging to different multiplets at the LHC (for $\sqrt{s} = 13$ TeV), as a function of the scalar mass, are shown in Fig.~\ref{fig:collprod}. It can be seen that the pair production cross section for the scalars is directly proportional to the charge of the scalars, therefore, in all cases, we see that $\sigma(pp \rightarrow \phi^{n\pm}\phi^{n\mp})>\sigma(pp \rightarrow \phi^{m\pm}\phi^{m\mp})$ for $n>m\geq 2$, where $n,m$ indicate the scalar charges. However, this is not true for the associated production that takes place via $W^\pm$ exchange. From Table~\ref{tab:couplings}, one can see that the numerical factors associated with $\phi^{\pm\pm}\phi^{\mp}W^\mp$ couplings for different scalar multiplets are greater or equal to those of  $\phi^{\pm\pm\pm}\phi^{\mp\mp}W^\mp$, therefore we obtain $\sigma(pp \rightarrow \phi^{n\pm}\phi^{(n-1)\mp}) \geq \sigma(pp \rightarrow \phi^{m\pm}\phi^{(m-1)\mp})$ for $n<m$ with ${n,m\geq 2}$.

\subsubsection{Decays}

In this section, we discuss the decay modes of doubly, triply and quadruply-charged scalars. For scenarios belonging to class-$\mathbf{B}$ which contain two new scalars, we take their masses to be similar so that one does not decay into the other. First, we focus solely on the decays of doubly-charged scalars, which can lead to interesting signals such as same-sign dilepton events, which are free of SM background. The effective interaction $h_{\alpha\beta}^{\Phi_i}$ between a doubly-charged scalar ($\Phi_i^{\pm\pm}$) belonging to the multiplet $\Phi_i$ and a pair of leptons of the same charge ($l_\alpha^{\pm} l_\beta^{\pm}$) can be obtained from the operator $LL\Phi_i \Phi_j$ by choosing a particular contraction. This coupling can be expressed in terms of the neutrino mass matrix elements which are proportional to $v_i v_j/\Lambda$ and would also depend on the mass ordering (normal or inverted).  It can be written as
\begin{equation}
h_{\alpha\beta}^{\Phi_i} = \kappa_i \, \frac{{(m_\nu)}_{\alpha\beta}}{v_i}\,,
\end{equation}
where $\kappa$ is a scenario-dependent numerical factor given in Table~\ref{tab:couplings} and $v_i$ is the VEV taken by the neutral component of the multiplet $\Phi_i$. Similarly, if the other multiplet $\Phi_j$ also contains a doubly-charged component and the tensor expansion in terms of the multiplet components contains the term $l^\mp l^\mp \Phi_j^{\pm\pm}\Phi_i^0$, we can simply make the substitution $i \r j$ in the above formula to obtain the respective effective coupling. Note the inverse dependence on the VEV of the scalar. A large scalar VEV would highly suppress these decays, and the dominant decay channel for the doubly-charged scalars would be into a pair of $W$ bosons, which is directly proportional to the square of the VEV. 

Furthermore, depending on the mass splitting between the scalar components, cascade decays ($\Phi^{\pm\pm}\rightarrow \Phi^{\pm}+X$) can become important, for example, decays into pions. 
As we show below in Section \ref{sec:ewpt}, the mass splittings between the scalars need to be small ($< 20$ GeV) in order to satisfy the SM data, whereas, in the case of CDF data \cite{CDF:2022hxs}, the splittings are constrained to be between $10 - 40$ GeV. Therefore, in the latter case, the cascade decays might be important for a range of VEVs. However, in both cases, we take the scalar masses to be much greater than the splittings, so the decay into solely scalar states are suppressed. 

\begin{figure}[!htb]
    \centering
    \includegraphics[width=0.49\linewidth]{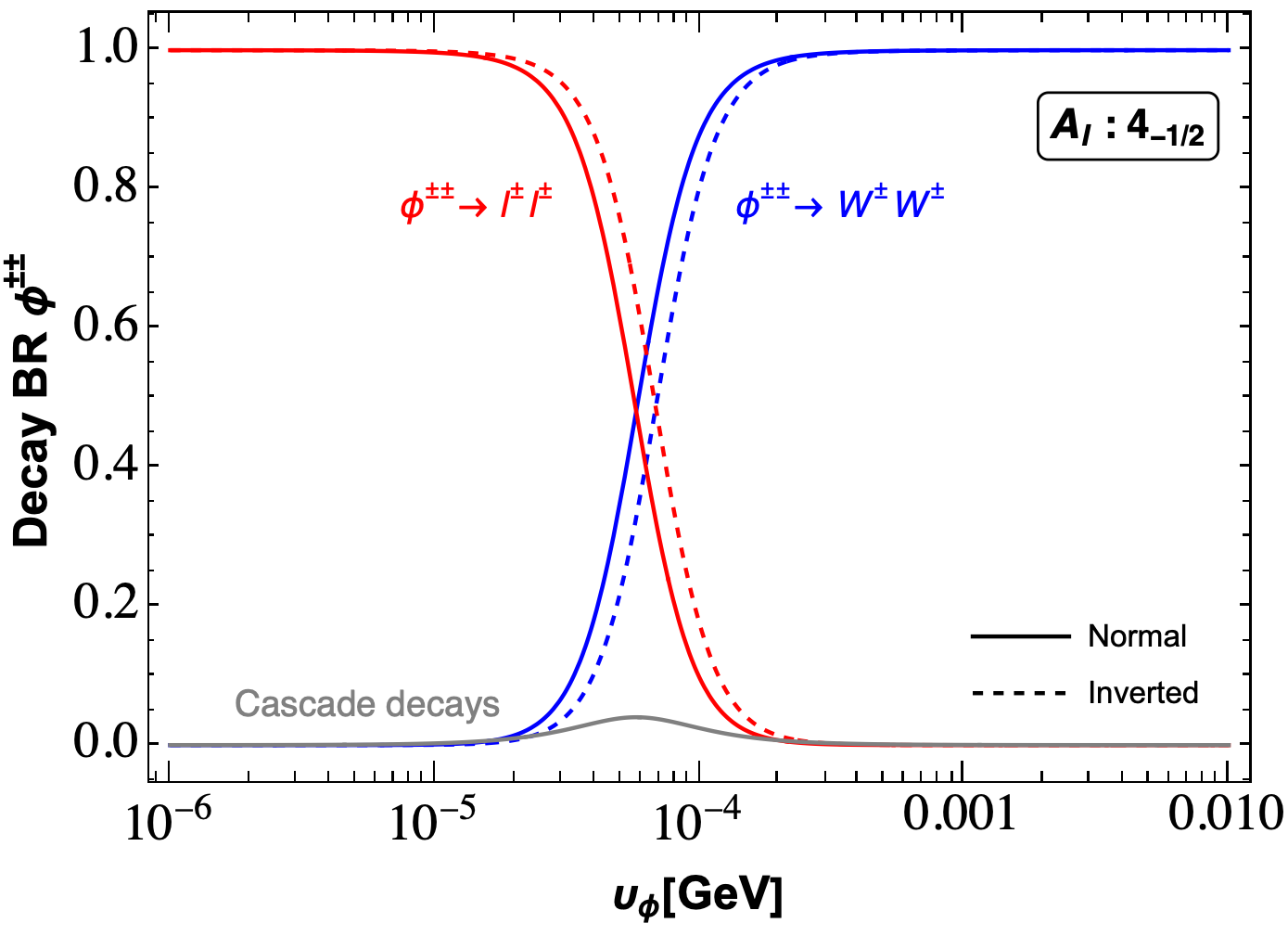}
    \includegraphics[width=0.49\linewidth]{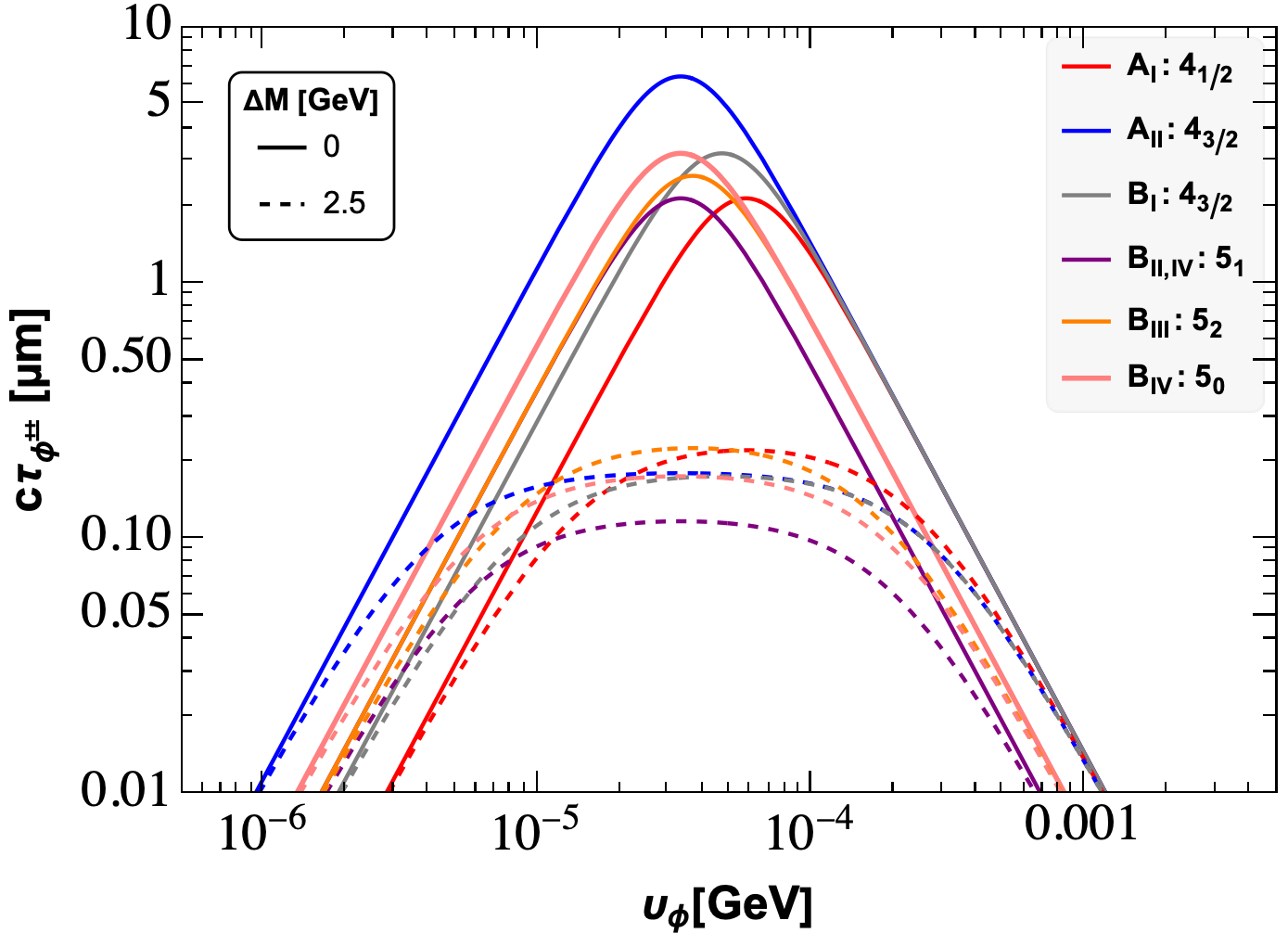}
    \caption{\textit{Left:} Branching ratio of the doubly-charged scalar belonging to $\mathbf{4}_{-1/2}$ in $\mathbf{A_I}$, with $M_\Phi = 500$ GeV and $\Delta M=1$ GeV. The solid and dashed curves represent the normal and inverted orderings of neutrino masses. The grey curves denote the cascade decays into pions, leptons and quarks. \textit{Right:} Dependence of the proper decay length of doubly-charged scalars on the scalar VEV for different scenarios. We take $M_{\Phi}=500$ GeV, $\Delta M= 0, 2.5$ GeV and consider normal ordering for neutrino masses.}
    \label{fig:decaydcs}
\end{figure}

The decay rates of doubly-charged scalars belonging to quadruplet and quintuplet representations into the different channels can be scaled by comparing to the decays of a doubly-charged scalar belonging to a scalar triplet ($\mathbf{3}_1$) as \cite{Ghosh:2018drw}
\begin{align}
    \Gamma (\dcs \r W^\pm W^\pm) &= S_{2W^\pm}^2 \frac{g^4 v_\Phi^2 M_{\dcs}^3}{16\pi M_W^4}\left(\frac{3M_W^4}{M_\dcs^4}-\frac{M_W^2}{M_\dcs^2}+\frac{1}{4}\right)\beta\left(\frac{M_W^2}{M_\dcs^2}\right)\,,\nonumber\\
    \Gamma (\dcs \r l_\alpha^\pm l_\beta^\pm) &= \frac{|{h_{\alpha\beta}|}^2 M_\dcs}{4\pi\, (1+\delta_{\alpha\beta})}\,,\quad \sum_{\alpha,\beta}\Gamma(\dcs \r l_\alpha^\pm l_\beta^\pm) = \kappa^2\frac{M_{\dcs}}{8\pi\, v_\Phi^2}\sum_{k=1}^3 m_k^2\,,\nonumber\\
    \Gamma (\dcs \r \Phi^\pm \pi^\pm) &= S_{\Phi^\pm W^\pm}^2 \frac{g^4 |V_{ud}|^2 \Delta M^3 f_\pi^2}{16 \pi M_W^4}\,,\nonumber\\
    \Gamma (\dcs \r \Phi^\pm l^\pm \nu_l) &= S_{\Phi^\pm W^\pm}^2 \frac{g^4 \Delta M^5}{240 \pi^3 M_W^4}\,,\nonumber\\
    \Gamma (\dcs \r \Phi^\pm q\bar{q'}) &= 3 \Gamma (\dcs \r \Phi^\pm l^\pm \nu_l)\,,
\end{align}
where $m_k$ is the individual mass of active neutrinos, $g$ is the $SU(2)$ coupling, $\Delta M$ is the mass splitting among the components and $S_{2W^\pm,\,\Phi^\pm W^\pm}$ is the scenario dependent factor related to the couplings $\dcs W^\mp W^\mp$ and $\dcs \Phi^\mp W^\mp$, respectively (see Table~\ref{tab:couplings}), $f_\pi =131$ MeV is the pion decay constant and $\beta(x)\equiv \sqrt{1-4x}$. We have also dropped the index $i$ and use $v_\Phi$ to denote the VEV of the scalar under consideration.
\begin{table}[!htb]
\centering
\begin{tabular}{|c|c|c|c|c|c|c|}
\cline{3-7}
\multicolumn{1}{c}{}&\multicolumn{1}{c}{} &\multicolumn{5}{|c|}{Couplings and scale factors}\\
\cline{1-7}
\textbf{Scenario}& \textbf{Scalar} & \thead{$|\kappa|$} & \thead{$S_{2W^\pm}$} & \thead{$S_{\Phi^\pm W^\pm}$} & \thead{$\Phi^{\pm\pm\pm}\Phi^{\mp\mp}W^\mp$} & \thead{$\Phi^{4\pm}\Phi^{3\mp}W^\mp$}\\ 
\hline\hline
$\mathbf{A_I}$ & $\mathbf{4}_{-1/2}$& $\sqrt{3}$& $\sqrt{6}$& $\sqrt{{3}/{2}}$ & $-$ & $-$\\ \hline
$\mathbf{A_{II}}$ &$\mathbf{4}_{-3/2}$& ${1}/{\sqrt{3}}$& $\sqrt{6}$& $\sqrt{2}$ & $\sqrt{{3}/{2}}$ & $-$\\ \hline
$\mathbf{B_{I}}$ &$\mathbf{4}_{-3/2}$& ${2}/{\sqrt{3}}$& $\sqrt{6}$& $\sqrt{2}$ & $\sqrt{{3}/{2}}$ & $-$\\ \hline
$\mathbf{B_{II}}$ &$\mathbf{5}_{-1}$& 1& $3\sqrt{2}$& $\sqrt{3}$ & $\sqrt{2}$ & $-$\\ \hline
$\mathbf{B_{III}}$ &$\mathbf{5}_{-2}$&$\sqrt{{3}/{2}}$& $2\sqrt{3}$&$\sqrt{3}$ & $\sqrt{3}$ & $\sqrt{2}$\\ \hline
$\mathbf{B_{IV}}$ &$\mathbf{5}_{-1}$& $1$& $3\sqrt{2}$&$\sqrt{3}$ & $\sqrt{2}$ & $-$\\ \hline
$\mathbf{B_{IV}}$ &$\mathbf{5}_{0}$& $\sqrt{{2}/{3}}$& $2\sqrt{3}$&$\sqrt{2}$ & $-$ & $-$\\ \hline
$\mathbf{B_{VI}}$ &$\mathbf{5}_{0}$& $\sqrt{6}$& $2\sqrt{3}$&$\sqrt{2}$ & $-$ & $-$\\ \hline
\end{tabular}
\caption{Numerical factors associated with the respective couplings that enter in the decay widths of the various charged scalars. The couplings with gauge bosons are computed with \texttt{SARAH} \cite{Staub:2013tta} and then scaled with that of the scalar triplet $\mathbf{3}_1$ for doubly-charged scalars in the fourth and fifth columns corresponding to $\dcs W^\mp W^\mp$ and $\dcs \Phi^\mp W^\mp$, respectively.}
\label{tab:couplings}
\end{table}
In the left (right) panel of Fig.~\ref{fig:decaydcs}  we show the branching ratio into different modes (proper decay length) of the doubly-charged scalar belonging to the multiplet $\mathbf{4}_{-1/2}$ of $\mathbf{A_I}$ (for all scenarios) as a function of the VEV of the multiplet. We take $M_\Phi = 500$ GeV and mass splitting $\Delta M=1\,(0,2.5)$ GeV in the left (right) panels. The grey curves on the left panel denote the cascade decays into pions, leptons and quarks.
It can be seen that, for small mass splitting, the cascade decays are suppressed, and the dominant decay modes are into gauge bosons and same-sign leptons. Neglecting cascade decays, the crossover from dominant decays into leptons and into gauge bosons takes place at a value of the VEV around
\begin{equation}\label{eq:VEVcrossover}
    v_{\dcs}^c \simeq 65{\rm~KeV}\,\left(\frac{\kappa}{S_{2W^\pm}}\right)^{1/2}\,\left(\frac{\sum_i m_i^2}{0.05^2 {\rm~eV}^2}\right)^{1/4}\,\left(\frac{500 {\rm~GeV}}{M_\dcs}\right)^{1/2}\,.
\end{equation}
The value of VEV at which this crossover takes place also corresponds to the point where the decay length of doubly-charged scalar is maximized. Thus, one can obtain an upper limit on the decay length of scalars. In the right panel of Fig.~\ref{fig:decaydcs} it can be seen that the maximum proper decay length ($c\tau$) that the doubly-charged scalars can achieve is less than $10~\mu{\rm m}$ for $M_\dcs =500$ GeV, corresponding to degenerate components of the multiplet. The decay length reduces for heavier masses as well as for larger mass splittings, as the cascade decays become dominant as shown by the dashed curves. It should be noted that CMS initiates the search for displaced vertices for a proper decay length of $\mathcal{O}(100\mu{\rm m})$ \cite{CMS:2021tkn}, and thus these scalars may not give any signals in those searches. However, prompt-lepton searches could be sensitive to them for $M_\dcs \gtrsim 200 {\rm~GeV}$.  

Next, we focus on the decays of the triply-charged scalars. Depending on the mass splitting among the components, the dominant channels for their decays would be three-body states such as $llW$ or $WWW$, and two-body decays into $\dcs W^{\pm\ast}$ and $\dcs \pi^\pm$ for $M_{\Phi^{\pm\pm\pm}}>M_\dcs$, which are similar to the decays of doubly-charged scalars into $\Phi^+ \pi^\pm$ (one can obtain the expressions using appropriate scaling factors). The decay widths of the triply-charged scalar present in $\mathbf{4}_{-3/2}$ of $\mathbf{A_{II}}$ for $M_\tcs \gg M_W$ are \cite{Arbelaez:2019cmj,Pan:2019wwv}
\begin{align}\label{eq:decaytcs}
    \Gamma(\tcs \r W^\pm W^\pm W^\pm) &= \frac{3g^6}{2048\pi^3}\frac{v_\Phi^2 M_\tcs^5}{M_W^6}\,,\nonumber\\
    \Gamma(\tcs \r W^\pm l^\pm l^\pm) &=\frac{g^2}{3072 \pi^3}\frac{M_\tcs^3\, \sum_i m_i^2}{v_\Phi^2\,M_W^2}\,.
\end{align}
Note that for large scalar masses, $M_\tcs \gg m_W$, the decay should be present in the gauge-less limit (Goldstone Boson Equivalence Theorem).\footnote{The expression for $\Gamma(\tcs \r W^\pm l^\pm l^\pm)$ in Eq.~\eqref{eq:decaytcs} differs by a factor of $2 (M_\tcs^2/M_W^2)$ from that in Refs.~\cite{Babu:2009aq,Ghosh:2017jbw}, where the decay rate vanishes in the gauge-less limit.}
The expressions for triply-charged scalars belonging to other multiplets can be obtained by rescaling the above expressions using the factors for couplings $\tcs\Phi^{\mp\mp} W^\mp$, $\dcs W^\mp W^\mp $ and $\dcs l^\mp l^\mp$, see Table~\ref{tab:couplings}.
\begin{figure}[!htb]
    \centering
    \includegraphics[width=0.48\linewidth]{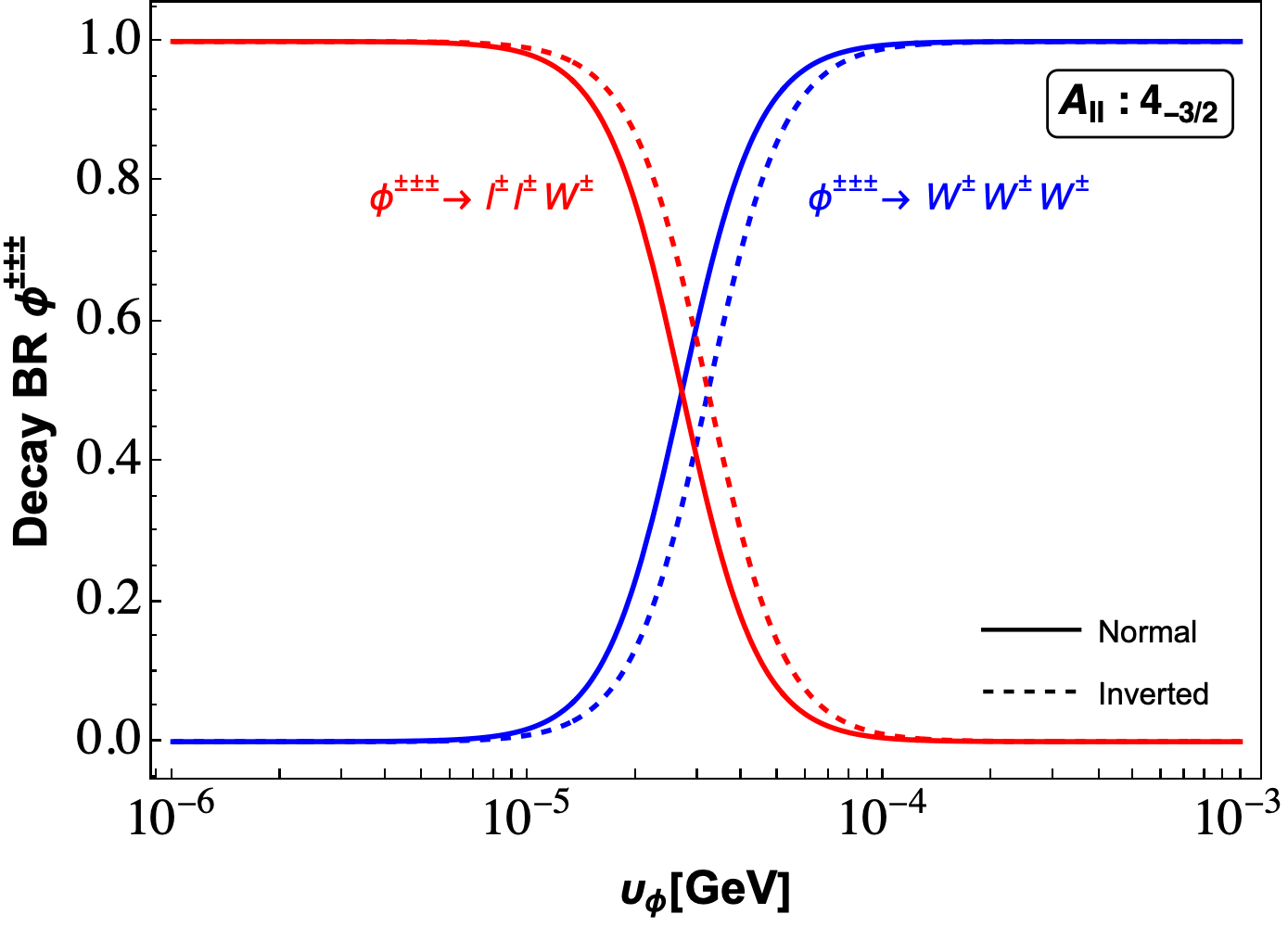}
    \includegraphics[width=0.48\linewidth]{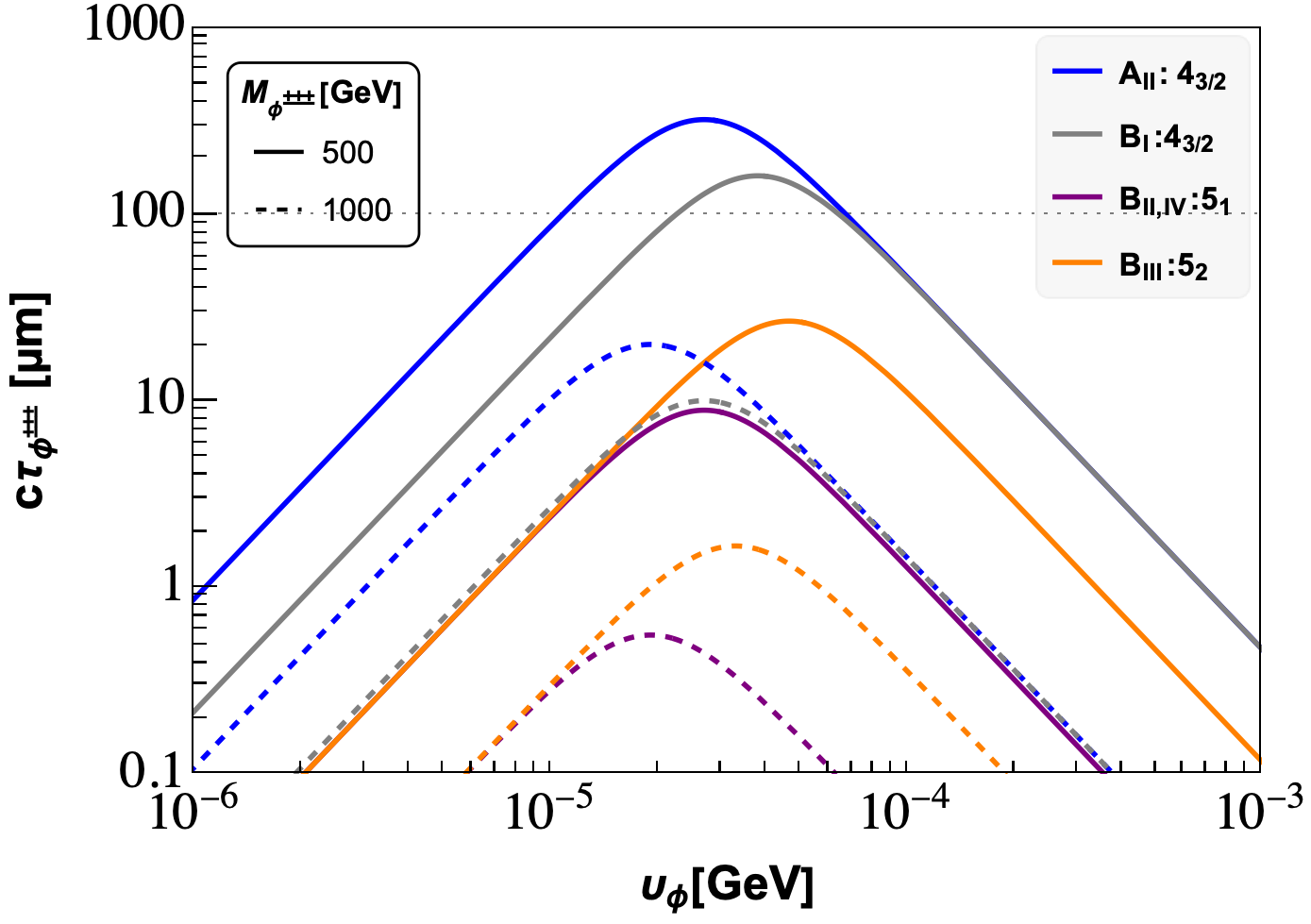}
    \caption{\textit{Left:} Same as the left panel of Fig.~\ref{fig:decaydcs} for triply-charged scalar decays in scenario $\mathbf{A_{II}}$ with $M_\Phi=500$ GeV. \textit{Right:} Dependence of the proper decay length of triply charged scalars on the scalar VEV in our scenarios for $M_\Phi=500,1000$ GeV, and normal ordering for neutrino masses.}
    \label{fig:decaytcs}
\end{figure}
Similar to the case of doubly-charged scalars, the decay length is maximized at the VEV corresponding to the crossover where decays to $WWW$ start dominating over decays to $llW$, which is given by
\begin{equation}\label{eq:VEV3crossover}
    v_{\phi^{\pm\pm\pm}}^c \simeq 55.4{\rm~KeV}\,\left(\frac{\kappa}{S_{2W^\pm}}\right)^{1/2}\,\left(\frac{\sum_i m_i^2}{0.05^2 {\rm~eV}^2}\right)^{1/4}\,\left(\frac{500{\rm~GeV}}{M_\dcs}\right)^{1/2}\,.
\end{equation}
In the left (right) panel of Fig.~\ref{fig:decaytcs}  we show the branching ratio into different modes (proper decay length) of the triply-charged scalar of scenario $\mathbf{A_{II}}$ ($\mathbf{A_{II}},\mathbf{B_{I,II,III,IV}}$) as a function of the VEV of the multiplet. Being three-body decays, these decay lengths $\sim \mathcal{O}(0.01-1{\rm~mm})$ are much larger than those of doubly-charged scalars, and may lead to displaced vertices.

Finally, we move our attention to the quadruply-charged scalar belonging to the multiplet $\mathbf{5}_2$. The decay width of these scalars are much smaller compared to other scalars in the scenarios due to phase-space suppression. Analogous to the cases discussed above, the dominant decay modes are the four body decays to $WWWW$ and $llWW$ which take place via a decay into a triply-charged scalar in association with a $W$ and the subsequent decays of the triply-charged scalar. The total decay width can be approximated by comparing to the three-body decays of triply-charged scalars as \cite{Arbelaez:2019cmj}
\begin{equation}
    \Gamma_{\rm tot}(\qcs)\sim \Gamma_{\rm tot}(\tcs)\,\frac{f(3)}{f(4)}\frac{g^2 M_{\qcs}^2}{M_W^2} \simeq 0.017\, \left(\frac{M_{\qcs}}{500 {\rm~GeV}}\right)^2\,  \Gamma_{\rm tot}(\tcs)\,,
\end{equation}
where $f(n)=4\,(4\pi)^{2n-3}\,(n-1)!(n-2)!$ accounts for the phase-space suppression for decays to $n$ body states. Hence, the decay length of quadruply-charged scalars is roughly 60 (15) times larger than that of $\tcs$ for masses of the order $500\, (1000)$ GeV, thus leading to $\mathcal{O}(1){\rm~cm}$ ($\mathcal{O}(1){\rm~mm}$) proper decay length. Therefore, scalars with masses$~< 1{\rm~TeV}$ can give rise to displaced vertices and can be searched for at the CMS detctor, whereas heavier quadruply-charged scalars may be looked for at prompt-lepton searches. We discuss the constraints from these searches below.

\subsubsection{Signatures}

The production and subsequent decays of multiply-charged scalars can lead to interesting signatures of new physics at the LHC. In Table~\ref{tab:collsig}, we collect the decay modes of various charged scalars and the signatures that can be obtained from pair and associated production of these scalars. As discussed above, the decay modes depend on the VEV of the corresponding scalar with leptonic decays dominating over the bosonic ones for low VEVs $v_\Phi<v_\Phi^c$ (see Eq.~\ref{eq:VEVcrossover}) and vice versa. If events like $l^\pm l^\pm W^\mp W^\mp$ are observed at the LHC, they can confirm LNV experimentally \cite{delAguila:2013yaa}.  
\begin{table}[!htb]
\begin{center}
\resizebox{\linewidth}{!}{
\begin{tabular}{|l|c|c|c|c|c|c| }
\hline
  Decays & $\Phi^{2-}\r 2l^-$ & $\Phi^{2-}\r 2W^-$ & $\Phi^{3-}\r 2l^-W^-$ & $\Phi^{3-}\r 3W^-$ & $\Phi^{4-}\r 2l^-2W^-$ & $\Phi^{4-}\r 4W^-$\\
\hline\hline
$\Phi^{2+}\r 2l^+$& $2l^+2l^-$ & $2l^+2W^-$ & $2l^+ 2l^- W^-$ & $2l^+ 3W^-$&\phantom{$-$}\ding{56} &\phantom{$-$}\ding{56}\\
\hline
$\Phi^{2+}\r 2W^+$&$2W^+ 2l^-$ &$2W^+ 2W^-$ & $2W^+ W^- 2l^-$ &$2W^+ 3W^-$ &\phantom{$-$}\ding{56} &\phantom{$-$}\ding{56}\\
\hline
$\Phi^{3+}\r 2l^+W^+$& $2l^+ 2l^- W^+$ &$2l^+2W^-W^+$ &$2l^+2l^-W^+W^-$ &$2l^+ 3W^- W^+$ &$2l^+ 2l^- 2W^-$ &$2l^+4W^- W^+$\\
\hline
$\Phi^{3+}\r 3W^+$& $3W^+2l^-$&$3W^+2W^-$ &$2l^-3W^+W^-$ &$3W^+ 3W^-$ &$2l^- 3W^+ 2W^-$ &$3W^+ 4W^-$\\
\hline
$\Phi^{4+}\r 2l^+2W^+$ &\ding{56} &\ding{56} &$2l^+2l^-2W^+W^-$ &$2l^+2W^+3W^-$ &$2l^+2l^-2W^+2W^-$ &$2l^+2W^+4W^-$ \\
\hline
$\Phi^{4+}\r 4W^+$ &\ding{56} &\ding{56} &$2l^-4W^+W^-$ &$4W^+3W^-$ &$2l^-4W^+2W^-$ &$4W^+4W^-$ \\
\hline
\end{tabular}}
\caption{Signatures of new physics from the pair and associated production of charged scalars and their subsequent decays to leptonic or bosonic modes.}
\label{tab:collsig}
\end{center}
\end{table}
Further leptonic decays of $W^{\pm}$ can lead to multi-lepton events with multiple same-sign leptons in the final state, which have a very small SM background. For example, the pair and associated production of quadruply-charged scalars can lead to final states with 0-8 leptons including same-sign dileptons ($SS2l$), tri-lepton (SS$3l$) and tetra-lepton events (SS$4l$). Moreover, the final states from the pair production of these multi-charged scalars can all lead to lepton flavor violating 4-lepton events of the form $l_i^\pm l_i^\pm l_j^\mp l_j^\mp$ and $l_i^\pm l_j^\pm l_j^\mp l_j^\mp$ $(i\neq j)$ from the on-shell/off-shell leptonic decays of the doubly-charged scalars, corresponding to the diagonal and off-diagonal matrix elements of the light neutrino mass matrix $m_\nu$.

The ATLAS and CMS collaborations search for doubly-charged scalars in multi-lepton finals states at $\sqrt{s}=13$ TeV, corresponding to an integrated luminosity of $139{\rm~ fb}^{-1}$ \cite{ATLAS:2022pbd} and $12.9{\rm~ fb}^{-1}$ \cite{CMS:2017pet}, respectively. The ATLAS search focuses on the pair production of doubly-charged scalars and their same-sign leptonic decays $\dcs \r l^\pm l'^{\pm}$ with $l,l'=e,\mu,\tau$ in two-, three-, four-lepton channels, considering only $e,\mu$ in the final state and assuming a 100\% branching ratio to leptons, with equal branching ratio to each possible leptonic channel. In the absence of any positive signal, limits can thus be imposed on the doubly-charged scalar pair production times the branching ratio to leptons to derive a lower bound on the doubly-charged scalar mass, which in the context of left-right symmetric Type-II models comes out to 1080 GeV \cite{ATLAS:2022pbd}. 
\begin{figure}[!htb]
\centering
\includegraphics[width=0.49\linewidth]{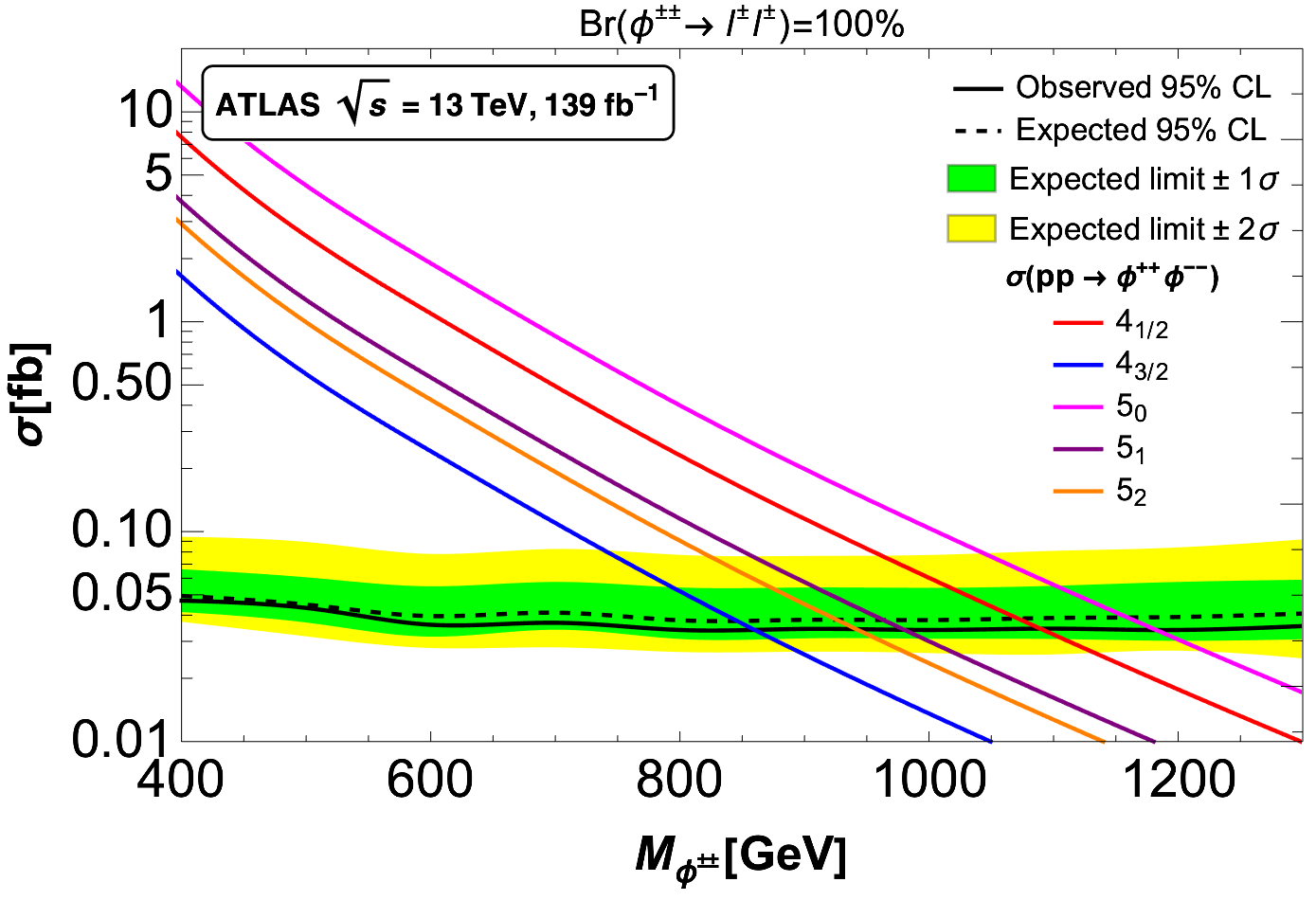}
\includegraphics[width=0.49\linewidth]{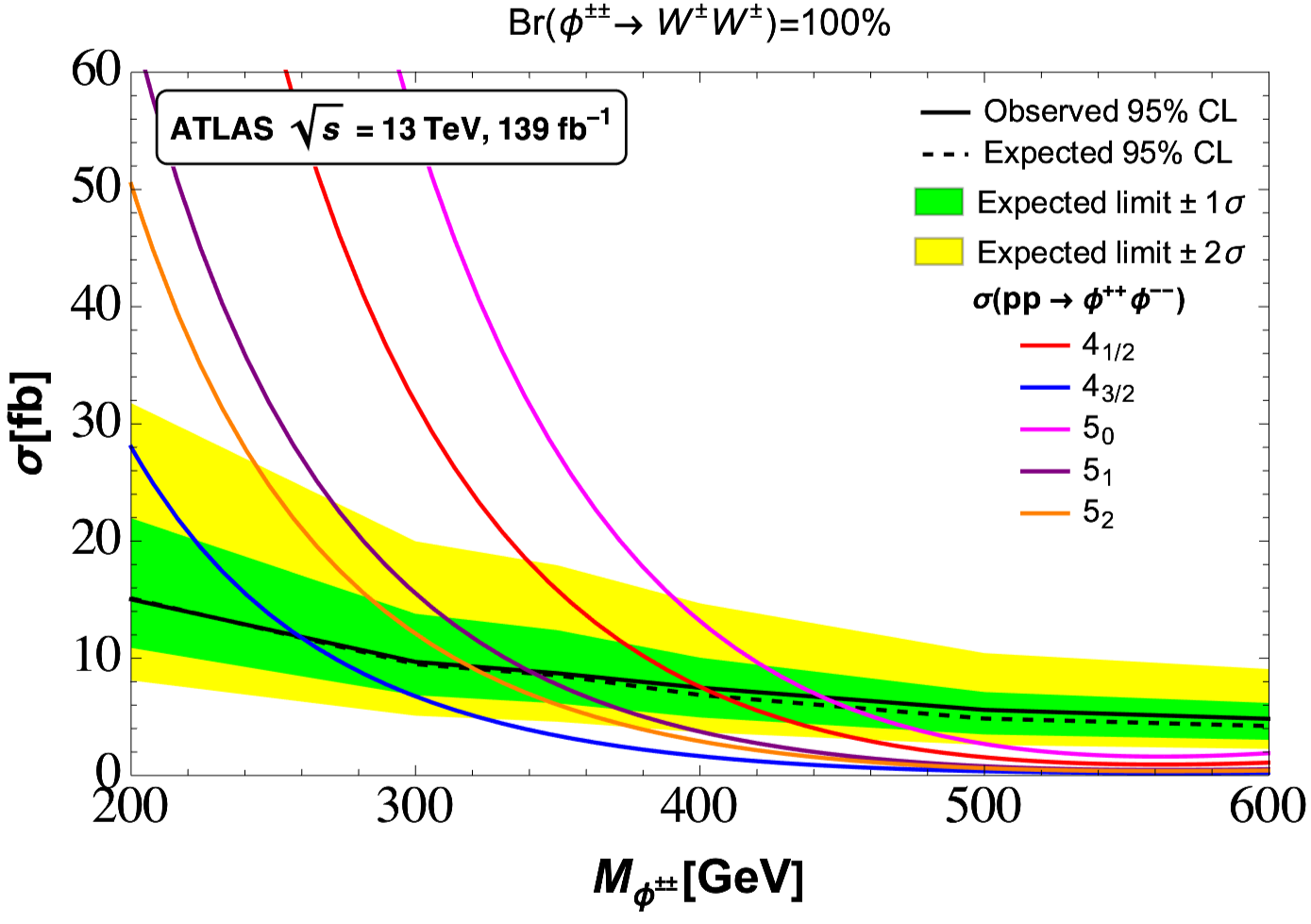}
\caption{The observed and expected 95\% CL upper limits on the production cross-section of doubly-charged scalars times their branching ratio into same-sign dileptons (\emph{left}) and to a pair of $W$ bosons (\emph{right}) for different multiplets. The region above the observed limit is excluded. The green and yellow band represent the expected exclusion curve within one and two standard deviations \cite{ATLAS:2022pbd,ATLAS:2021jol}.}
\label{fig:searchdoub}
\end{figure}
Given that the doubly-charged scalars of our scenarios exhibit similar decays, we can extend the ATLAS analysis to derive a limit on $\sigma(pp \r \dcs \Phi^{\mp\mp}) \times {\rm BR}(\dcs \r l^\pm l^\pm)$ and thus on $M_\dcs$ belonging to different multiplets,\footnote{Since the mass splittings among the various components of the multiplets is expected to be small from EWPTs, obtaining a bound on the doubly-charged scalar mass also gives an idea about the mass ranges of the other components of the multiplet.} assuming that their induced VEV is extremely small so that ${\rm BR}(\dcs \r l^\pm l^\pm)=100\%$. In the left plot of Fig.~\ref{fig:searchdoub}, we compare the theoretical pair production of doubly-charged scalars belonging to different multiplets (see Fig.~\ref{fig:collprod}) with the ATLAS results. The solid (dashed) lines represent the observed (expected) 95\% CL upper limits on the $\dcs$ pair production as a function of $M_\dcs$, and the green and yellow bands correspond to $\pm 1\sigma$ and $\pm 2\sigma$ uncertainty around the expected limit. It can be seen that $M_\dcs < 1090$ GeV is excluded for the doubly-charged scalars belonging to the multiplet $\mathbf{4}_{1/2}$.

The ATLAS collaboration also searched for doubly- and singly-charged Higgs decaying into vector bosons in multi-lepton final states at $\sqrt{s}=13$ TeV with an integrated luminosity of 139 ${\rm fb}^{-1}$, considering pair production of doubly-charged Higgs and its associated production with a singly-charged scalar in the context of the Type-II seesaw model \cite{ATLAS:2021jol}. Similarly, in the absence of any deviation from SM prediction, we can extend the analysis to the doubly-charged scalars of the different multiplets to obtain naive bounds on their masses, assuming that ${\rm BR}(\dcs \r W^\pm W^\pm)=100\%$, i.e., in the region of higher scalar VEVs. We compare the theoretical pair production cross-sections of the doubly-charged scalars with their results and show it in the right plot of Fig.~\ref{fig:searchdoub}, where the lines and bands have the same meaning as discussed above. 
\begin{table}[!htb]
\begin{center}
{\small
\begin{tabular}{|c|c|c| }
\cline{2-3}
\multicolumn{1}{c}{} &\multicolumn{2}{|c|}{\bf Limits on $M_\dcs~[{\rm GeV}]$}\\
\cline{1-3}
\hline
{\bf Multiplet} & BR$(\dcs \r l^\pm l^\pm)=100\%$ & BR$(\dcs \r W^\pm W^\pm)=100\%$\\
\hline\hline
$\mathbf{4}_{1/2}$ & 1090 & 400\\
\hline
$\mathbf{4}_{3/2}$ & 860 & 260\\
\hline
$\mathbf{5}_{0}$ & 1180 & 440\\
\hline
$\mathbf{5}_{1}$ & 980 & 340\\
\hline
$\mathbf{5}_{2}$ & 940 & 320\\
\hline 
\end{tabular}}
\caption{95\% CL exclusion limits on the mass of doubly-charged scalars using the ATLAS results at $\sqrt{s}=13$ TeV with $139{\rm~fb}^{-1}$ of integrated luminosity for different $SU(2)_L$ multiplets, under the assumption that the branching ratio to same-sign leptons or a pair of gauge bosons is 100\%. Therefore, the actual limit is expected to lie between these extreme cases.} 
\label{tab:bounddcs}
\end{center}
\end{table}
The lower bound on $M_\dcs$ can be obtained similarly to the case above and are found to be much less stringent, for example, $M_\dcs < 400$ GeV is excluded for the doubly-charged scalars belonging to the multiplet $\mathbf{4}_{1/2}$. The exclusion limits on $M_\dcs$ belonging to different multiplets from both these ATLAS searches are summarized in Table~\ref{tab:bounddcs}. These exclusion limits suffice to provide a rough estimate of the mass scales of the new scalars. These searches may help in distinguishing the doubly-charged scalars from different multiplets, since, depending on the presence of other highly-charged components in the multiplets and the mass hierarchy among them, the reconstruction of $M_\dcs$ from the same-sign dilepton invariant mass may not always be possible. A study to discriminate lepton number violating scalars at the LHC, assuming that there are no excitations with $Q>2$ was done in Ref.~\cite{delAguila:2013yaa}.

The search for higher-charged ($Q>2$) scalars in the scenarios and the potential reach of the LHC in discovering them requires a more dedicated collider study in the context of a specific model and is beyond the scope of current work. We refer the reader to Refs.~\cite{Bambhaniya:2013yca,Ghosh:2018drw} for the LHC phenomenology of scenario $\mathbf{A_{II}}$.

\subsection{Electroweak Precision Tests at loop level}\label{sec:ewpt}

The addition of heavy $SU(2)$ scalar multiplets affect the oblique parameters (namely $S,~T$ and $U$) that parameterize the effect of new physics on EW parameters. The new scenarios can be constrained by demanding these corrections to be small so that they do not distort the EW predictions of the SM. The oblique parameters are given by \cite{Peskin:1991sw}
\begin{equation}
    \begin{aligned}
     \alpha S &\equiv 4e^2 \,\frac{d}{dp^2}l\left[\Pi_{33}(0)-\Pi_{3Q}(0)\right]_{p^2=0}\,,\nonumber\\
     \alpha T &\equiv \frac{e^2}{s_W^2 c_W^2 M_Z^2}\,\left[\Pi_{11}(0)-\Pi_{33}(0)\right]\,,\nonumber\\
     \alpha U &\equiv 4e^2 \,\frac{d}{dp^2}  \left[\Pi_{11}(0)-\Pi_{33}(0)\right]_{p^2=0}\,,
    \end{aligned}
\end{equation}
where $\alpha$ is the fine structure constant, $s_W=\sin{\theta_W},c_W=\cos{\theta_W}$, and $\Pi_{IJ}(I,J=1,3,Q)$ are the vacuum polarisation amplitudes.

The $\rho$ parameter, which can be written in terms of $T$ as $\rho \equiv 1 + \alpha T$, is equal to 1 at tree level in the SM, but can receive further corrections at the one-loop level. In our case, the leading contribution to $T$ at tree level due to addition of new scalar multiplets with isospin $N_i$ and hypercharge $Y_i$ is given by Eq.~\eqref{conic}. It can be seen that, except for scenarios containing just scalar doublets, $\Delta \rho \neq 1$ at tree level for larger scalar multiplets obtaining VEVs, hence breaking the custodial symmetry, i.e. $m_W \neq m_Z c_W$.

\begin{figure}[!htb]
\centering
\includegraphics[scale=0.36]{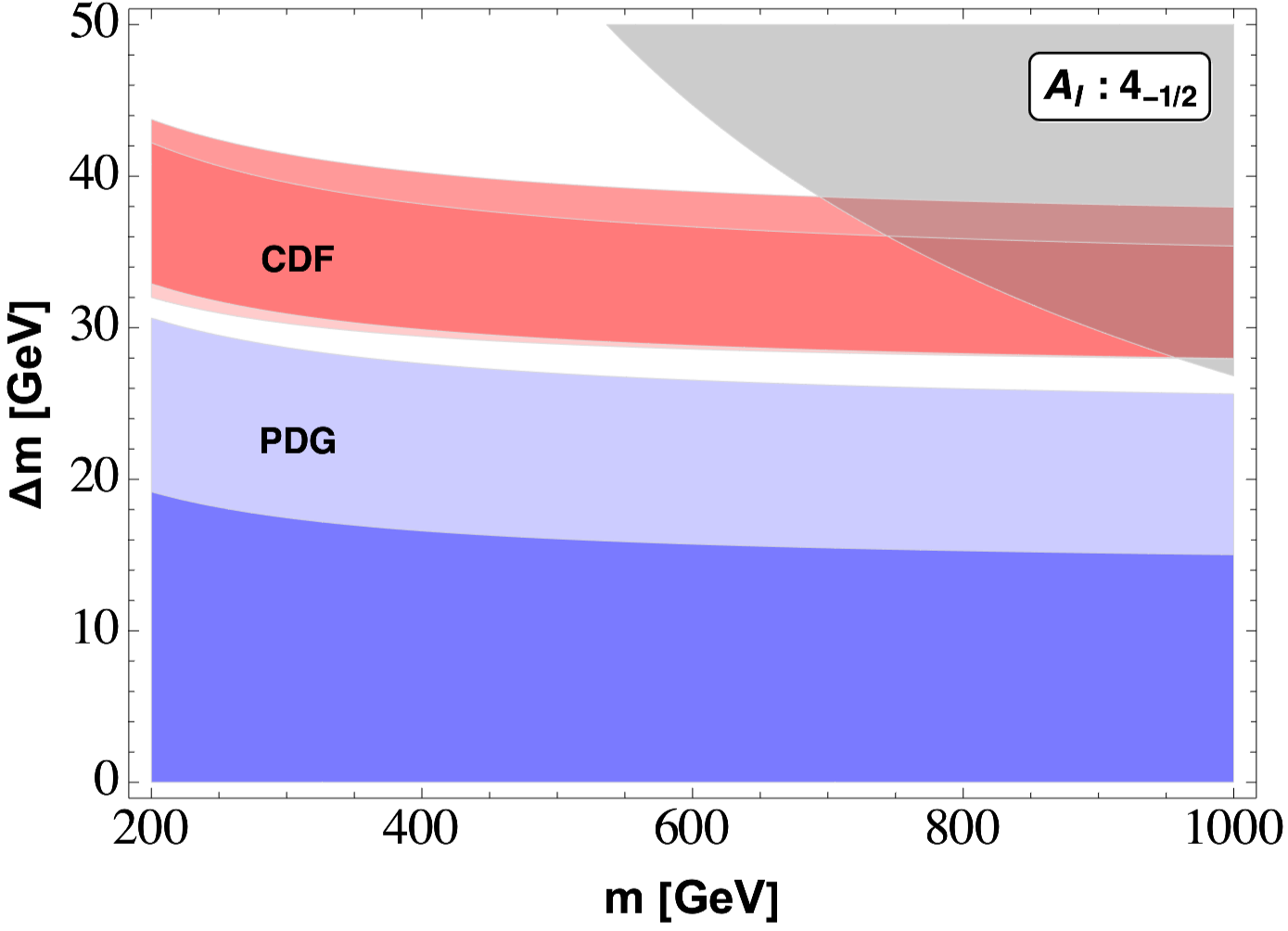}
\includegraphics[scale=0.36]{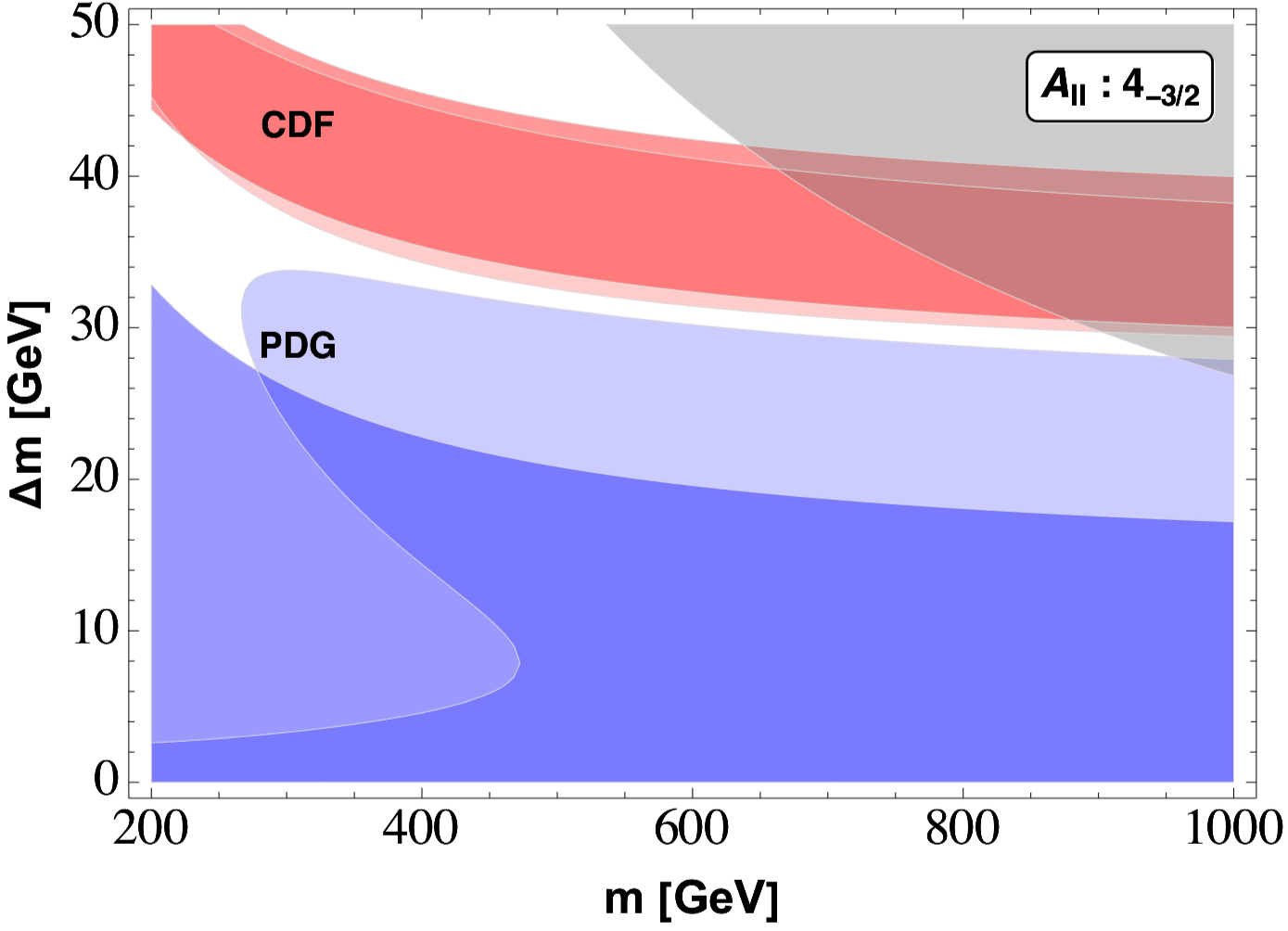}
\includegraphics[scale=0.36]{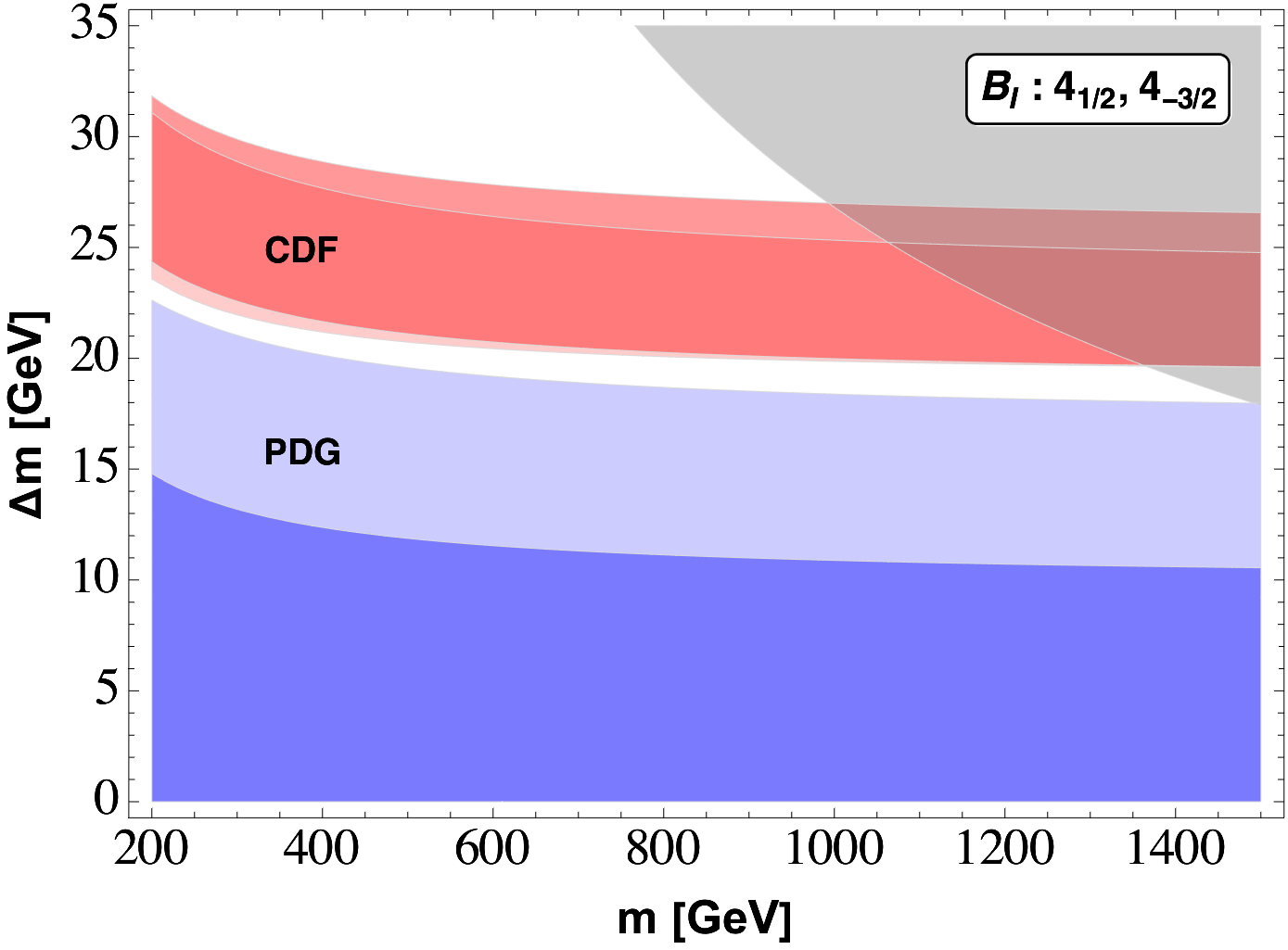}
\includegraphics[scale=0.36]{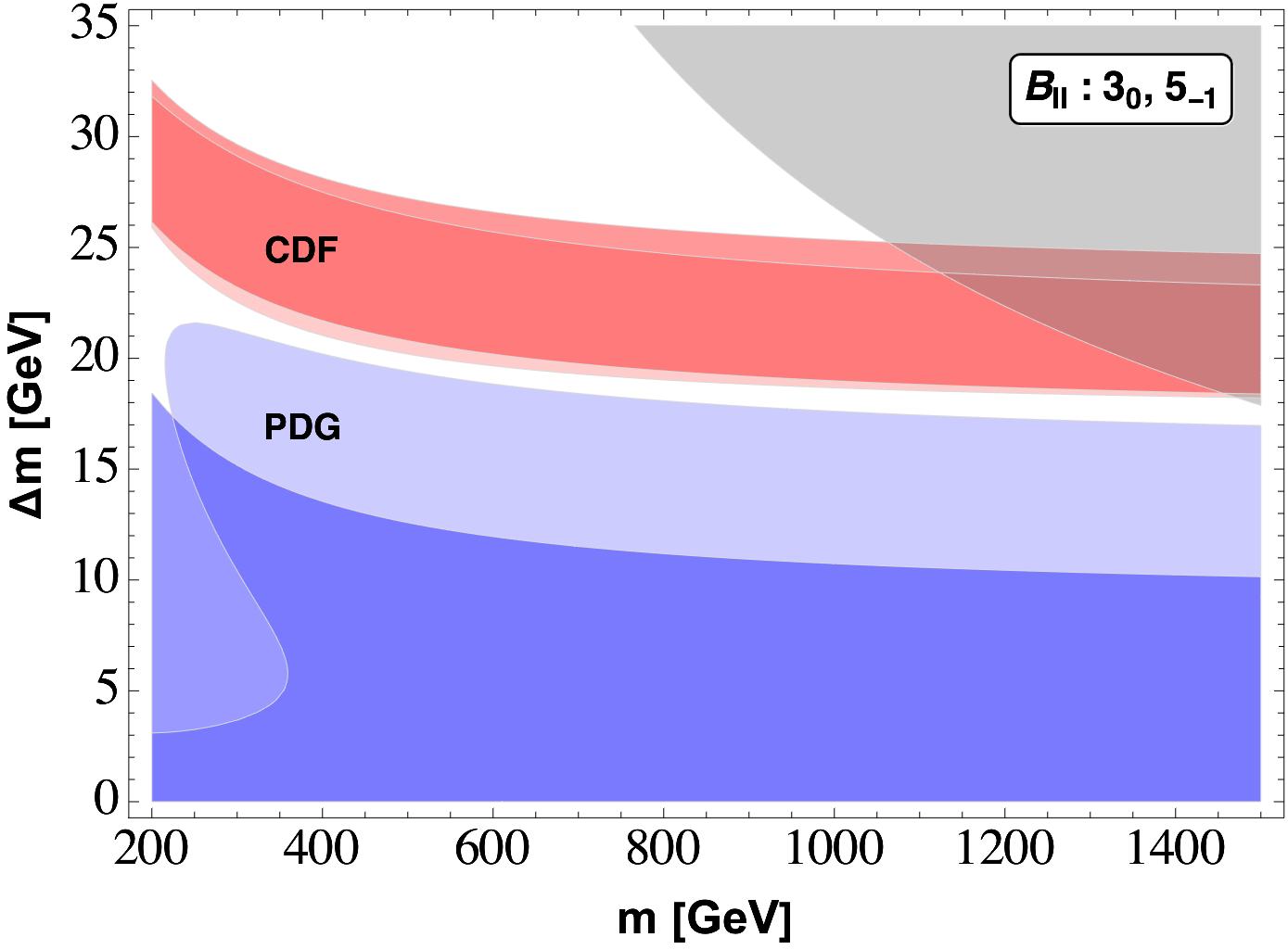}
\includegraphics[scale=0.36]{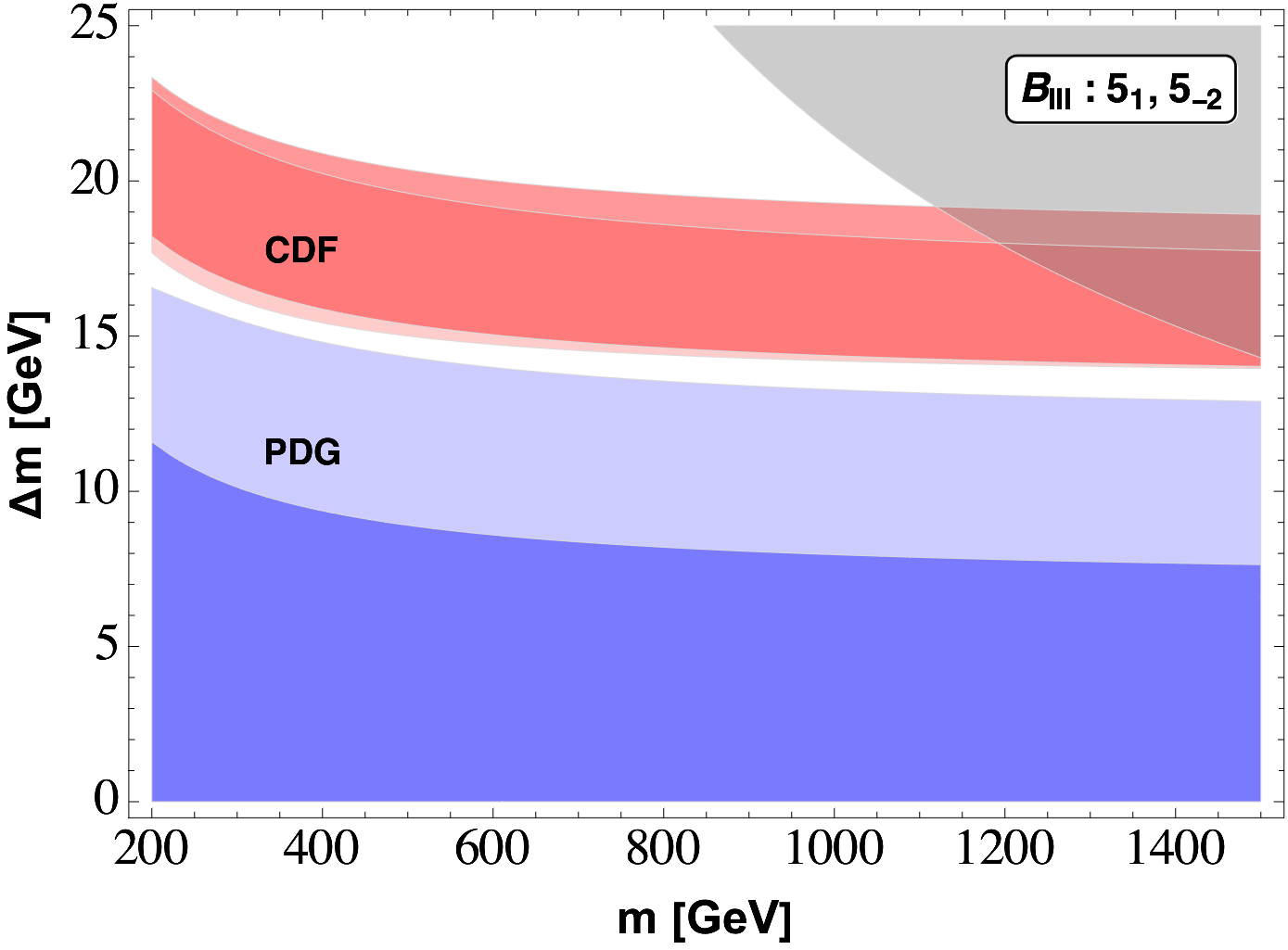}
\includegraphics[scale=0.36]{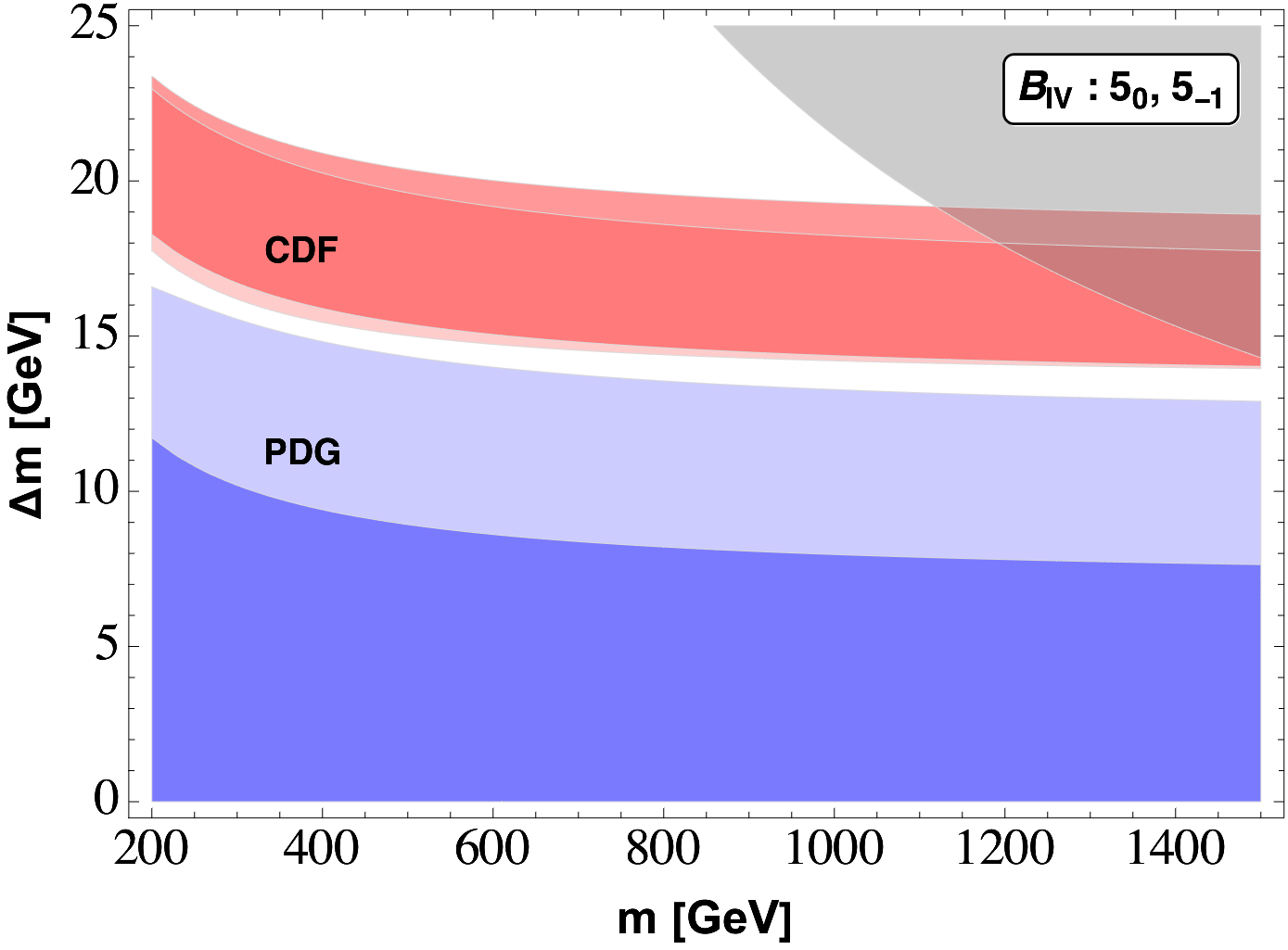}
\includegraphics[scale=0.36]{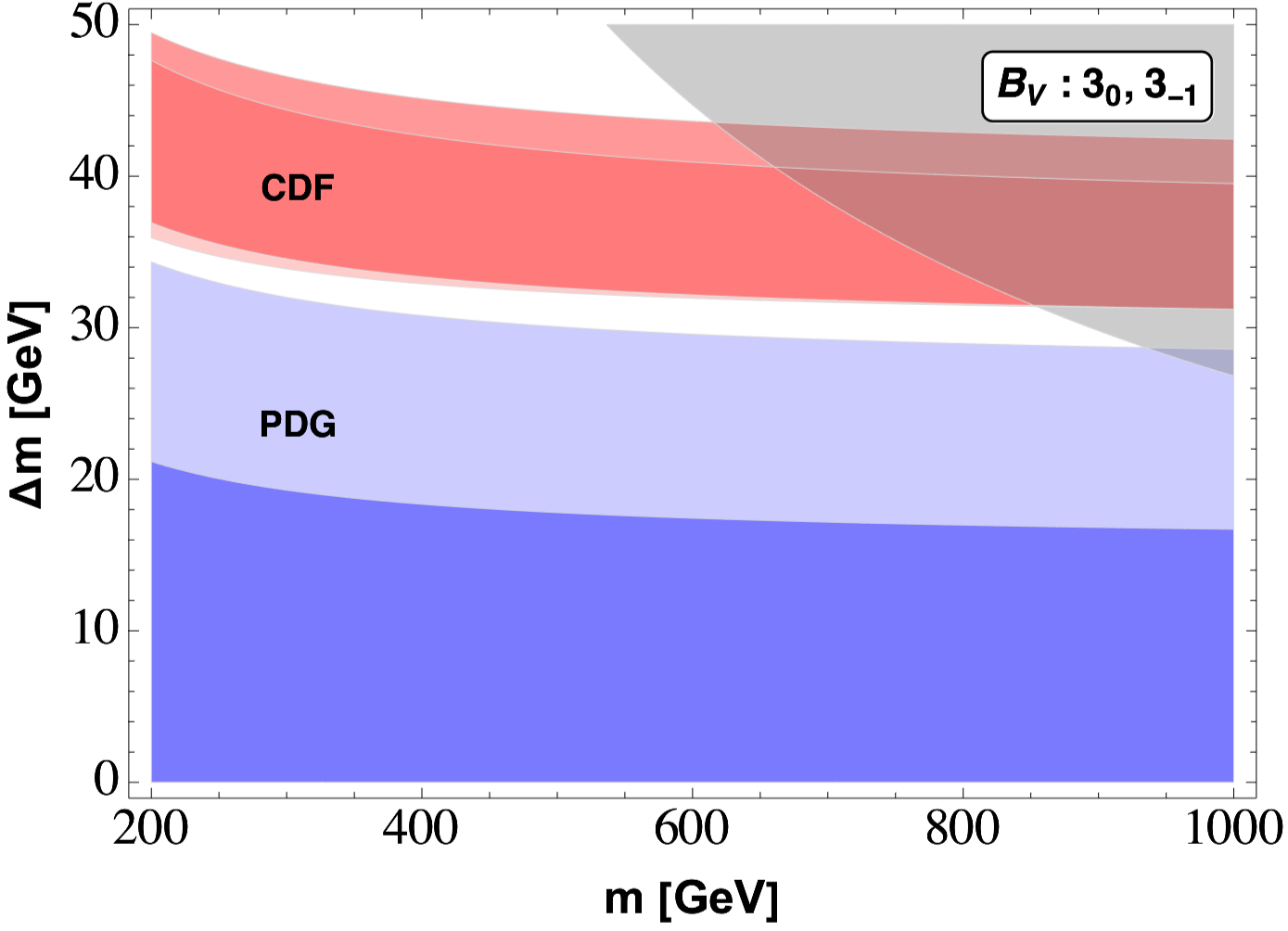}
\includegraphics[scale=0.36]{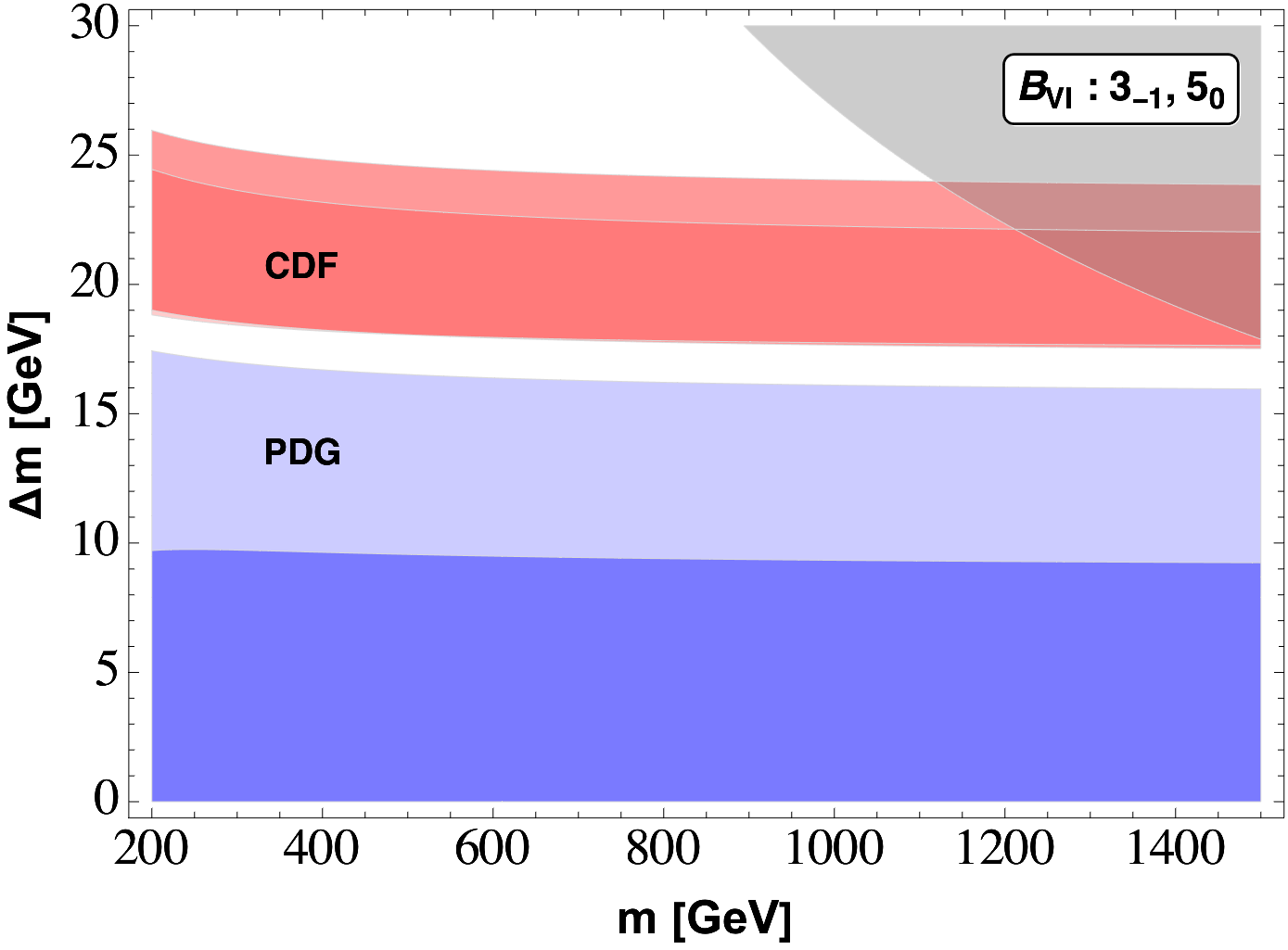}
\caption{Constraints on the scalar multiplets of the scenarios from EWPT, the experimental measurement of $M_W$ and perturbativity. The light blue (red) region indicates the parameter space allowed by EWPT from the PDG (CDF), whereas the dark red (blue) region shows the parameter space allowed by the new (old) measurement of $M_W$. The gray region is excluded from the perturbativity of the scalar couplings, i.e., $\lambda \leq \sqrt{4\pi}$.}
\label{fig:obfit}
\end{figure}

In scenarios where custodial symmetry is broken, there are complications in the calculation of oblique parameters at loop level. The $T$ parameter is divergent at one-loop level \cite{Gunion:1990dt},   whereas computing $S$ and $U$ at one loop leads to either a gauge-dependent result or they become divergent as the radiative corrections that one gets in the new framework are qualitatively very different from the SM ones \cite{Jegerlehner:1991ed,Albergaria:2021dmq}. However, since we constrain the VEVs so that the $\rho$ parameter is close to one, we get $v_i \ll v = 174 {\rm~GeV}$, as discussed in Section~\ref{sec:rhoSec}. Therefore, the corrections to the oblique parameters can be estimated by making use of the general formulas presented in Ref.~\cite{Lavoura:1993nq}, where the following assumptions are made: \textit{i)} the VEVs of the complex scalars are negligible, \textit{ii)} the scalars do not mix with themselves or any other scalars in the theory. Both assumptions are justified for our analysis as we work in the limit $v_i \ll v$, where $i$ labels the new scalar added to the SM. Further, we take $U=0$ as it is typically suppressed, which greatly improves the precision on $S$ and $T$.  The contribution of a scalar multiplet to the oblique parameters $S$ and $T$ is given by \cite{Lavoura:1993nq}:
\begin{align}\label{eq:oblpar}
T &= \frac{1}{16\pi c_W^2 s_W^2 M_Z}\,\sum_{I_3=-I}^{+I}(I^2-I_3^2+I+I_3)\theta_{+}(M_{I_3},M_{I_3 -1})\nonumber\,,\\
S &= -\frac{Y}{3\pi}\,\sum_{I_3=-I}^{+I} I_3 \ln\frac{M_{I_3}}{\mu^2}-\frac{2}{\pi}\,\sum_{I_3=-I}^{+I}(I_3 c_W^2-Y s_W^2)^2 \xi\left(\frac{M_{I_3}}{M_Z},\frac{M_{I_3}}{M_Z}\right)\nonumber\,,
\end{align}
where $\mu$ is a arbitrary mass parameter used in dimensional regularization, $Y$ is the hypercharge, $I$ is the weak isospin, $I_3$ is the third component of isospin for a multiplet with $(2I+1)$ components and $M_{I_3}$ denotes the mass of the scalar component corresponding to it, with
\begin{align}
\theta_+ (x,y) &= \frac{(y-x)^2}{3x}\,,\nonumber\\
\xi(x,y) &\simeq \frac{1}{15}\frac{1}{4x-1}-\left(\frac{2(y-x)}{15}+\frac{1}{21}\right)\left(\frac{1}{4x-1}\right)^2\,, 
\end{align}
for ${|{y-x}|}/{x}\ll 1$ \cite{Mandal:2022zmy}. 
We also discuss the implications of the addition of new scalar multiplets in alleviating the CDF anomaly related to the measurement of the $W$ boson mass\footnote{{A similar study to explain the W-boson mass anomaly in terms of one loop effect of a general $SU(2)_L$ scalar multiplet and scalar extensions violating custodial symmetry was done in Ref.~\cite{Wu:2022uwk,Wu:2023utc,Song:2022jns}.}}. The correction to $W$ mass in terms of these oblique parameters (taking $U=0$) is given by \cite{Maksymyk:1993zm}
\begin{equation}\label{eq:wmass}
m_W \simeq m_W^{\rm SM}\left[1-\frac{\alpha}{4(1-2s_W^2)}(S-2(1-s_W^2)T)\right]\,.
\end{equation}
The values of $S$ and $T$ and the correlation $\rho_{ST}$ from the current global fit of electroweak precision data using the old $M_W$ (PDG 2022 \cite{ParticleDataGroup:2022pth}) and new CDF value of $M_W$ \cite{Lu:2022bgw} are shown in Table~\ref{tab:obfit}.
\begin{table}[!htb]
\begin{center}
{\small
\begin{tabular}{|c|c|c| }
\hline
 & {\bf PDG 2022} &  {\bf CDF 2022}  \\
\hline\hline
$S$	& $-0.01\pm 0.07$ 	& $0.14\pm 0.08$ \\
\hline
$T$	& $0.04 \pm 0.06$ 	& $0.26 \pm 0.06$ 	\\
\hline
$\rho_{ST}$ & 0.92 & 0.93 \\
\hline 
\end{tabular}
\caption{Values of $S$, $T$ and the correlation $\rho_{ST}$ for $U=0$ allowed by the EW fit using the old and the new data from the PDG and the CDF Collaboration, respectively.}
\label{tab:obfit}
}
\end{center}
\end{table}
For the analysis below, we parameterize a scalar multiplet with $(2I+1)$ components as $\Phi=(\Phi_I, \Phi_{I-1}, \ldots, \Phi_{-I})^T$ and take the masses of the components as
\begin{equation}
M_{\Phi_{-I}}=m,\,M_{\Phi_{-I+1}}=m+\Delta m,\ldots, M_{\Phi_{I}}=m+2I\,\Delta m\,,
\end{equation}
assuming that the mass splittings among the components are equally spaced, with $\Delta m \ll m$. In scenarios of class-$\mathbf{B}$, where two new scalars are involved, for the ease of analysis we take the mass of the lightest component to be the same, $m$, and assume equal mass gaps for both multiplets, $\Delta m$. Thus, using Eqs.~\eqref{eq:oblpar} and \eqref{eq:wmass}, and the values given in Table~\ref{tab:obfit}, we can constrain the mass of the lightest component $m$ and the mass splitting $\Delta m$ by means of a two-parameter $\chi^2$ analysis. In Fig.~\ref{fig:obfit} we show the EWPT bounds (at 95\% C.L.) in the $m-\Delta m$ plane along with the bounds to reproduce the previous and the new measurement of the $W$ boson mass. The light blue (light red) region indicates the parameter space allowed by EWPT from the PDG (CDF measurement), whereas the dark blue (dark red) shows the parameter space allowed by the old (new) measurement of $M_W$. 
Note that the mass splitting among the components is proportional to the Higgs VEV and the quartic coupling $\lambda$. The mass gaps can then be parameterized as
\begin{equation}
    \Delta m \sim \mathcal{O}(0.1)\, \lambda\, \frac{v^2}{m}\,,
\end{equation}
where $\lambda \leq \sqrt{4\pi}$ is bounded by perturbativity. We show this bound in the top right gray region in Fig.~\ref{fig:obfit}. 

From the above estimates, it can be seen that multiplets belonging to higher representation require smaller mass splittings in order to be consistent with electroweak fits (EWPT and $m_W$). The allowed splittings are larger for the new CDF fit, as expected, since the new measurement indicates a larger value of $S$ and $T$ which is strongly disfavored by SM and is in tension with the PDG data. In class-$\mathbf{B}$ scenarios, the CDF data points to a mass splitting of around 40 - 50 GeV for $\mathbf{B_{V}}$, 20 - 30 GeV for $\mathbf{B_{I,II,VI}}$ and 15 - 25 GeV for $\mathbf{B_{III,IV}}$. Further, if the CDF data turns out to be correct, perturbativity excludes the new scalar masses to be larger than $\mathcal{O}(1){\rm~TeV}$.

\section{Conclusions} \label{sec:conc} 

The origin of neutrino masses remains an open question. In the SM, with just the Higgs doublet, the Weinberg operator seems a very plausible indication for the Majorana nature of neutrinos. However, if this is the mechanism by which neutrinos acquire their tiny masses, it will be very difficult to test it. On the other hand, new scalar multiplets may be present at energies not far from the EW scale, and they give rise to new Weinberg-like operators. If these new scalars take VEVs, necessarily suppressed ($\lesssim \mathcal{O}(1)$ GeV) due to constraints from EWPTs, neutrino masses may be generated for smaller LNV scales.

In this work we study the possible Weinberg-like operators with new electroweak scalars. We consider scenarios where neutrino masses induced by the SM Weinberg operator are suppressed. These requirements leave us with $2\,(6)$ possibilities with $1\, (2)$ new scalar multiplets up to quintuplet $SU(2)$ representations. We have found that, in some cases, naturally small induced VEVs are generated. Their UV completions lead to new seesaw scenarios, which are studied in a companion paper~\cite{Giarnetti:2023osf}, where we also study the one-loop contributions to neutrino masses and the phenomenology induced by the heavy fermions. 

By construction, the new scalars are expected to lie not far from the EW scale, i.e., at TeV energies. Therefore, they may be produced at colliders. If the new VEVs are small enough ($\lesssim \mathcal{O}(100)$ keV), the neutrino mass matrix may be reconstructed from doubly-charged decays. Let us also mention that in these scenarios the electroweak phase transition may be first order, which in turn could explain the baryon asymmetry of the universe via electroweak baryogenesis. Gravitational waves may provide a way to probe this scenario. A detailed study of this possibility would be interesting to pursue.

We conclude that, if a positive signal is observed at colliders and/or a deviation is measured in EWPTs, this may point to new scalars at the TeV scale, which may well be related to the origin of neutrino masses as analysed in this work.

\vspace{0.3cm}

\acknowledgments

We are grateful to Andrea Di Iura and Claudia Hagedorn for their participation in the early stages of this project. We also thank Anirban Karan, Emanuela Musumeci, Arcadi Santamaria and Avelino Vicente and for useful discussions. All Feynman
diagrams were generated using the Ti\textit{k}Z-Feynman package for
\LaTeX~\cite{Ellis:2016jkw}.

JHG and DV are partially supported by the ``Generalitat Valenciana'' through the GenT Excellence Program (CIDEGENT\slash 2020\slash 020) and by the Spanish ``Agencia Estatal de Investigación'', MICINN\slash AEI (10.13039\slash 501100011033) grants PID2020-113334GB-I00 and PID2020-113644GB-I00. JHG is also supported by the ``Consolidación Investigadora'' Grant CNS2022-135592 funded by the Spanish ``Agencia Estatal de Investigación'', MICINN\slash AEI (10.13039\slash 501100011033) and by ``European Union NextGenerationEU/PRTR''.

\appendix

\section{$n>5$ Weinberg-like operators with SM Higgs doublets}
\label{sec:d>5}
As long as $m^2_\Phi \gg v^2$, one can integrate-out the new multiplets and at low energies Weinberg-like operators with just Higgs doublets at dimension $n$ are generated \cite{Anamiati:2018cuq}, of the form
\begin{align}
\mathcal{L}^{(0)}_{n} &=\frac{{C_n^{(0)}}}{2}\, \mathcal{O}^{(0)}_{n} + {\rm H.c.} \nonumber\\
&=\frac{{C_n^{(0)}}}{2}\, (LH)_{\bf 1} (L H)_{\bf 1} (H^\dagger H)^{\frac{n-5}{2}} + {\rm H.c.}. \label{eq:WeinbergHiggs}
\end{align}
We see that the $n=5$ case corresponds to the dimension-$5$ Weinberg operator. After EWSB, these Weinberg-like operators generate neutrino masses at tree level. This can also be understood from the fact that new scalars acquire an induced VEV, as discussed in Section~\ref{sec:inducedvevs}. 

Let us first consider the class-\textbf{A} scenarios. The effective neutrino mass for $\mathbf{A_I}$ reads
\begin{equation}  \label{eq:A1Higgs}
    (m_{\nu})_{\alpha\beta}= \xi_1\, ({\tilde{C}_5^{(1)}})_{\alpha\beta}\,\lambda_{8} \frac{v^4}{2m_{\Phi_1}^2}+\xi_1^2\, ({C^{(2)}_5})_{\alpha\beta}\,\lambda_{8}^2 \frac{v^6}{2m_{\Phi_1}^4}\,,
\end{equation}
where $\xi_{1}$ is the numerical factor associated with the contraction of the fields (which we report in Table~\ref{tab:coefficients3}) in the potential term linear in $\Phi_1$ and $ {\tilde{C}_5^{(1)}}$ is the symmetric part of the Wilson coefficients matrix associated to $\mathcal{O}_5^{(1)}$, as mentioned below Eq.~{\ref{eq:mnueft}}. The mass dimensions of the two effective operators entering the neutrino masses in this case are $n=7$ and $n=9$, respectively. 

For the $\mathbf{A_{II}}$ scenario, on the other hand, the neutrino mass matrix reads 
\begin{equation} \label{eq:A2Higgs}
    (m_{\nu})_{\alpha\beta}= \xi_1\, ({C^{(2)}_5})_{\alpha\beta}\, \lambda_{6} \frac{v^4}{2m_{\Phi_1}^2} \,,
\end{equation}
that corresponds to $n=7$ operator.
\begin{table}[!htb]
\centering
\begin{tabular}{c|ccc|c|}
\cline{2-5}
                                      & \multicolumn{1}{c|}{$\xi_1$}       & \multicolumn{1}{c|}{$\xi_2$}   & $\xi_{12}$    & \textbf{n}            \\ \hline
                                      
\multicolumn{1}{|c|}{$\mathbf{A_I}$} & \multicolumn{1}{c|}{$1/\sqrt{3}$ } 
& \multicolumn{1}{c|}{\ding{56}}          &  \multicolumn{1}{c|}{\ding{56}}              & 7, 9        \\ \hline
\multicolumn{1}{|c|}{$\mathbf{A_{II}}$} & \multicolumn{1}{c|}{$-1$ } 
& \multicolumn{1}{c|}{\ding{56}}          &  \multicolumn{1}{c|}{\ding{56}}              & 7       \\ \hline
\multicolumn{1}{|c|}{$\mathbf{B_{I}}$} & \multicolumn{1}{c|}{$1/\sqrt{3}$}  & \multicolumn{1}{c|}{$-1$}      & $1/\sqrt{3}$  & 7, 9, 11, 13 \\ \hline
\multicolumn{1}{|c|}{$\mathbf{B_{II}}$} & \multicolumn{1}{c|}{$-1/\sqrt{2}$} & \multicolumn{1}{c|}{\ding{56}} & $-1/\sqrt{2}$ & 9            \\ \hline
\multicolumn{1}{|c|}{$\mathbf{B_{III}}$} & \multicolumn{1}{c|}{\ding{56}}     & \multicolumn{1}{c|}{\ding{56}} & $-1/2$        & 7            \\ \hline
\multicolumn{1}{|c|}{$\mathbf{B_{IV}}$} & \multicolumn{1}{c|}{\ding{56}}     & \multicolumn{1}{c|}{\ding{56}} & $-\sqrt{3/8}$ & 7            \\ \hline
\multicolumn{1}{|c|}{$\mathbf{B_{V}}$} & \multicolumn{1}{c|}{$-1/\sqrt{2}$ } 
& \multicolumn{1}{c|}{$1$}          &  \multicolumn{1}{c|}{$-1/\sqrt{2}$}              & 9, 11        \\ \hline
\multicolumn{1}{|c|}{$\mathbf{B_{VI}}$} & \multicolumn{1}{c|}{$-1$}              & \multicolumn{1}{c|}{\ding{56}} &              \multicolumn{1}{c|}{$1/\sqrt{6}$} & 9            \\ \hline
\end{tabular}
\caption{\label{tab:coefficients3}List of coefficients from the contractions of potential terms which induce the small BSM VEVs in class-{\bf A} and {\bf B} scenarios. The last column shows the dimensions of the higher order operators allowing for small neutrino masses.}
\end{table}

\begin{figure}[!htb]\centering
    \begin{tikzpicture}[scale=0.7, transform shape,every text node part/.style={align=center}]
        \begin{feynman}
          \node [blob,draw=black, pattern color=black] (a);
            \node [above left=1.5cmof a] [dot] (b);
            \node [below = 0.7cmof b] (bb);
            \node [right=0.3 cmof bb] (name1) {\Large{$\Phi_1$}};
            \vertex [below left=2.5cmof a] (c) {\Large{$\nu_L$}};
            \vertex [above=1.7cmof b] (2l) {\Large{$\braket{H}$}};
            \vertex [above left=1.7cmof b] (1l) {\Large{$\braket{H}$}};
            \vertex [left=1.7cmof b] (3l) {\Large{$\braket{H}$}};
            \node [above right=1.5 cmof a][dot] (e);
            \node [below = 0.7cmof e] (ee);
            \node [left=0.3cmof ee] (name2) {\Large{$\Phi_2$}};
            \vertex [above=1.7cmof e] (2r) {\Large{$\braket{H}$}};
            \vertex [above right=1.7cmof e] (1r) {\Large{$\braket{H}$}};
            \vertex [right=1.7cmof e] (3r) {\Large{$\braket{H}$}};
            \vertex [below right=2.5cmof a] (f) {\Large{$\nu_L^c$}};
            \diagram* {
                (c) -- [fermion,thick] (a);
                (a) -- [scalar,thick] (b),
                (b) -- [scalar,thick, insertion=1] (1l),
                (b) -- [scalar,thick, insertion=1] (2l),
                (b) -- [scalar,thick, insertion=1] (3l),
                (a) -- [scalar,thick] (e),
                (e) -- [scalar,thick, insertion=1] (1r),
                (e) -- [scalar,thick, insertion=1] (2r),
                (e) -- [scalar,thick, insertion=1] (3r),
                (f) -- [anti fermion,thick] (a);
            };
            \end{feynman}
    \end{tikzpicture}
    \hspace{0.8cm}
    \begin{tikzpicture}[scale=0.7, transform shape,every text node part/.style={align=center}]
        \begin{feynman}
          \node [blob,draw=black, pattern color=black] (a);
            \node [above left=1.5cmof a] [dot] (b);
            \node [below = 0.7cmof b] (bb);
            \node [right=0.3 cmof bb] (name1) {\Large{$\Phi_1$}};
            \vertex [below left=2.5cmof a] (c) {\Large{$\nu_L$}};
            \vertex [above=1.7cmof b] (2l) {\Large{$\braket{H}$}};
            \vertex [above left=1.7cmof b] (1l) {\Large{$\braket{H}$}};
            \vertex [left=1.7cmof b] (3l) {\Large{$\braket{H}$}};
            \node [right=1.5 cmof a] (eold);
            \node [above=0.6cmof eold][dot](e);
            \node [below = 0.06cmof eold] (ee);
            \node [left=0.45cmof ee] (name2) {\Large{$\Phi_2$}};
            \vertex [right=1.7cmof e] (2r) {\Large{$\braket{H}$}};
            \vertex [below right=1.7cmof e] (1r) {\Large{$\braket{H}$}};
            \node [above =1.3cmof e] (fold);
            \node [left =0.2cmof fold][dot] (f);
            \node [left=0.35cmof f] (ff);
            \node [below=0.65cmof ff] (name3) {\Large{$\Phi_1$}};
            \vertex [above=1.7cmof f] (1u) {\Large{$\braket{H}$}};
            \vertex [right=1.7cmof 1u] (2u) {\Large{$\braket{H}$}};
            \vertex [left=1.7cmof 1u] (3u) {\Large{$\braket{H}$}};
            \vertex [below right=2.5cmof a] (g) {\Large{$\nu_L^c$}};
            \diagram* {
                (c) -- [fermion,thick] (a);
                (a) -- [scalar,thick] (b),
                (b) -- [scalar,thick, insertion=1] (1l),
                (b) -- [scalar,thick, insertion=1] (2l),
                (b) -- [scalar,thick, insertion=1] (3l),
                (a) -- [scalar,thick] (e),
                (e) --[scalar,thick,insertion=1] (1r),
                (e) --[scalar,thick,insertion=1] (2r),
                (e) --[scalar,thick] (f),
                (f) -- [scalar,thick, insertion=1] (1u),
                (f) -- [scalar,thick, insertion=1] (2u),
                (f) -- [scalar,thick, insertion=1] (3u),
                (g) -- [anti fermion,thick] (a);
            };
            \end{feynman}
    \end{tikzpicture}
    \caption{Example of neutrino mass generation due to induced VEVs. We show the case of ${\bf II_b}$ scenario, where neutrino masses are generated at tree level via dimension $n=9$ (\textit{left}) and $n=11$ (\textit{right}) operators.}
    \label{fig:dim>5Diagrams}
    \end{figure}
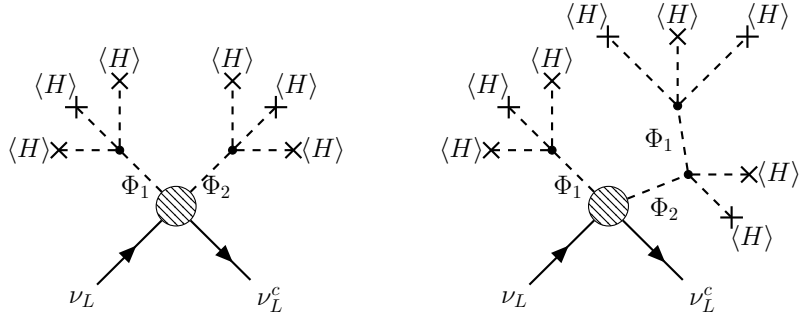

Let us now move to the $\mathbf{B}$ scenarios. The $\mathbf{B_{I}}$ scenario is the most complicated one, since with two quadruplets we can have three types of BSM Weinberg-like operators, namely $\mathcal{O}_5^{(1)}$, $\mathcal{O}_5^{(2)}$ and $\mathcal{O}_5^{(3)}$. Moreover, just like the previous case, we can induce both VEVs using the SM Higgs via quartic terms, or we can induce one of the two with the mixed term. If both new VEVs are induced by the Higgs, then neutrino masses read
\begin{eqnarray} \label{eq:B1Higgs1}
\nonumber    (m_{\nu})_{\alpha\beta}&=&\xi_1 \left([ { \tilde{C}^{(1)}_5}]_1\right)_{\alpha\beta} \lambda_{5} \frac{v^4}{2 m_{\Phi_1}^2}+\xi_2\, \left([ { \tilde{C}^{(1)}_5}]_2\right)_{\alpha\beta} \lambda_{6} \frac{v^4}{2 m_{\Phi_2}^2}+\\ 
    &+&\xi_1^2\left([ { C^{(2)}_5}]_1\right)_{\alpha\beta} \lambda_{5}^2 \frac{v^6}{4 m_{\Phi_2}^4}+\xi_1\xi_2\,( { \tilde{C}^{(3)}_5})_{\alpha\beta} \lambda_{5}\lambda_{6} \frac{v^6}{4 m_{\Phi_1}^2m_{\Phi_2}^2}\, ,
\end{eqnarray}
where it is clear that the Weinberg operator $\Phi_2\Phi_2HH$ does not exist due to the -3/2 hypercharge of the second quadruplet. The effective dimension $n$ of the four operators are $n=7$ for the first two operators, and $n=11$ for the third and fourth one.
If only $v_1$ is induced by the SM Higgs VEV while $v_2$ is induced via the mixing, neutrino masses read
\begin{eqnarray} \label{eq:B1Higgs2}
  \nonumber  (m_{\nu})_{\alpha\beta}&=&\xi_1 \left([ { \tilde{C}^{(1)}_5}]_1\right)_{\alpha\beta} \lambda_{5} \frac{v^4}{2 m_{\Phi_1}^2}+\xi_{12}\xi_1 \left([ { \tilde{C}^{(1)}_5}]_2\right)_{\alpha\beta} \lambda_{1} \lambda_{5} \frac{v^6}{4 m_{\Phi_1}^2m_{\Phi_2}^2}+\\ 
    &+&\xi_1^2\left([ { C^{(2)}_5}]_1\right)_{\alpha\beta} \lambda_{5}^2 \frac{v^6}{4m_{\Phi_2}^4}+\xi_{12}\xi_1^2\,( { \tilde{C}^{(3)}_5})_{\alpha\beta} \lambda_{5}^2\lambda_{1} \frac{v^8}{8 m_{\Phi_1}^4 m_{\Phi_2}^2}\, ,
\end{eqnarray}
where the effective dimension is $n=7$ for the first operator, $n=9$ for the second and third operators and $n=11$ for the last one.
On the other hand, if only $v_2$ is induced by the Higgs VEV we have a more suppressed contribution from the third term and neutrino masses are
\begin{eqnarray} \label{eq:B1Higgs3}
  \nonumber  (m_{\nu})_{\alpha\beta}&=&\xi_{12}\xi_2^2 \left([ { \tilde{C}^{(1)}_5}]_1\right)_{\alpha\beta} \lambda_{1}\lambda_{6} \frac{v^6}{4m_{\Phi_1}^2 m_{\Phi_2}^2}+\xi_2 \left([ { \tilde{C}^{(1)}_5}]_2\right)_{\alpha\beta} \lambda_{6} \frac{v^4}{2m_{\Phi_2}^2}+\\ 
    &+&\xi_{12}^2\xi_2^2\left([ { C^{(2)}_5}]_1\right)_{\alpha\beta} \lambda_{1}^2 \lambda_{6}^2\frac{v^{10}}{16m_{\Phi_2}^4 m_{\Phi_1}^4}+\\ \nonumber
    &+&\xi_{12}\xi_2^2\,( { \tilde{C}^{(3)}_5})_{\alpha\beta} \lambda_{6}^2\lambda_{1} \frac{v^8}{8m_{\Phi_1}^2 m_{\Phi_2}^3}\, .
\end{eqnarray}
Here the neutrino masses are obtained, in order, from $n=9$,  $n=7$, $n=13$ and $n=11$ effective operators. Depending on the value of the Wilson coefficients $ { C^{(i)}_5}$, the dominant contribution can come from different terms. As an example, we show the Feynman diagram related to $n=9$ (left) and $n=11$ (right) operators in $\mathbf{B_{I}}$ in Fig. \ref{fig:dim>5Diagrams}. 

Let us now consider the $\mathbf{B_{II}}$ and $\mathbf{B_{VI}}$ scenarios. In this case we have only one possibility, namely that $v_1$ is induced by the SM Higgs while $v_2$, coming from a quintuplet, can only be induced by the mixed term. We have $n=9$ operator from which neutrino masses can be written as
\begin{equation} \label{eq:B2Higgs}
    (m_{\nu})_{\alpha\beta}=\xi_{12}\xi_1^2\,  ( { \tilde{C}^{(3)}_5})_{\alpha\beta}\,\mu_{2}^2 \lambda_1 \, \frac{v^6}{8m_{\Phi_1}^4 m_{\Phi_2}^2}\,.
\end{equation}

For the scenarios, $\mathbf{B_{III}}$ and $\mathbf{B_{IV}}$, as discussed in the previous section, there does not exist a mechanism which ensures that both the VEVs are naturally small. However, neutrino masses may still arise from $n=7$ operator, for instance, $v_2$ induced by $v_1$ (i.e., $v_2\ll v_1$),
\begin{equation} \label{eq:B3Higgs}
    (m_{\nu})_{\alpha\beta}=\xi_{12}\,  \left( { \tilde{C}^{(3)}_5}\right)_{\alpha\beta} \lambda_{1} \frac{v^2}{ 2 m_{\Phi_2}^2}v_1^2\,.
\end{equation}
In the opposite case the neutrino mass matrix can be obtained with the substitutions $v_1\to v_2$ and $m_{\Phi_2}\to m_{\Phi_1}$.

Finally, for the $\mathbf{B_{V}}$, neutrino masses could arise from a $n=9$ or $n=11$ operator, depending on how the new multiplets VEVs are induced. If both of the VEVs are induced from SM Higgs one, we obtain a dimension-$9$ operator, which gives the following neutrino mass
\begin{equation}
    (m_{\nu})_{\alpha\beta}= \xi_1\xi_2\, ({\tilde{C}^{(3)}_5})_{\alpha\beta}\, \mu_{2}\mu_{3} \frac{v^4}{4 m_{\Phi_1}^2 m_{\Phi_2}^2} \,,
\end{equation}
where $\tilde C_5^{(3)}$ is the symmetric part of the Wilson coefficients matrix associated to $\mathcal{O}_5^{(3)}$, $\xi_1$ and $\xi_2$ are the numerical factors (see Table~\ref{tab:coefficients3}) related to the contractions of the $\Phi_1 H^2$ and $\Phi_2 H^2$ terms, respectively. On the other hand, if one of the two VEV is induced by the SM Higgs while the second one is induced by the mixed term $\Phi_1\Phi_2 H^2$, neutrino masses arise from a dimension-$11$ operator and they read
\begin{equation}
    (m_{\nu})_{\alpha\beta}= \xi_1^2\xi_{12}\, ({\tilde{C}^{(3)}_5})_{\alpha\beta}\, \mu_{2}^2\lambda_{1} \frac{v^6}{8 m_{\Phi_1}^4 m_{\Phi_2}^2} \,,
    \label{massIIadim11}
\end{equation}
where now $\xi_{12}$ is the numerical factor related to the contraction of the mixed potential term. In this case we expect $v_1\gg v_2$. If the second scalar multiplet VEV $v_2$ is the one induced directly by the Higgs, neutrino masses are obtained from Eq.~\eqref{massIIadim11} with the substitutions $\xi_1\to\xi_2$, $\mu_{2}\to\mu_{3}$ and $m_{\Phi_1}\leftrightarrow m_{\Phi_2}$.

All the numerical coefficients arising from the contraction of the potential terms that induce the small VEVs are reported, as already mentioned, in 
Table~\ref{tab:coefficients3} along with the mass dimension $n$ at which the neutrino masses are generated.

 \section{Tensor contractions}
 \label{sec:contractions}

We define here in tensor notation the lepton number violating potential terms  in $\mathbf{B_{I}}$ scenario (see Table~\ref{tab:potentials}), calling for simplicity $\Phi_1\equiv\Phi$ and $\Phi_2\equiv\Delta$
\begin{equation}
    \begin{aligned} \label{eq:tensorB1}
         \Delta^\ast\Phi^\ast H\,\Delta=\Delta^{\ast ijk}\Phi^{\ast pqr}H_i \Delta_{kqr}\epsilon_{jp}\,,\quad& (\Delta^\ast\Phi^\ast H\,\Delta)^\prime=\Delta^{\ast ijk}\Phi^{\ast pqr} H_s\,\Delta_{tkr}\epsilon^{st}\epsilon_{ip}\epsilon_{jq}\\ \Phi^\ast\Phi^\ast HH=\Phi^{\ast ijk}\Phi^{\ast pqr}H_i H_p\epsilon_{jq}\epsilon_{kr}\,,\quad& 
         \Phi^{\ast}\Phi^{\ast}H\Phi=\Phi^{\ast ijk}\Phi^{\ast pqr}H_i \Phi_{kqr}\epsilon_{jp}\\
         HHH\Delta=\epsilon^{ij}\epsilon^{kp}\epsilon^{lm}H_iH_kH_l\Delta_{jpm}\,,\quad &HH\Phi\Delta=H_iH_j\Phi_{lmn}\Delta_{pqr}\epsilon^{il}\epsilon^{jp}\epsilon^{mq}\epsilon^{nr}\\
         H\Phi\Phi\Delta=H_i\Phi_{jkl}\Phi_{mnp}\Delta_{qrs}\epsilon^{ij}\epsilon^{km}\epsilon^{lq}\epsilon^{nr}\epsilon^{ps}\,,\quad&
         \Phi^\ast H^\ast H\,H=\Phi^{\ast ijk}H^p H_i H_j \epsilon_{kp}
         \\
         \Phi\Phi\Phi\Delta=\Phi_{ijk}\Phi_{pqr}\Phi_{stu}\Delta_{mno}&\epsilon^{ip}\epsilon^{jq}\epsilon^{kr}\epsilon^{sm}\epsilon^{tn}\epsilon^{uo}\,.
    \end{aligned}
\end{equation}

\bibliographystyle{JHEP}
\bibliography{Weinbergs}
\end{document}